\documentclass[11pt]{article}
\usepackage[utf8]{inputenc}
\usepackage[T1]{fontenc}
\usepackage{lmodern}
\usepackage{amsmath,amssymb,amsthm,mathtools}
\usepackage[letterpaper,margin=1in]{geometry}
\usepackage{booktabs,array,tabularx}
\usepackage{float}
\usepackage{xcolor,tikz}
\usepackage[section]{placeins}
\usepackage{enumitem}
\usepackage{microtype}
\usepackage{parskip}
\usepackage{titlesec}
\usepackage[numbers]{natbib}
\usepackage{needspace}
\usetikzlibrary{arrows.meta,positioning,calc,patterns,patterns.meta,fit,backgrounds}
\usepackage{etoolbox}
\usepackage{hyperref}
\usepackage[capitalise,nameinlink,noabbrev]{cleveref}

\hypersetup{breaklinks=true,bookmarksnumbered=true,bookmarksdepth=2,colorlinks=true,linkcolor=blue,citecolor=blue,urlcolor=blue,pdfborder={0 0 0},
	pdftitle={Bounded Knapsack in Truly Subquadratic Time},
	pdfauthor={Huanyi Zhou},pdfsubject={Algorithms and fine-grained complexity}}
\titlespacing*{\section}{0pt}{1.5\baselineskip}{1\baselineskip}
\titlespacing*{\subsection}{0pt}{1.1\baselineskip}{.5\baselineskip}
\setlist[itemize]{leftmargin=2em,itemsep=.6em}
\makeatletter
\patchcmd{\l@section}{1.0em}{.4em}{}{\PackageError{manuscript}{Contents spacing patch failed}{}}
\makeatother
\allowdisplaybreaks[2]
\newcommand{\wm}{w_{\max}}
\newcommand{\pmx}{p_{\max}}
\newcommand{\Ot}{\widetilde O}
\newcommand{\Z}{\mathbb Z}
\newcommand{\R}{\mathbb R}
\newcommand{\Q}{\mathbb Q}

\newcommand{\OPT}{\mathrm{OPT}}
\newcommand{\MinConv}{\textnormal{\scshape MinConv}}

\newcommand{\conv}{\operatorname{conv}}

\newcommand{\vol}{\operatorname{vol}}
\newcommand{\rank}{\operatorname{rank}}

\theoremstyle{plain}
\newtheorem{theorem}{Theorem}[section]
\newtheorem{lemma}[theorem]{Lemma}

\newtheorem{corollary}[theorem]{Corollary}

\theoremstyle{definition}
\newtheorem{definition}[theorem]{Definition}
\newtheorem{remark}[theorem]{Remark}
\newtheorem{example}[theorem]{Example}
\makeatletter
\newcommand{\bkalgcaption}[2]{%
	\leftskip=8.5pt\relax
	\rightskip=8.5pt plus 1fil\relax
	\noindent\textbf{#1:}\enspace#2\par
}
\newcommand{\fs@bkalg}{%
	\let\@fs@capt\bkalgcaption
	\def\@fs@pre{\hrule height.8pt\kern1.5pt}%
	\def\@fs@mid{\kern2pt\hrule height.65pt\kern1pt}%
	\def\@fs@post{\kern4pt\hrule height.8pt}%
	\let\@fs@iftopcapt\iftrue
}
\makeatother
\floatstyle{bkalg}
\newfloat{algorithm}{tbp}{loa}
\floatstyle{plain}
\floatname{algorithm}{Algorithm}
\crefname{algorithm}{Algorithm}{Algorithms}
\newcounter{bkalgline}
\newcount\bkalgdepth
\newsavebox{\bkalgbody}
\newsavebox{\bkalgblock}
\newlength{\bkalgindent}
\newcommand{\bkalgopen}{%
	\par
	\setbox\bkalgblock=\vtop\bgroup
	\advance\hsize by -\bkalgindent
	\advance\bkalgdepth by 1\relax
	\ignorespaces
}
\newcommand{\bkalgclose}{%
	\par\kern1pt\egroup
	\hbox{%
		\kern4.5pt
		\rlap{\lower\dp\bkalgblock\vbox to \dimexpr\ht\bkalgblock+\dp\bkalgblock\relax
			{\vfil\hrule width6.5pt height.6pt}}%
		\vrule width.65pt height\ht\bkalgblock depth\dp\bkalgblock
		\kern\dimexpr\bkalgindent-5.15pt\relax
		\box\bkalgblock
	}%
	\prevdepth=3pt\relax
	\ignorespaces
}
\newenvironment{algorithmic}[1][0]{%
	\setcounter{bkalgline}{0}%
	\setbox\bkalgbody=\vbox\bgroup
	\normalsize
	\advance\hsize by -25.5pt
	\parindent=0pt\relax
	\parskip=0pt\relax
	\leftskip=0pt\relax
	\rightskip=0pt plus 1fil\relax
	\emergencystretch=.75em\relax
	\bkalgdepth=0\relax
	\ignorespaces
}{%
	\par\egroup
	\nointerlineskip\hbox{\kern17pt\box\bkalgbody}%
}
\newcommand{\bkalgnextline}{\par\hangindent=.5em\relax\noindent}
\newcommand{\State}{%
	\bkalgnextline
	\stepcounter{bkalgline}%
	\llap{\footnotesize\bfseries\arabic{bkalgline}%
		\kern\dimexpr5.5pt+\bkalgindent*\bkalgdepth\relax}%
	\ignorespaces
}
\newcommand{\Statex}{\bkalgnextline\ignorespaces}
\newcommand{\Require}{\Statex\textbf{Input:}\enspace\ignorespaces}
\newcommand{\Ensure}{\Statex\textbf{Output:}\enspace\ignorespaces}
\newcommand{\Begin}{\State\textbf{begin}\bkalgopen}
\newcommand{\End}{\bkalgclose}
\newcommand{\If}[1]{\State\textbf{if} #1 \textbf{then}\bkalgopen}
\newcommand{\ElsIf}[1]{\bkalgclose\State\textbf{else if} #1 \textbf{then}\bkalgopen}
\newcommand{\Else}{\bkalgclose\State\textbf{else}\bkalgopen}
\newcommand{\EndIf}{\bkalgclose}

\newcommand{\ForAll}[1]{\State\textbf{for each} #1 \textbf{do}\bkalgopen}
\newcommand{\EndFor}{\bkalgclose}
\newcommand{\While}[1]{\State\textbf{while} #1 \textbf{do}\bkalgopen}
\newcommand{\EndWhile}{\bkalgclose}
\newcommand{\Return}{\textbf{return}\enspace}
\newcommand{\Call}[2]{\textsc{#1}\ensuremath{(#2)}}

\usetikzlibrary{arrows.meta,calc,positioning,decorations.pathreplacing}

\definecolor{bkFigInk}{HTML}{243647}
\definecolor{bkFigBlue}{HTML}{245D7A}
\definecolor{bkFigPale}{HTML}{E9F2F7}
\definecolor{bkFigAmber}{HTML}{875112}
\definecolor{bkFigCream}{HTML}{FBF2DF}
\definecolor{bkFigGrid}{HTML}{CDD5DB}
\definecolor{bkFigGuide}{HTML}{8495A4}
\definecolor{bkFigFog}{HTML}{F2F5F7}

\tikzset{
	bkfig/.style={font=\small, text=bkFigInk, draw=bkFigInk,
		line width=.55pt, >=Latex},
	bkpanel/.style={font=\small\bfseries, anchor=west},
	bkstep/.style={draw=bkFigInk, fill=white, rounded corners=1.4pt,
		inner xsep=2pt, inner ysep=3pt, minimum height=9mm, align=center},
	bkedge/.style={-{Latex[length=1.65mm]}, draw=bkFigInk},
	bkflowlabel/.style={font=\footnotesize, align=center},
	bkdot/.style={circle, fill=bkFigBlue, inner sep=1.25pt},
}

\newcommand{\ClipFigure}{%
	\begin{figure}[H]
		\centering
		\begin{tikzpicture}[bkfig,x=.68cm,y=1.11cm]
			\node[bkpanel] at (-4,1.60) {(a) Coefficient box};
			\node[bkpanel] at (8,1.60) {(b) Physical crop};
			\draw[fill=bkFigCream,draw=bkFigAmber,densely dashed] (-4,-1) rectangle (4,1);
			\draw[fill=bkFigPale,draw=bkFigBlue,line width=.75pt] (-4,-.1) rectangle (4,.1);
			\foreach \j in {-4,...,4}{
				\foreach \k in {-1,1}
				\fill[bkFigAmber] (\j,\k) circle[radius=1.10pt];
				\fill[bkFigBlue] (\j,0) circle[radius=1.32pt];
			}
			\node[bkflowlabel,text=bkFigAmber] at (0,.58) {27 lattice sites};
			\node[bkflowlabel,text=bkFigBlue,anchor=north] at (0,-1.17)
			{Blue band \(R=[-4,4]\times[-\varepsilon,\varepsilon]\)};
			\draw[bkedge] (4.48,0) -- node[bkflowlabel,above=2pt] {crop} (7.52,0);
			\begin{scope}[shift={(12,0)}]
				\draw[draw=bkFigGuide,densely dashed] (-4,-1) rectangle (4,1);
				\draw[fill=bkFigPale,draw=bkFigBlue,line width=.75pt] (-4,-.1) rectangle (4,.1);
				\foreach \j in {-4,...,4}{
					\foreach \k in {-1,1}{
						\draw[draw=bkFigAmber,line width=.55pt]
						($(\j,\k)+(-1.22pt,-1.22pt)$) -- ($(\j,\k)+(1.22pt,1.22pt)$)
						($(\j,\k)+(-1.22pt,1.22pt)$) -- ($(\j,\k)+(1.22pt,-1.22pt)$);
					}
					\fill[bkFigBlue] (\j,0) circle[radius=1.32pt];
				}
				\node[bkflowlabel,text=bkFigBlue] at (0,.58) {9 eligible lattice sites};
				\node[bkflowlabel,text=bkFigBlue,anchor=north] at (0,-1.17)
				{Retain witnessed entries in $R$};
			\end{scope}
		\end{tikzpicture}
		\caption{Geometric cropping on \(\mathbb Z^2\) with \(\varepsilon=1/10\).
			The coefficient box contains 27 lattice sites, and the physical window
			\(R\) contains nine. The box is \(1/\varepsilon\) times as tall as \(R\).
			Only witnessed entries in \(R\) pass to query-structure construction.}
		\label{fig:clip}
	\end{figure}%
}

\newcommand{\DispatchFigure}{%
	\begin{figure}[H]
		\centering
		\begin{tikzpicture}[bkfig,x=1cm,y=1cm]
			\tikzset{
				bkDispatchStep/.style={bkstep,text width=3.42cm,inner xsep=3pt,
					inner ysep=3pt,minimum height=13mm},
				bkDispatchLabel/.style={bkflowlabel,fill=white,inner sep=2pt}
			}
			\node[bkDispatchStep,fill=bkFigPale,text width=4.1cm]
			(batch) at (5.9,0) {\textbf{Stored batch}\\One pool, merged queries};
			\node[bkDispatchStep,fill=bkFigPale,text width=5.24cm]
			(partition) at (7.88,-2.03)
			{\textbf{Split by target}\\Group by face and mark anchors};
			\node[bkDispatchStep] (terminal) at (0,-4.25)
			{\textbf{Terminal test}\\Test the zero target};
			\node[bkDispatchStep] (face) at (3.94,-4.25)
			{\textbf{Face restriction}\\Fix forced endpoints};
			\node[bkDispatchStep] (thin) at (7.88,-4.25)
			{\textbf{Thin anchor}\\Branch on endpoints};
			\node[bkDispatchStep] (deep) at (11.82,-4.25)
			{\textbf{Deep anchor}\\Test the trial pool};
			\coordinate (first) at (5.9,-.90);
			\draw (batch.south) -- (first);
			\draw[bkedge] (first) -| (terminal.north);
			\draw[bkedge] (first) -| (partition.north);
			\fill[bkFigInk] (first) circle[radius=.7pt];
			\node[bkDispatchLabel,anchor=south] at (2.75,-.90) {Rank zero};
			\node[bkDispatchLabel,anchor=west] at (7.98,-1.13) {Positive rank};
			\coordinate (second) at (7.88,-3.16);
			\draw (partition.south) -- (second);
			\draw[bkedge] (second) -| (face.north);
			\draw[bkedge] (second) -- (thin.north);
			\draw[bkedge] (second) -| (deep.north);
			\fill[bkFigInk] (second) circle[radius=.7pt];
			\node[bkDispatchLabel,anchor=south,inner sep=1pt] at (5.80,-3.16) {Boundary targets};
			\node[bkDispatchLabel,anchor=south,inner sep=1pt] at (9.94,-3.16) {Interior anchors};
		\end{tikzpicture}
		\caption{Dispatching a stored batch. At rank zero the zero target is
			tested directly. At positive rank, a boundary query is restricted to
			its minimal face by fixing forced endpoints. A face query in atlas
			mode reuses its stored source. Interior queries are assigned to anchors and handled
			by a thin branch or the deep trial in \cref{sec:deep-lift}.}
		\label{fig:dispatch}
	\end{figure}%
}

\colorlet{bkFigSeedTile}{bkFigPale!76!bkFigBlue}
\colorlet{bkFigNextTile}{bkFigCream!89!bkFigAmber}

\newcommand{\TransferExampleFigure}{%
	\begin{figure}[H]
		\centering
		\begin{tikzpicture}[bkfig,x=1cm,y=1cm]
			\tikzset{exhead/.style={font=\footnotesize\bfseries,anchor=west},
				exrow/.style={font=\footnotesize,anchor=west},
				exfine/.style={font=\footnotesize},
				exgrid/.style={draw=bkFigGrid,line width=.45pt}}
			\fill[bkFigFog,rounded corners=2pt] (0,4.80) rectangle (15.1,5.70);
			\node[exhead] at (.19,5.43) {Parent allowances \((t,S(t))\)};
			\node[exrow] at (.19,5.02) {\((8,10),\ (9,12),\ (10,15)\)};
			\draw[exgrid] (6.1,4.96) -- (6.1,5.54);
			\node[exhead] at (6.32,5.43) {Removal entries \((v,D(v),\text{witness})\)};
			\node[exrow] at (6.32,5.02)
			{\((0,0,\emptyset),\ (2,3,\{a\}),\ (4,7,\{b\}),\ (6,10,\{a,b\})\)};
			
			\fill[bkFigFog] (0,3.92) rectangle (15.1,4.53);
			\fill[bkFigPale] (8.71,.97) rectangle (9.93,4.53);
			\fill[bkFigBlue] (8.71,1.95) rectangle (9.93,2.43);
			\draw[exgrid] (0,3.92) -- (15.1,3.92);
			\draw[exgrid] (3.83,2.54) -- (15.1,2.54);
			\draw[exgrid] (0,1.91) -- (15.1,1.91);
			\draw[exgrid] (0,.97) -- (15.1,.97);
			
			\node[exhead] at (.19,4.20) {Target \(y=t-v\)};
			\node[exrow]  at (.19,3.70) {From \(t=8\)};
			\node[exrow]  at (.19,3.23) {From \(t=9\)};
			\node[exrow]  at (.19,2.76) {From \(t=10\)};
			\node[exhead] at (.19,2.19) {Best allowance \(S_R(y)\)};
			\node[exhead] at (.19,1.44) {Filter result};
			
			\foreach \y/\x/\first/\second/\third/\best/\code in {
				2/4.44/{0}/{\cdot}/{\cdot}/{0}/{\times},
				3/5.66/{\cdot}/{2}/{\cdot}/{2}/{=},
				4/6.88/{3}/{\cdot}/{5}/{5}/{\times},
				5/8.10/{\cdot}/{5}/{\cdot}/{5}/{\times},
				6/9.32/{7}/{\cdot}/{8}/{8}/{\checkmark},
				7/10.54/{\cdot}/{9}/{\cdot}/{9}/{\times},
				8/11.76/{10}/{\cdot}/{12}/{12}/{\times},
				9/12.98/{\cdot}/{12}/{\cdot}/{12}/{\varnothing},
				10/14.20/{\cdot}/{\cdot}/{15}/{15}/{\times}} {
				\node[font=\small\bfseries] at (\x,4.20) {\(\y\)};
				\node at (\x,3.70) {\(\first\)};
				\node at (\x,3.23) {\(\second\)};
				\ifnum\y=6\relax
				\node[text=bkFigBlue,font=\bfseries\boldmath] at (\x,2.76) {\(\third\)};
				\node[text=white,font=\small\bfseries] at (\x,2.19) {\(\best\)};
				\node[text=bkFigBlue] at (\x,1.44) {\(\code\)};
				\else
				\node at (\x,2.76) {\(\third\)};
				\node[font=\small\bfseries] at (\x,2.19) {\(\best\)};
				\node[text=bkFigAmber] at (\x,1.44) {\(\code\)};
				\fi
			}
			\node[exfine,anchor=west] at (.12,.40)
			{\textcolor{bkFigAmber}{\(\times\)} off-lattice\qquad
				\textcolor{bkFigAmber}{\(=\)} equals the cost limit\qquad
				\textcolor{bkFigAmber}{\(\varnothing\)} LP infeasible\qquad
				\textcolor{bkFigBlue}{\(\checkmark\)} retained};
		\end{tikzpicture}
		\caption{The source rows give \(S(t)-D(t-y)\), with a dot for
			an absent removal entry. The column maxima give the residual
			allowances. The child has two weight-\(3\) copies of costs \(2\)
			and \(4\), and the bottom row records the filter decisions.}
		\label{fig:transfer-example}
	\end{figure}%
}

\newcommand{\CosetExampleFigure}{%
	\begin{figure}[H]
		\centering
		\begin{tikzpicture}[bkfig,x=1cm,y=1cm]
			\tikzset{exaxis/.style={draw=bkFigGuide,line width=.65pt,-{Latex[length=1.2mm]}},
				exleg/.style={font=\footnotesize,anchor=west},
				exvect/.style={-{Latex[length=1.8mm]},line width=1.05pt}}
			\node[bkpanel] at (.0,3.36) {(a) Columns of the parent pool};
			\node[bkpanel] at (7.17,3.36) {(b) Residual targets and lattice membership};
			\draw[draw=bkFigGrid,line width=.45pt] (6.84,.03) -- (6.84,3.12);
			\begin{scope}[shift={(.87,.82)},x=1.20cm,y=.68cm]
				\draw[exaxis] (-.20,0) -- (3.20,0);
				\draw[exaxis] (0,-.22) -- (0,3.28);
				\foreach \i in {0,1,2,3}{
					\foreach \j in {0,1,2,3}{
						\fill[bkFigGrid] (\i,\j) circle[radius=1.03pt];
					}
				}
				\draw[exvect,draw=bkFigBlue] (0,0) -- (2,0);
				\draw[exvect,draw=bkFigBlue] (0,0) -- (0,1);
				\draw[exvect,draw=bkFigAmber,densely dashed] (0,0) -- (1,1);
				\node[font=\small,text=bkFigBlue,anchor=north] at (1.64,-.08) {\((2,0)\)};
				\node[font=\small,text=bkFigBlue,anchor=south west] at (.03,1.06) {\((0,1)\)};
				\node[font=\small,text=bkFigAmber,anchor=south west] at (1.05,.96) {\((1,1)\)};
			\end{scope}
			\node[exleg] at (.36,.03) {Solid pair spans \(\Gamma\). All three span \(\mathbb Z^2\).};
			\begin{scope}[shift={(7.72,.64)},x=1.40cm,y=.68cm]
				\foreach \i in {0,1,2,3,4}{
					\foreach \j in {0,1,2,3}{
						\ifodd\i
						\draw[draw=bkFigGuide,line width=.65pt,fill=white] (\i,\j) circle[radius=1.7pt];
						\else
						\fill[bkFigBlue] (\i,\j) circle[radius=1.7pt];
						\fi
					}
				}
				\draw[exvect,draw=bkFigAmber,shorten >=3.2pt,shorten <=3.2pt] (3,2) --
				node[sloped,above,font=\footnotesize,text=bkFigAmber] {\(-(1,1)\)} (2,1);
				\draw[draw=bkFigAmber,line width=.85pt,fill=white] (3,2) circle[radius=2.45pt];
				\fill[bkFigBlue] (2,1) circle[radius=2.55pt];
				\node[font=\small,text=bkFigAmber,anchor=south west] at (3.04,2.10) {\(t=(3,2)\)};
				\node[font=\small,text=bkFigBlue,anchor=north east] at (2.00,.90) {\(y=(2,1)\)};
			\end{scope}
			\fill[bkFigBlue] (7.52,.03) circle[radius=1.7pt];
			\node[exleg] at (7.64,.03) {\(\Gamma=2\mathbb Z\times\mathbb Z\)};
			\draw[draw=bkFigGuide,line width=.7pt,fill=white] (11.25,.03) circle[radius=1.7pt];
			\node[exleg] at (11.37,.03) {odd first coordinate};
		\end{tikzpicture}
		\caption{Deleting \((1,1)\) from the pool changes the column lattice
			from \(\mathbb Z^2\) to its index-two sublattice \(\Gamma\).
			With an empty removal witness the residual target is \(t\notin\Gamma\).
			Selecting the removed copy yields \(y=t-(1,1)\in\Gamma\), so this
			target passes the lattice test before its LP value is checked.}
		\label{fig:coset-example}
	\end{figure}%
}

\newcommand{\PlanarFigure}{%
	\begin{figure}[H]
		\centering
		\begin{tikzpicture}[bkfig,x=1cm,y=1cm]
			\node[bkpanel] at (.02,4.40) {(a) Tiles in target space};
			\node[bkpanel] at (7.67,4.40) {(b) Pivot in price space};
			\draw[draw=bkFigGrid,line width=.45pt] (7.15,.30) -- (7.15,4.15);
			
			\begin{scope}[shift={(.84,.51)},x=1.63cm,y=1.15cm]
				\fill[bkFigFog] (0,0) -- (2,0) -- (3,1) -- (3,3) -- (1,3) -- (0,2) -- cycle;
				\fill[bkFigSeedTile] (1,0) -- (1,1) -- (2,2) -- (2,1) -- cycle;
				\fill[bkFigNextTile] (2,1) rectangle (3,2);
				\draw[draw=bkFigGrid,line width=.55pt]
				(0,1) -- (1,1) -- (2,2) -- (3,2)
				(1,0) -- (1,1)
				(1,0) -- (2,1) -- (2,2)
				(2,1) -- (3,1)
				(0,1) -- (1,2) -- (2,2) -- (2,3)
				(1,2) -- (1,3);
				\draw[draw=bkFigInk,line width=.8pt]
				(0,0) -- (2,0) -- (3,1) -- (3,3) -- (1,3) -- (0,2) -- cycle;
				\draw[draw=bkFigBlue,densely dashed,line width=.7pt] (.4,-.1) rectangle (2.6,2.1);
				\draw[draw=bkFigBlue,line width=.8pt] (1.4,.9) rectangle (1.6,1.1);
				\draw[draw=bkFigAmber,line width=1.25pt] (2,1) -- (2,2);
				\node[bkflowlabel,text=bkFigBlue,anchor=south east] at (2.61,2.11) {\(\widehat K\)};
				\node[bkflowlabel,text=bkFigInk,anchor=east] at (2.96,2.80) {\(Z\)};
				\fill[bkFigBlue] (1.5,1) circle[radius=1.45pt];
				\fill[bkFigBlue] ({5/3},.70) ++(-1.3pt,-1.3pt) rectangle ++(2.6pt,2.6pt);
				\draw[draw=bkFigBlue,thin] (1.19,.83) -- (1.47,.985);
				\node[bkflowlabel,text=bkFigBlue,anchor=east] at (1.22,.83) {\(b\)};
				\draw[draw=bkFigBlue,thin] (1.18,1.20) -- (1.4,1.1);
				\node[bkflowlabel,text=bkFigBlue,anchor=east] at (1.19,1.20) {\(K\)};
				\draw[draw=bkFigBlue,thin] (1.40,.49) -- ({5/3-.03},.67);
				\node[bkflowlabel,text=bkFigBlue,anchor=east] at (1.43,.49) {seed \(t_0\)};
			\end{scope}
			\fill[fill=bkFigSeedTile,draw=bkFigBlue,thin] (.86,.055) rectangle (1.14,.205);
			\node[bkflowlabel,anchor=west] at (1.21,.13) {seed tile};
			\fill[fill=bkFigNextTile,draw=bkFigAmber,thin] (3.02,.055) rectangle (3.30,.205);
			\node[bkflowlabel,anchor=west] at (3.36,.13) {neighboring tile};
			
			\begin{scope}[shift={(8.24,.51)},x=1.27cm,y=1.15cm]
				\draw[bkedge,draw=bkFigGuide] (0,0) -- (4.90,0)
				node[right,font=\footnotesize] {\(\pi_x\)};
				\draw[bkedge,draw=bkFigGuide] (0,0) -- (0,3.05)
				node[left,font=\footnotesize] {\(\pi_y\)};
				\fill[bkFigNextTile] (3,2) -- (4,2) -- (4,1) -- cycle;
				\draw[draw=bkFigGrid,densely dashed,line width=.65pt] (.55,2) -- (4.68,2);
				\draw[draw=bkFigGrid,densely dashed,line width=.65pt] (2.08,2.92) -- (4.72,.28);
				\draw[draw=bkFigAmber,line width=.82pt] (4,.45) -- (4,2.88);
				\draw[draw=bkFigInk,line width=.65pt] (3,2) -- (4,2) -- (4,1) -- cycle;
				\draw[bkedge,draw=bkFigBlue,line width=1.0pt,shorten <=2pt,shorten >=2pt] (3,2) -- (4,2);
				\fill[bkFigBlue] (3,2) circle[radius=1.65pt];
				\fill[bkFigAmber] (4,2) circle[radius=1.65pt];
				\node[bkflowlabel,text=bkFigBlue,anchor=south east] at (2.95,2.12) {\(\pi=(3,2)\)};
				\node[bkflowlabel,text=bkFigAmber,anchor=west] at (4.07,2.24) {\(\pi'=(4,2)\)};
				\node[bkflowlabel,text=bkFigAmber,anchor=south] at (4,2.87) {\(\pi_x=4\)};
				\node[bkflowlabel,anchor=north west] at (.58,1.98) {\(\pi_y=2\)};
				\node[bkflowlabel,anchor=north east] at (3.54,1.40) {\(\pi_x+\pi_y=5\)};
			\end{scope}
		\end{tikzpicture}
		\caption{An atlas pivot with horizontal costs \(1,4\), vertical costs
			\(2,7\), and one \((1,1)\) column of cost \(5\).
			The feasible zonogon \(Z\) contains the query window \(K\) around
			\(b=(3/2,1)\), whose optimal fractional cost is \(9/2\).
			For this example take the padding \(D=[-1,1]^2\). The dashed window
			\(\widehat K=K+D\) contains the seed \(t_0=(5/3,7/10)\) in the same tile. Crossing the highlighted
			target-space facet fixes the diagonal label to one.
			The price ray keeps the vertical label tied at \(\pi_y=2\).
			Its first blocker \(\pi_x=4\) identifies the adjacent tile.}
		\label{fig:planar}
	\end{figure}%
}

\title{Bounded Knapsack in Truly Subquadratic Time}
\author{Huanyi Zhou\thanks{Tsinghua University. \href{mailto:zhouhuan24@mails.tsinghua.edu.cn}{zhouhuan24@mails.tsinghua.edu.cn}.}}
\date{}

\begin{document}
	\maketitle
	\begin{abstract}
		Bounded Knapsack asks how many copies of each item to take
		to maximize profit under a capacity constraint. Given \(n\) records
		with binary-encoded multiplicities, we return optimal counts in
		randomized time with a fixed power saving in the maximum item
		weight \(\wm\) or maximum item profit \(\pmx\).
		
		Fix \(1<\rho<2\) and suppose integer \((\min,+)\)-convolution on
		length-\(m\) arrays can be computed with failure probability at most
		\(\eta\) in \(\Ot(m^\rho)\) time, up to polynomial factors in
		the entry bit length and \(\log(1/\eta)\).
		Put \(f(\rho)=4(3-\rho)/(8-3\rho)\).
		For any fixed \(\epsilon>0\), on inputs whose integers have
		magnitude polynomial in \(n+\min(\wm,\pmx)\), our algorithm runs in time
		\[
		\Ot\bigl(n+\min(\wm,\pmx)^{f(\rho)+\epsilon}\bigr).
		\]
		It returns optimal counts with high probability. If convolution
		saves \(\delta=2-\rho\) in its exponent, knapsack retains
		\(\delta/(1+3\delta/2)\).
		With the integer convolution exponent \(1.9601397040925\),
		a separate parameter choice gives
		\(\Ot(n+\min(\wm,\pmx)^{1.961})\), including for 0--1 Knapsack.
		
		We keep the copies compressed. Witnessed convolutions generate
		residual queries, and local linear programs filter them by cost.
		Resilient lattice cores allow recursive states to coalesce.
		Progression reserves support rounding, and a thin affordable fiber
		yields a lift that reduces the normalized column volume. This
		volume bounds the number of LP regions a grid lookup examines.
	\end{abstract}
	\newpage
	\begingroup
	\small
	\setlength{\parskip}{0pt}
	\tableofcontents
	\endgroup
	\newpage
	
	\section{Introduction}\label{sec:intro}
	
	In Bounded Knapsack, record \(i\) specifies a positive integer weight
	\(w_i\), an integer profit \(p_i\), and a binary-encoded nonnegative
	multiplicity \(u_i\). Given a nonnegative integer capacity \(C\), we
	seek a count vector attaining
	\[
	\max\left(\sum_{i=1}^{n}p_i x_i
	\ \middle|\
	\sum_{i=1}^{n}w_i x_i\le C,\quad
	x\in\Z^n,\quad 0\le x_i\le u_i\ (1\le i\le n)\right).
	\]
	Here \(n\) counts records. Write \(\wm=\max_{1\le i\le n}w_i\) and
	\(\pmx=\max\{1,\max_{1\le i\le n}p_i\}\). Records with nonpositive
	profits can be discarded. When every \(u_i=1\), this is 0--1 Knapsack.
	
	The classical dynamic program is indexed by capacity~\cite{Bel57}.
	Even two records of weights one and two can encode enough copies
	to fill a large capacity. Parameterizing by \(\wm\) allows the work
	beyond reading the records to depend on the largest single weight.
	The parameter \(\pmx\) plays the corresponding role for profits.
	
	A sequence of proximity and dynamic-programming results culminated
	in deterministic near-quadratic algorithms for both parameterizations
	of Bounded Knapsack and 0--1 Knapsack~\cite{Bringmann24,Jin24}.
	The \((\min,+)\)-convolution hypothesis ruled out a fixed power
	saving in the joint parameter \(n+\wm\)~\cite{CMWW19,KPS17}.
	The breakthroughs for 3SUM and all-pairs shortest paths (APSP)
	give truly subquadratic integer \((\min,+)\)-convolution and remove
	that conditional barrier~\cite{AV26}. The exchange range from
	standard proximity is \(O(\wm^2)\)~\cite{PRW21}, so a table over
	that entire range still has a quadratic size bound in \(\wm\).
	
	\subsection{Main result}
	
	We use the bit-cost model of \cref{sec:model}, with \(\Ot\) absorbing
	polylogarithmic factors. The convolution interface is specified in
	\cref{def:minconv}.
	
	\Needspace{11\baselineskip}
	\begin{theorem}[from integer convolution to Bounded Knapsack]
		\label[theorem]{thm:main}
		Suppose integer \((\min,+)\)-convolution on length-\(m\) arrays has
		running time \(\Ot(m^\rho)\) with exponent \(1<\rho<2\) in the sense
		of \cref{def:minconv}. Put
		\[
		f(\rho)=\frac{4(3-\rho)}{8-3\rho}.
		\]
		For every fixed \(\epsilon>0\), Bounded Knapsack admits randomized
		algorithms returning optimal counts in times
		\[
		\Ot\bigl(n+\wm^{f(\rho)+\epsilon}\bigr)
		\qquad\text{and}\qquad
		\Ot\bigl(n+\pmx^{f(\rho)+\epsilon}\bigr),
		\]
		for the weight and profit parameterizations, respectively. For either
		parameter \(d\in\{\wm,\pmx\}\), assume the input integers have
		magnitude \((n+d+2)^{O(1)}\). The success probability is at least
		\(1-(n+d+2)^{-a}\) for any fixed \(a>0\).
	\end{theorem}
	
	With \(\delta=2-\rho\), the saving \(2-f(\rho)\) is
	\(\sigma=\delta/(1+3\delta/2)\). The loss \(\delta-\sigma\) is quadratic
	in a small \(\delta\), and any fixed \(\epsilon<\sigma\) gives truly
	subquadratic dependence on the item parameter. The maximum-profit
	bound uses the complement transformation of~\cite[Section~4]{PRW21}
	directly on the multiplicities.
	
	The weighted-cache 3SUM decision algorithm
	of~\cite{Wang26} has exponent
	\(\alpha=1.920279408185\) on polynomially bounded integers.
	\Cref{app:minconv} converts decision queries to witnessed integer
	convolution with exponent \(\rho_0=1+\alpha/2=1.9601397040925\).
	\Cref{subsec:transfer-exponent} optimizes the internal parameters
	for this particular \(\rho_0\).
	
	\begin{corollary}[a fixed power saving]\label[corollary]{cor:concrete}
		Under the input assumptions of \cref{thm:main},
		Bounded Knapsack, and hence 0--1 Knapsack, has randomized algorithms
		running in \(\Ot(n+\wm^{1.961})\) and \(\Ot(n+\pmx^{1.961})\)
		time, with optimal count vectors as output and the success
		probability of \cref{thm:main}.
	\end{corollary}
	
	Using the \(1.9992\) integer 3SUM exponent of~\cite[Theorem~2]{AV26}
	instead gives convolution exponent \(1.9996\). The transfer theorem
	already yields \(\Ot(n+d^{1.9997})\) for \(d=\wm\) or \(d=\pmx\).
	
	The fixed power saving improves on the weight term
	\(\wm^2/2^{\Omega(\sqrt{\log\wm})}\) in the questions of
	Bringmann and Jin~\cite[Section~8]{Bringmann24}\cite[Section~1.4]{Jin24}.
	The input term here is \(\Ot(n)\), whereas Jin's weight-parameter
	question specifies \(O(n)\). The corollary also gives
	\(\Ot(n+(\wm+\pmx)^{2-\delta'})\) for Bounded Knapsack and
	\(O((n+\wm+\pmx)^{2-\delta'})\) for 0--1 Knapsack for a fixed
	\(\delta'>0\), as asked in those works.
	
	\subsection{Earlier algorithms and the parameter gap}
	
	Tighter proximity bounds and cheaper correction updates drive the
	improvements in small-weight knapsack (\cref{tab:history}).
	\begin{table}[H]
		\centering\small
		\setlength{\tabcolsep}{5pt}
		\renewcommand{\arraystretch}{1.12}
		\begin{tabularx}{\linewidth}{@{}l l l >{\raggedright\arraybackslash}X@{}}
			\toprule
			Work & Variant & Time in \(n,\wm\) & New algorithmic idea \\
			\midrule
			\cite{Tam09} & Bounded & \(O(n^3\wm^2)\) & Capacity-independent residual from a coordinatewise LP proximity box \\
			\cite{EW18} & Bounded & \(\Ot(n\wm^2)\) & Steinitz reordering bounds total LP--integer deviation independently of \(n\) \\
			\cite{BHSS18} & Bounded & \(\Ot(n\wm^2\min(n,\wm))\) & Modular exchanges fix copies from a greedy prefix at reduced capacity \\
			\cite{PRW21} & Bounded & \(O(n+\wm^3)\) & Separate addition and removal profiles over an \(O(\wm^2)\) weight range \\
			\cite{Jin23} & 0--1 & \(\Ot(n+\wm^{5/2})\) & Erd\H{o}s--S\'ark\"ozy exchanges bound the number of changed weight classes \\
			\cite{HX24} & Bounded & \(\Ot(n+\wm^{5/2})\) & Bounded-supply Frobenius exchanges isolate \(\Ot(\wm)\) candidate copies \\
			\cite{CLMZ24} & Bounded & \(\Ot(n+\wm^{12/5})\) & Multiplicity-sensitive equal-sum exchanges give separate correction budgets for groups of weights \\
			\cite{Bringmann24} & Bounded & \(\Ot(n+\wm^2)\) & Partition with correction weight times distinct weight count bounded by \(\Ot(\wm^2)\) per part \\
			\cite{Jin24} & 0--1 & \(O(n+\wm^2\log^4(\wm+2))\) & Dyadic rank proximity and pruning of dominated witness extensions \\
			\cite{Vaughn26} & Bounded & \(O(n+\wm^2\log^3(\wm+2))\) & Kneser-based exchange bound and DP organized by correction scale \\
			This work & Bounded & \(\Ot(n+\wm^{1.961})\) & Witnessed convolution on lattice boxes, phase-wide accounting, and volume-bounded LP indexing \\
			\bottomrule
		\end{tabularx}
		\caption{Representative exact knapsack bounds and the new
			techniques behind them. For bounded instances, \(n\) counts
			records with binary-encoded multiplicities.
			The bounds of~\cite{BHSS18,HX24} and the last
			row allow randomization. The bound of~\cite{Vaughn26} counts
			arithmetic and comparison operations. The last row has the input
			and success guarantees of \cref{thm:main,cor:concrete}.}
		\label{tab:history}
	\end{table}
	
	Early algorithms restrict the search near a greedy or fractional
	solution~\cite{Pis99,Tam09,EW18,BHSS18}. Tamir applies coordinatewise
	LP proximity from~\cite{CGST86}, and the signed
	exchanges~\cite{EW18} bound the total number of changed copies independently of
	\(n\). Polak, Rohwedder, and W\k{e}grzycki use that proximity to split
	corrections into additions and removals, with earlier concave updates
	for equal-weight classes~\cite{KP04,AT19,PRW21}.
	
	Subsequent proximity bounds track the weights that change or assign
	separate budgets to groups of weights. The latter approach uses
	dense subset sums~\cite{BW21,CLMZ24} and leads to Bringmann's
	partition theorem, with the same concave updates~\cite[Theorem~4]{Bringmann24}.
	Independently, Jin refines support-sensitive proximity and witness
	propagation~\cite{DMZ23,Jin23,Jin24}. He and Xu use a bounded-supply
	Frobenius argument to restrict the candidate copies~\cite[Theorem~6]{HX24}.
	Vaughn refines the exchange and DP bounds in the partition
	approach~\cite[Theorem~4 and Section~3]{Vaughn26}.
	
	Several bounds use different parameters. The small-\(n\) 0--1
	algorithm of~\cite{HX24} takes \(\Ot(n^{3/2}\min(\wm,\pmx))\) time.
	Bounded monotone convolution yields bounds in capacity and optimal
	profit, as well as \(\Ot(n+(\wm+\pmx)^{3/2})\) for Unbounded
	Knapsack~\cite{BC22}. Near-convex convolution uses bounded numerical
	deviation~\cite{BC23}. Rectangular monotone convolution and balancing
	yield further mixed-parameter bounds~\cite{BDP24}. The monotone
	primitive of~\cite{CDXZ22} has a deterministic \(m^{3/2+o(1)}\)-time
	version~\cite{JPSX26}. The barely-monotone bounds still use numerical
	and monotonicity promises~\cite{HMS26}, and higher-order BSG
	relates structured variants~\cite{FJX26}. Weak approximation
	allows simultaneous relaxation of profit and capacity with
	subquadratic dependence on inverse accuracy~\cite{CLMZ25}.
	
	The unrestricted convolution input comes from the fixed power
	savings for 3SUM and APSP. Alman and Vassilevska Williams use triangles
	in sparse lopsided graphs and selected-entry matrix products~\cite{AV26}.
	Karingula and Lovett simplify the selected-entry construction~\cite{KL26}.
	Majdoddin improves the thin product and uses adaptive counting to
	preserve its saving in the reductions~\cite{Maj26}. Fischer, Polak,
	and Schmidt transfer fixed savings for integer 3SUM to real 3SUM
	via Frank--Tardos compression and additive structure~\cite{FPS26}.
	The subset-convolution approach of~\cite{CIMY26} also handles all-numbers
	reporting. Our numerical bound uses the weighted-cache decision
	algorithm of~\cite{Wang26} through the reduction in \cref{app:minconv}.
	
	The geometric ingredients come from additive combinatorics and
	integer programming. Common generalized arithmetic progressions
	describe subset sums in the inverse theorem of~\cite{CFP23}, and
	dense multidimensional subset sums fill lattice regions inside a
	zonotope under the hypotheses of~\cite{CHMZ26}. Chen, Mao, and Zhang
	use progression reserves and recursive lifts for small-item Subset
	Sum~\cite{CMZFeas26}. Jin and Mao use the inverse theorem and quotient
	cap coverings~\cite{JM26}. The bounded integer optimization framework
	of~\cite{CMZOpt26} adds LP prices, affordable regions, and resilient
	lattice restrictions.
	
	The revised framework~\cite[Theorem~1.1]{CMZOptRev26} gives a
	deterministic \(N^{2d+\epsilon}\) parameter term for \(d\) rows,
	up to cost logarithms and in addition to input and LP costs. Its
	one-row exponent is \(2+\epsilon\). Its grouped DP computes cropped
	transfers with attaining sources~\cite[Lemma~11.2]{CMZOptRev26}.
	Both versions certify a chosen source at every supporting
	price~\cite[Lemma~8.4 and Section~12.2]{CMZOpt26}\cite[Lemma~9.4 and
	Section~12.1]{CMZOptRev26}. This provides the witness and price
	invariants for the recursion used here.
	
	\subsection{Why a faster convolution does not suffice}
	
	The capacity reduction of~\cite{CMWW19} gives a randomized
	\(\Ot(n+C^{2-\delta})\) algorithm for 0--1 Knapsack for some
	\(\delta>0\) with the convolution bound of~\cite{AV26}.
	Binary multiplicities allow \(C\) to be much larger than \(\wm\).
	Even a quadratic exchange window in \(\wm\) is too wide for this
	capacity bound to give a subquadratic guarantee in \(\wm\).
	The reductions underlying the conditional lower bound encode
	convolution as knapsack. The algorithmic question here is how to
	reduce knapsack to convolution with lengths controlled by \(\wm\).
	
	For Unbounded Knapsack, Axiotis and Tzamos give a small-item reduction
	for one target~\cite{AT19}. Deng, Mao, and Zhong give an all-target
	algorithm in \(\Ot(n+T(\wm)+C)\) when convolution of length \(m\)
	costs \(T(m)\)~\cite[Theorem~5.2]{DMZ23}. Their witness propagation
	uses the small support of an optimum and the unlimited supply of
	each item. With bounded supply, each extension must respect the
	remaining labels. Bringmann's bounded-to-0--1 reduction and
	Vaughn's updates expand kernels of up to \(O(\wm^2)\)
	copies~\cite{Bringmann24,Vaughn26}, while the smaller candidate set
	of~\cite{HX24} still incurs dynamic-programming work. Jin's hinted
	updates pay for the table range and the hint-set sizes~\cite{Jin24}.
	Vaughn's updates pay for the correction range and the number of
	distinct weights~\cite{Vaughn26}. These costs remain even with a
	faster unrestricted convolution routine.
	
	\subsection{The algorithmic contribution}
	
	We apply greedy proximity on compressed records and solve the
	resulting exact-weight problem using the lattice recursion
	of~\cite{CMZFeas26,CMZOpt26}. Our transfer replaces grouped DP with
	witnessed convolution in integer-coordinate boxes. A geometric
	crop controls the windows passed to the LP oracle. The generating
	state pays for the whole batch, including rejected candidates.
	
	We coalesce equal lattices across an arithmetic phase. The
	certificate at every supporting price lets one attaining source
	per retained query serve both recovery and the analysis of the
	sum of state masses and their \(\rho\)-th powers. For LP queries,
	a grid indexes tiles of the local LP subdivision by target. Each
	lift increases the rank by one and reduces the normalized column
	volume, tightening the bound on tiles per cell. A common strip
	bounds the heights tested for each lifted residual. These bounds
	control the work of constructing and testing transfers while
	retaining disjoint labeled assignments.
	
	\Cref{sec:overview} follows these steps on concrete examples.
	\Cref{sec:model,sec:global} give the compressed input reduction and
	the recursive accounting. The local LP oracles, integer transfer
	tables, and deep reductions are developed in
	\cref{sec:geometry,sec:transfer,sec:deep}. \Cref{sec:runtime}
	combines their costs and gives the profit-parameter reduction.
	
	\section{Technical Overview}\label{sec:overview}
	
	A recursive state groups exact-weight queries on one pool of
	labeled copies. A transfer fixes choices on the removed labels
	and sends residual queries to the surviving pool.
	
	A decision query \((t,s)\) asks for an integral selection from a pool \(I\)
	with target \(t\) and cost below \(s\). At the root, each column
	\(a_i\) is a scalar weight. Before perturbation, \(c_i\) is the
	negative profit, with cost zero for slack.
	A lift appends a coordinate to \(a_i\). Write \(A_Ix=\sum_{i\in I}a_ix_i\)
	and use
	\[
	\Lambda_I=\sum_{i\in I}\Z a_i,\qquad
	Z_I=\sum_{i\in I}[0,a_i],\qquad
	f_I(t)=\min\left(
	c^\top x\ \middle|\ A_Ix=t,\quad x\in[0,1]^I
	\right).
	\]
	The lattice \(\Lambda_I\) tests integer compatibility. The zonotope \(Z_I\)
	and the linear-programming (LP) value \(f_I\) test fractional feasibility
	and cost. Set \(f_I(t)=+\infty\) outside \(Z_I\).
	
	\subsection{Proximity and recursive transfers}
	
	The exchange argument for small weights~\cite{PRW21,Bringmann24}
	confines an optimum to an interval of
	\(O(\wm)\) copies around the greedy prefix count in each weight class.
	We store each interval on the original records and add one slack
	record to impose an equality. The witnessed integer convolution of
	\cref{sec:model} returns an attaining pair of indices at each finite
	entry. These pairs connect a recursive solution to its predecessors.
	
	Suppose a step enumerates the permitted choices on \(I\setminus R\) and
	passes to pool \(R\). If \(D(v)\) is the minimum removal cost recorded at
	displacement \(v\), and \(S(t)\) is the current allowance at \(t\), the
	residual allowance at \(y\) is
	\[
	S_R(y)=\max_{\substack{t-v=y\\ S(t),D(v)\text{ finite}}}
	\bigl(S(t)-D(v)\bigr).
	\]
	Absent pairs give no entry. With \(D^\leftarrow(v)=D(-v)\), the
	witnessed \((\min,+)\)-convolution gives \(S_R=-((-S)\star D^\leftarrow)\).
	After cropping, retain entries with \(y\in\Lambda_R\cap Z_R\) and
	\(f_R(y)<S_R(y)\), together with their attaining removal choices.
	
	\begin{example}[a transfer with competing predecessors]
		Let the removal pool have copies \(a,b\) with weight--cost pairs
		\((2,3),(4,7)\). Pool \(R\) has two weight-\(3\) copies of costs \(2,4\).
		Use allowances \(S(8)=10\), \(S(9)=12\), and \(S(10)=15\).
		At \(y=6\), removing \(a\) from target \(8\) offers allowance \(7\), and
		removing \(b\) from target \(10\) offers \(8\). We keep the latter.
		The child takes both copies for cost \(6<8\). Reattaching \(b\) gives
		target \(10\) at cost \(13<15\). In contrast, \(y=9\) passes the lattice
		test but lies outside \(Z_R=[0,6]\). At \(y=3\), the fractional cost
		equals the allowance \(2\), so the strict query is rejected.
		\Cref{fig:transfer-example} shows all the entries.
	\end{example}
	
	\TransferExampleFigure
	
	A \emph{source} records the child pool, witnesses, and a window
	covering compatible candidates and their nearby offsets. The
	symmetric affordable region of \((t,s)\) consists of offsets \(z\)
	satisfying
	\[
	f_I(t+z)+f_I(t-z)\le 2s .
	\]
	Following the geometric decomposition of~\cite{CMZFeas26,CMZOpt26},
	faces force endpoints and interior targets group around anchors. Thin
	anchors remove columns large in a narrow direction. Deep anchors
	provide room for a core and reserve. \Cref{sec:global} bounds the
	source windows in all three cases.
	
	\subsection{Locating a local linear program}
	
	At a feasible target \(t\) in rank \(r\), the LP dual is
	\[
	f_I(t)=\max_{\pi\in\R^r}\left(
	\pi\cdot t+\sum_{i\in I}\min(0,c_i-\pi\cdot a_i)
	\right).
	\]
	At an optimal price \(\pi\), strict negative reduced costs select
	copies and strict positive ones omit them. Within a type, the selected
	copies form a prefix in cost order. A common perturbation makes the
	tight columns linearly independent. In rank two, two tight columns give
	a parallelogram of targets sharing the same price. In \cref{fig:planar},
	at \(\pi=(3,2)\) the horizontal copy of cost \(1\) is fixed at one,
	while the vertical copy of cost \(2\) and the diagonal copy of cost \(5\)
	are free. Taking half of each free copy reaches \((3/2,1)\) at cost \(9/2\).
	
	A source builds its atlas by finding one tile and crossing facets
	in a padded version of the query window \(K\). The volume \(V(K,I)\),
	measured in lattice units, bounds the number of tiles built
	(\cref{lem:atlas-navigation}). A pivot changes only a constant number
	of type prefixes. Each tile stores its price, its tight labels, and
	a persistent prefix pointer for recovering a basic optimum.
	For the grid atlas, an actual-column basis defines the grid and
	the target determines a single bucket.
	For example, the tiles \([0,1]^2\) and \([1,2]\times[0,1]\) both
	register at key \((1,0)\). The target \((1,1/2)\) uses that key and
	passes the closed containment test for either tile.
	
	In these coordinates, every column has bounded length for fixed rank.
	Each tile meets a constant number of buckets, and the tiles registered
	in one bucket fit inside a constant enlargement of the bucket.
	Their disjoint interiors and minimum lattice volume bound the number
	of entries by \(O_r(\Psi(I))\). The normalized column volume
	\(\Psi(I)\) measures the symmetric hull in units of the lattice
	covolume and controls the size of one bucket
	(\cref{lem:grid-atlas}). Blocks of affine sign tests search the
	bucket for a containing tile, including on shared faces.
	\Cref{lem:lift-volume,lem:geometry-total} use the decrease in
	\(\Psi(I)\) to bound query costs across lift levels.
	
	\subsection{Integer transfer tables and resilient lattice cores}
	
	A lattice box with few entries can extend far beyond a thin physical
	window (\cref{fig:clip}). A mixed-radix code can alias sums with
	overflowing coordinates. Minimizing digit sum before cost and
	checking the winning digit sum against the target reduces each
	orthant table to witnessed all-capacities knapsack~\cite[Corollary~14]{CMWW19}.
	We crop the output to the physical window before building the
	LP oracle (\cref{sec:transfer}).
	
	To find a small deletion on a compressed pool, store the live
	types in a balanced tree. Each node stores a Hermite normal form
	(HNF) basis for its subtree lattice. A lattice character is a
	homomorphism to \(\Q/\Z\). If it vanishes on that basis, skip the
	whole subtree. Otherwise descend to an actual nonzero leaf and
	count all of its available copies at once. A large multiplicity
	can reject a character without expanding its labels. If the exact
	support is small, deleting it already leaves a closed pool in
	the survivor lattice. The same tree finds actual independent
	columns, so both the lattice test and its returned labels are
	constructive. The resulting lattice screens the residual targets,
	as illustrated in \cref{fig:coset-example}.
	
	\CosetExampleFigure
	
	\subsection{Deep anchors, progression reserves, and lifts}
	
	At a deep anchor, normalizing at its supporting price gives nonnegative
	costs \(|c_i-\pi\cdot a_i|\) for flips away from preferred endpoints.
	We use a generalized arithmetic progression for the cheap
	columns~\cite{CMZFeas26,CMZOpt26}. A small reserve covers an integer box in
	progression coordinates. Its labels can repair the error when a basic
	fractional solution is rounded, and the price normalization pays for
	the reserve within the query's allowance.
	
	The trial pool consists of this reserve and a remainder. If no trial
	candidate survives, we reject the anchor's queries. The anchor gap is
	its allowance minus its fractional optimum. If the normalized allowances
	of the surviving queries are all below one quarter of that gap, we restore
	every copy whose column lies in the trial's proper sublattice and
	recompute the removal table. The same sublattice specifies the same restored pool,
	so these states can merge. The charging argument needs one predecessor
	valid at every supporting price of a buffered residual, because the
	price used to count its ancestors can differ from the price used to
	construct the trial. \Cref{sec:deep} proves this certificate for the
	removal choice actually kept by convolution.
	
	A trial with high allowance instead gives an affordable fiber, the set
	of progression points projecting to its old target within budget.
	Fixed-dimensional integer optimization either finds an integral point,
	which the reserve repairs to a selection, or supplies a thin integral
	direction~\cite{CMZOpt26}. In the latter case we append that direction
	to the columns as one new coordinate. The proof gives a short common
	strip of heights for the retained witnesses and their query buffers. This bounds the
	number of heights searched at each old target and multiplies the
	normalized column volume by at most \(O(1/\nu)\), where
	\(\nu=\lceil\wm^\theta\rceil\) for a fixed scale exponent \(\theta>0\).
	
	\subsection{From local costs to the running time}
	
	\begin{example}[two outcomes of a trial transfer]
		Consider the local transfer and filtering step at price zero.
		The trial pool has \(2m\) weight-\(2\) labels of cost zero, where
		\(m\ge2\). The two excluded labels \(a,b\) have weight--cost pairs
		\((1,c),(2,1)\). At target \(2m+1\) with allowance \(32\), the
		compatible trial targets are \(2m\) with allowance \(32-c\)
		(witness \(\{a\}\)) and \(2m-2\) with allowance \(31-c\)
		(witness \(\{a,b\}\)). Both have fractional value zero.
		
		For \(c=40\), both targets fail the strict LP test. The generator
		still pays for the removal table, the source window covering both
		targets and their offsets, its rank-one atlas, and the two LP tests.
		The charge is made to the parent anchor under
		\cref{lem:source-halo,thm:ledger}.
		
		For \(c=28\), both allowances are positive and below \(32/4\).
		Restoring to \(2\Z\) puts \(b\) back into the pool. Recomputing
		on the complement \(\{a\}\) yields only the compatible target
		\(2m\), with allowance \(4\) and witness \(\{a\}\). The old witness
		for \(2m-2\) used \(b\) and is invalid for the new complement.
		For a deep-anchor restoration, \cref{lem:trial-restoration}
		proves this recomputation preserves the decision.
		\Cref{ex:price-extension,lem:restored-collar,lem:phase-mass}
		explain how the resulting source is traced when the supporting
		price changes.
	\end{example}
	
	Write \(X=\nu\wm\). The global bound of \cref{thm:ledger} gives
	\(\Ot(X\wm^\beta)\) for the sum of source volumes and query counts,
	for any fixed \(\beta>0\). Convolutions cost \(\Ot(X^\rho\wm^\beta)\).
	Lattice searches, progression construction, and terminal reserve DP
	fit within \(\Ot(\nu^2\wm^{1+\beta})\).
	
	An atlas created after \(\ell\) lifts has rank \(r\le\ell+1\) and
	column volume \(\Psi(I)=O_D(\wm/\nu^\ell)\), with \(D\) a fixed rank
	bound. A block of width \(h\)
	costs about \(h^r\) to rebuild, while a query visits about
	\(1+\Psi(I)/h\) blocks. Balancing these costs gives an overhead of
	\(1+\Psi(I)^{r/(r+1)}\), up to an arbitrarily small power loss.
	The early grid levels benefit from low rank, and the later levels
	benefit from small column volume. At intermediate levels the saving
	from exponent two balances terms of orders \(1/(\ell+2)\) and
	\((\ell+2)\theta\), giving a saving of order \(\sqrt\theta\).
	For a small convolution saving \(\delta=2-\rho\), taking \(\theta\)
	of order \(\delta^2\) makes the geometric saving first order in
	\(\delta\) and costs only \(O(\delta^2)\) in the convolution exponent.
	\Cref{subsec:transfer-exponent} balances the convolution, structural,
	and oracle costs for all \(\rho\).
	
	\section{Preliminaries and Input Reduction}\label{sec:model}
	
	The input reduction produces one exact-weight optimization problem on
	compressed copies. The convolution interface below combines its
	residual queries while retaining the assignments that attain their costs.
	
	For \(n\ge1\), write \([n]=\{1,\ldots,n\}\). We use \(\Ot\) to
	suppress polylogarithmic factors. The bit bound \(B\) covers the
	input integers and the indices up to \(n+\wm+2\). We use the
	random-access bit-cost model, charging for exact binary arithmetic
	on \(O(B)\)-bit integers and rational
	numbers of polynomial encoding length. The rank and geometric
	constants are independent of \(n\) and \(\wm\).
	
	A \emph{label} identifies one virtual copy, and a \emph{run} represents
	an interval of copies from one input record. A \emph{pool} \(I\) is a
	versioned set of labels. A permanent \emph{type} groups copies of the
	same input weight. In each state its labels share a column from which
	the algorithm can read that weight. For \(i\in I\), write \(a_i\) for
	the stored column and \(c_i\) for the stored integer cost.
	Write \(A_Ix=\sum_{i\in I}a_ix_i\), and define
	\[
	\Lambda_I=\sum_{i\in I}\Z a_i,\qquad
	P_I=\conv\bigl(\{0\}\cup\{\pm a_i\mid i\in I\}\bigr),\qquad
	Z_I=\sum_{i\in I}[0,a_i].
	\]
	For \(r=\dim\operatorname{span}_{\R}\Lambda_I>0\), the normalized
	column volume is
	\[
	\Psi(I)=\frac{r!\operatorname{vol}_r(P_I)}{2^r\det\Lambda_I},
	\]
	where \(\det\Lambda_I\) is the covolume in the column span.
	Volumes, norms, and relative interiors are taken in the relevant
	span. Every stored column has positive input-weight readout, so a
	rank-zero pool is empty. For this pool, set \(\Psi(I)=1\), and both
	\(P_I\) and \(Z_I\) are \(\{0\}\).
	
	For a target \(t\), let
	\[
	f_I(t)=\min\left(c^\top x\ \middle|\ A_Ix=t,\quad x\in[0,1]^I\right).
	\]
	We use the convention \(\min(\varnothing)=+\infty\).
	A \emph{strict query} \((t,s)\) asks for a selection \(x\in\{0,1\}^I\)
	with \(A_Ix=t\) and \(c^\top x<s\). Every yes answer passes the
	lattice and LP tests \(t\in\Lambda_I\) and \(f_I(t)<s\). For a
	surviving query, its \emph{fractional gap} is \(s-f_I(t)>0\).
	
	\begin{example}[a compatible target without an integral witness]
		Two labels have columns \(2,3\) and costs \(0,0\). Then
		\[
		\Lambda_I=\Z,\qquad Z_I=[0,5],\qquad
		\{A_Ix\mid x\in\{0,1\}^2\}=\{0,2,3,5\}.
		\]
		The query \((1,1)\) passes both tests because \(1=3-2\) lies in
		the lattice and the LP can take half of the weight-\(2\) label
		at cost zero. No integral selection reaches \(1\).
	\end{example}
	
	\subsection{A proximity reduction with binary multiplicities}
	
	Discard records with zero multiplicity, nonpositive profit, or weight
	above \(C\), and set their output counts to zero. If all remaining
	copies fit, take them all. Otherwise use the proximity reduction
	of~\cite[Section~6]{Bringmann24} directly on the multiplicity runs.
	
	\begin{lemma}[compressed proximity kernel]\label[lemma]{lem:kernel}
		In \(\Ot(n)\) arithmetic and comparison operations on \(O(B)\)-bit integers, one can either return an optimal count vector or compute fixed counts \(f\), retained multiplicities \(r\), and a residual capacity \(C'\) such that
		\[
		\begin{gathered}
			0\le f_i,\quad 0\le r_i,\quad f_i+r_i\le u_i,\qquad
			\sum_{\substack{1\le i\le n\\w_i=a}}r_i\le 4\wm \quad (a\in[\wm]),\\
			0\le C'\le \wm^2(\wm+1)+\wm-1,\qquad
			\OPT(w,p,u,C)=p^\top f+\OPT(w,p,r,C').
		\end{gathered}
		\]
		Every residual feasible count vector \(y\) lifts to the original feasible vector \(f+y\). Adding one weight-one, zero-profit record with multiplicity \(C'\) makes the residual constraint an equality and gives \(U=O(\wm^3)\) virtual labels, represented by at most \(n+1\) records.
	\end{lemma}
	
	The proof is given in \cref{app:kernel}.
	
	\begin{example}[keeping a short interval of copies]\label[example]{ex:compressed-kernel}
		Set \(N=10^{12}\). Take the records \((3,7,N)\) and \((3,5,5)\)
		with capacity \(3(N+1)+1\). The greedy prefix takes \(g_3=N+1\)
		copies of weight three. The truncation in \cref{app:kernel} sets
		\[
		L_3=g_3-2\wm=N-5,\qquad
		R_3=\min(N+5,g_3+2\wm)=N+5.
		\]
		Fixing \(L_3\) copies leaves two runs of five copies each and
		residual capacity \(19\). The counts are obtained by intersecting
		the retained interval \((L_3,R_3]\) with the two original runs.
	\end{example}
	
	Before perturbation, a copy of record \(i\) costs \(-p_i\), and a
	slack label costs zero. The common integer perturbation of
	\cref{lem:perturb} preserves the ordering of distinct integral
	selection costs. Throughout the recursion \(c_i\) denotes this
	perturbed cost in the stored orientation. Since \(U=O(\wm^3)\), the
	root costs and strict half-integral thresholds have \(O(B)\) bits.
	
	\subsection{Integer convolution with witnesses}
	
	\begin{definition}[integer convolution exponent]\label[definition]{def:minconv}
		For arrays \(A,B'\) of lengths \(1\le\ell_A,\ell_B\le m\) with integer
		or \(+\infty\) entries, their \((\min,+)\)-convolution
		\(D=A\star B'\) is given by
		\[
		D[k]=\min_{\substack{0\le i<\ell_A,\ 0\le j<\ell_B\\i+j=k}}
		\bigl(A[i]+B'[j]\bigr)
		\qquad (0\le k<\ell_A+\ell_B-1).
		\]
		A witnessed convolution also returns an attaining pair \((i,j)\)
		for each finite \(D[k]\). For \(b\)-bit finite entries, an exponent
		\(\rho\) means that the computation uses
		\(\Ot(m^\rho)\operatorname{poly}(b+\log(1/\eta))\) bit operations
		and has joint error probability at most \(\eta\).
	\end{definition}
	
	\begin{lemma}[recovering convolution witnesses]\label[lemma]{lem:minconv-witness}
		Any algorithm computing the convolution values in
		\cref{def:minconv} gives an algorithm of the same exponent that
		also returns witnesses.
	\end{lemma}
	
	\begin{proof}
		Replace each finite \(A[i]\) by \(mA[i]+i\) and each finite \(B'[j]\)
		by \(mB'[j]\), leaving infinite entries unchanged. For a finite
		output the new value is \(mD[k]+i_*\), where \(i_*\) is the least
		first index among minimizing pairs. Indeed, the values before the
		perturbation are integral and \(0\le i<m\), so the low digit
		breaks ties without changing the minimizing cost. Euclidean division
		by \(m\) recovers \(D[k]\) and \(i_*\), and the second index is
		\(k-i_*\). This uses one value call and adds \(O(\log(m+2))\) bits.
	\end{proof}
	
	\begin{lemma}[integer convolution from 3SUM]\label[lemma]{lem:minconv}
		Witnessed integer convolution in the sense of \cref{def:minconv} has exponent \(\rho_0=784055881637/400000000000=1.9601397040925\).
	\end{lemma}
	
	The proof is given in \cref{app:minconv}.
	
	\section{The Recursive Algorithm and Source Accounting}
	\label{sec:global}
	
	At each binary-search threshold the compressed kernel gives one
	rank-one query. The recursion processes batches on a common pool
	and returns a feasible selection when the root query succeeds.
	
	Line and planar atlases serve sources born with zero or one lift.
	Grid atlases serve later sources until \(k_{\rm scan}\) lifts, after
	which new sources use grouped LP. A face restriction routes its
	queries through the answering source's affine map. Rank drops
	preserve the accumulated lift count. The scale and scan threshold
	are set before recursion as in \cref{subsec:transfer-exponent}.
	\Cref{fig:dispatch} shows the dispatch on a stored batch.
	
	\DispatchFigure
	
	\Cref{alg:bounded} gives the outer binary search and the count recovery.
	
	\begin{algorithm}[H]
		\caption{\textsc{BoundedKnapsack}}
		\label{alg:bounded}
		\begin{algorithmic}[1]
			\Require Records \((w_i,p_i,u_i)_{i=1}^n\) and capacity \(C\)
			\Ensure An optimal count vector \(x\) on the joint success event
			\State \Call{BoundedKnapsack}{(w_i,p_i,u_i)_{i=1}^n,C}
			\Begin
			\State Apply \cref{lem:kernel} to obtain \(x^*\) or \((f,r,C')\).
			\If{the kernel returns \(x^*\)}
			\State \Return \(x^*\)
			\EndIf
			\State Build \(I_0\) from \((w_i,p_i,r_i)_{i=1}^n\) and slack record \((1,0,C')\).
			\State Use the prescribed scale and scan threshold \((\theta,k_{\rm scan})\).
			\State Set the buffers and perturb the costs using the uniform reboxing bound.
			\State \(D_0\gets\sum_{i\in I_0}|c_i|\) using compressed cost sums.
			\State Binary-search \(z\in[-D_0,D_0]\cap\Z\) for the least accepted \(z_*\).
			\Statex At each test, build the root atlas for \((C',z+1/2)\) and process its phases and exits.
			\State Retrieve a witness at \((C',z_*+1/2)\) and trace its attaining sources.
			\State Recover \(x\) by adding \(f\), sweeping original runs, and dropping slack.
			\State \Return \(x\)
			\End
		\end{algorithmic}
	\end{algorithm}
	
	\subsection{States, sources, and transfers}
	\label{subsec:source-states}
	
	The batches arise from the horizontal reductions
	of~\cite[Sections~6--10 and~12]{CMZOpt26}. We analyze them at fixed
	rank and suppress polynomial factors in the common bit bound.
	
	A state consists of an immutable pool \(I\), an arithmetic depth
	\(j\), an accumulated lift count \(\ell\), and a batch \(\mathcal Q\)
	of strict queries. The batch succeeds if at least one \(q=(t_q,s_q)\in
	\mathcal Q\) has an integral witness. Each retained query satisfies
	\(t_q\in\Lambda_I\) and \(g_q=s_q-f_I(t_q)>0\). Its source stores
	the offsets and witness needed to extend a solution to the parent.
	
	Arithmetic depth increases on proper restorations. The other
	reductions preserve it, and coalescence takes the maximum incoming
	depth. Fix decreasing positive scales \(H_j\) and \(\xi_j<1\).
	For a retained query define
	\[
	Q_q=\{z\mid f_I(t_q+z)+f_I(t_q-z)\le 2s_q\},
	\qquad S_q=Q_q\cap H_j P_I .
	\]
	The buffer and its mass are
	\[
	D(I,j)=\Lambda_I\cap
	\bigcup_{q\in\mathcal Q}\bigl(t_q+(\xi_jQ_q\cap H_jP_I)\bigr),
	\qquad M(I,j)=|D(I,j)|.
	\]
	Since \(0<\xi_j<1\), convexity of \(Z_I\) puts every buffered point
	in \(Z_I\). The buffer contains all the centers, so \(M(I,j)\)
	bounds the number of distinct targets.
	
	A \emph{source} records the child pool version, the birth depth and lift
	count, the affine coordinate map, a rational query window \(K\), and the attaining
	witnesses used by the transfer. For each compatible target \(t\) of
	the batch, its window covers \(t+H_{\mathrm{birth}}P_I\), including
	offsets that can fall outside the feasible domain. For a pool of
	rank \(r\ge1\), the construction charge is
	\[
	V(K,I)=1+
	\frac{\operatorname{vol}_r(K+6P_I)}{\det\Lambda_I}.
	\]
	A nonempty rank-zero window has charge one.
	The source pointer stays with its query if the query joins a larger
	batch. A later depth can only shrink the halo that the pointer must
	serve.
	
	A generalized arithmetic progression (GAP) is the image of an integer
	box under an affine map, and is proper when that map is injective on
	the box. The progression coordinates have a linear readout to the
	old columns. A \emph{reserve} is a disjoint labeled subpool whose
	subset sums fill a translated box in those coordinates.
	\Cref{sec:compressed-reserve} constructs the reserve and explains
	how to recover an actual subset for an old-coordinate deficit.
	
	A lift appends an integral coordinate to each surviving column.
	The old equation is recovered by projection before reboxing and
	by a stored coordinate map after reboxing. An arithmetic phase is a
	maximal part of the recursion in which the stored coordinates are
	fixed and only proper full-rank lattice restorations are followed.
	We use column Hermite normal form (HNF) in one fixed basis to
	identify lattices inside a phase.
	
	Boundary queries restrict the pool to their minimal faces. Interior
	queries are assigned to anchors, whose windows determine whether a
	thin transfer or a deep trial applies. \Cref{alg:process-state}
	gives the dispatch, and \cref{sec:deep} proves the deep branch's
	witness guarantees.
	
	\Cref{alg:generate-source} applies to thin, trial, restoration,
	and lift sources. At rank zero a retained batch has buffer \(\{0\}\)
	and mass one. Its empty child selection is extended through the
	saved source and endpoint template. A face inherits its source
	through the affine embedding in \cref{subsec:source-halo}.
	
	\begin{algorithm}[H]
		\caption{Generate and filter one source}
		\label{alg:generate-source}
		\begin{algorithmic}[1]
			\Require Parent anchor, allowance table \(S\), nonface transfer, child state \((R,j',\ell')\)
			\Ensure Retained queries with attaining witnesses and source pointers
			\Begin
			\State Build the witnessed removal table \(D\) in integer coordinates.
			\State \(A\gets-\bigl((-S)\star D^{\leftarrow}\bigr)\), where \(D^{\leftarrow}(v)=D(-v)\).
			\State Decode and crop the finite entries. For a lift, enumerate the strip heights.
			\State Let \(\mathcal C\) be the resulting triples \((y,a,\omega)\) of target, allowance, and witness.
			\State \(\mathcal B\gets\{(y,a,\omega)\in\mathcal C\mid y\in\Lambda_R\}\)
			\If{\(\mathcal B=\emptyset\)}
			\State \Return \(\emptyset\)
			\EndIf
			\If{\(\operatorname{rank}(\Lambda_R)=0\)}
			\State At \(0\), retain the greatest positive allowance with its source, if any.
			\State \Return the resulting batch, with unit window charge.
			\EndIf
			\State Construct the query window \(K\) covering \(y+H_{j'}P_R\) for each entry of \(\mathcal B\).
			\If{\(\ell'\ge k_{\rm scan}\)}
			\State Use the grouped LP oracle on \(R\).
			\ElsIf{\(\ell'<2\)}
			\State Build the line or planar atlas on \(K\) in the current rank.
			\Else
			\State Build the grid atlas on \(K\) with its own block width \(h\).
			\EndIf
			\State \(\mathcal Q'\gets\emptyset\)
			\ForAll{\((y,a,\omega)\in\mathcal B\)}
			\If{\(f_R(y)<a\)}
			\State Insert \((y,a,\omega)\) with its source pointer into \(\mathcal Q'\).
			\EndIf
			\EndFor
			\State \Return \(\mathcal Q'\)
			\End
		\end{algorithmic}
	\end{algorithm}
	
	\subsection{Efficient anchor marking}
	
	For an interior query \(v\), use a rational parallelotope \(R_v\)
	with
	\[
	S_v\subseteq R_v\subseteq C_r S_v,
	\]
	where \(C_r\ge 2r\) depends only on the rank. We give its construction
	below. Scan the distinct interior targets by decreasing gap, breaking
	ties by target key. Each unmarked query becomes an anchor. It marks
	and claims all unmarked queries in \(t_v+\eta R_v\) by enumerating the
	lattice points of this window and looking up their target keys. Let
	\(\mathcal V\) be the resulting set of anchors.
	
	\begin{lemma}[marking and locality]
		\label[lemma]{lem:marking}
		Take \(0<\eta\le 1/(100 C_r)\), \(b=\eta/4\), and
		\(C_r\eta\le\gamma\le 1/100\). Assume \(\xi_j\ge b\). Put
		\(\mu_v=|(t_v+(b/2)S_v)\cap\Lambda_I|\) for each anchor.
		Then \(\sum_{v\in\mathcal V}\mu_v\le M\). Every query \(q\) assigned to \(v\)
		satisfies
		\[
		t_q-t_v\in\gamma S_v,\qquad g_q\le g_v,\qquad
		Q_q\subseteq \frac{1+\gamma}{1-\gamma} Q_v .
		\]
		The lattice enumeration for marking has total size \(O_r(M)\).
	\end{lemma}
	
	\begin{proof}
		Write \(\mathcal E\) for the epigraph of \(f_I\).
		For \(p=(t,s)\in\mathcal E\) let
		\(\mathcal M_p=(\mathcal E-p)\cap(p-\mathcal E)\).
		Its horizontal projection is \(Q_p\) because the two reflected vertical
		coordinates can be chosen exactly when the sum of the two fractional
		values is at most \(2s\).
		If \(p'=p+d\) for \(d\in\delta\mathcal M_p\), convex combinations
		with \(p\pm d/\delta\) give
		\[
		(1-\delta)\mathcal M_p\subseteq\mathcal M_{p'}
		\subseteq(1+\delta)\mathcal M_p.
		\]
		The gap at \(p'\) is at least \((1-\delta)g_p\) by concavity of
		\(s-f_I(t)\). Applying concavity to
		\(p=(p'+\delta(p-d/\delta))/(1+\delta)\) also bounds this gap by
		\((1+\delta)g_p\).
		At a fixed horizontal target, replacing a positive gap \(g\) by
		\(g'\) enlarges the projected body by at most
		\(\max(1,g'/g)\). This follows by scaling the sublevel set of the
		convex function
		\(u\mapsto f_I(t+u)+f_I(t-u)-2f_I(t)\), which is zero at \(u=0\).
		A marked target \(q\) admits a lift to \(p'\in v+\gamma\mathcal M_v\)
		at horizontal target \(t_q\). Applying the gap comparison at \(t_q\)
		proves the displayed locality bound.
		
		Suppose \(t_v+bS_v\) and \(t_q+bS_q\) intersect for two anchors
		with \(v\) earlier in the order. Lift the common horizontal point
		separately into \(v+b\mathcal M_v\) and \(q+b\mathcal M_q\).
		At the common horizontal target, the second lifted gap is at most
		\((1+b)/(1-b)\) times the first. Combining the two projected-body
		comparisons gives
		\(Q_q\subseteq((1+b)/(1-b))^2 Q_v\), and intersecting with \(H_jP_I\)
		gives the same bound on \(S_q\). Thus
		\(t_q-t_v\in b(1+((1+b)/(1-b))^2)S_v\subseteq\eta R_v\).
		The earlier anchor would have marked \(q\). The smaller paid
		windows are consequently disjoint and belong to the parent buffer.
		
		For any rank-\(r\) lattice \(\Lambda\), \(t\in\Lambda\), an
		origin-symmetric convex body \(S\), \(C>0\), and integer \(k\ge1\),
		\begin{equation}\label{eq:lattice-dilation}
			|(t+CS)\cap\Lambda|
			\le k^r |(t+(2C/k)S)\cap\Lambda|.
		\end{equation}
		In each occupied coset modulo \(k\Lambda\), fix a point \(z\)
		of \((t+CS)\cap\Lambda\). The map \(x\mapsto t+(x-z)/k\)
		injects the points of that coset into the smaller set. For the
		marking window use \(C=\eta C_r\) and \(k\) with \(2C/k\le b/2\).
		Since \(b=\eta/4\), this gives \(O_r(\mu_v)\) points per anchor.
		Fixed-dimensional integer feasibility on the explicit facets
		enumerates them with a polynomial bit factor per output and window.
	\end{proof}
	
	A separation query for \(S_v\) first enforces \(H_jP_I\), so the
	two LP queries at \(t_v\pm z\) lie in the source halo. A domain
	separator handles an infeasible offset. If both offsets are feasible,
	the sum of their supporting affine functions separates any violation
	of the \(Q_v\) inequality.
	
	Positive gap at an interior target makes \(S_v\) full-dimensional.
	The facets of \(H_jP_I\), \(Z_I\), and the piecewise-affine functions
	\(f_I(t_v\pm z)\) have polynomial encoding length by fixed-dimensional
	determinant bounds on columns and endpoint sums. Hence \(S_v\) has
	rational inner and outer radius bounds of polynomial bit length.
	Start with a basis \(v_1,\ldots,v_r\in S_v\) from the inner box.
	With \(A=[v_1\ \cdots\ v_r]\), maximize each coordinate of
	\(A^{-1}z\) over \(S_v\) in both signs by strong separation and exact
	optimization~\cite{GLS88}, using an optimizing vertex at each step.
	Replace a basis vector if the corresponding
	absolute coordinate exceeds \(2\). Each replacement more than
	doubles \(|\det A|\), and the radius bounds give a polynomial number
	of replacements. At termination take \(R_v=A[-2,2]^r\), which contains
	\(S_v\) and lies in \(2rS_v\) by convexity and symmetry.
	
	For the thin test, use the facets of \(R_v\) and test
	\(C_rK_*P_I\subseteq R_v\). A passing test proves
	\(K_*P_I\subseteq S_v\). If the test fails at a facet \(\lambda\),
	the light columns \(H_j a_i\in R_v\) obey
	\[
	|\lambda a_i|\le h_{R_v}(\lambda)/H_j
	< (C_r K_*/H_j)h_{P_I}(\lambda),
	\]
	where \(h_K\) denotes the support function of \(K\).
	In the same-rank case the light hull therefore occupies a central
	slab of relative volume \(O_r(K_*/H_j)\). The lattice covolume
	cannot decrease in such a restriction. The slab bound follows by
	comparing the maximal, central section of \(P_I\) with the double
	cone from this section to the two extreme supporting points.
	
	Fix a supporting price \(\pi\) and a basic optimum \(x^0\) at
	the anchor \(t_v\). The normalized LP value
	\(F_v(t)=f_I(t)-f_I(t_v)-\pi\cdot(t-t_v)\) is nonnegative
	and vanishes at \(t_v\). For \(z\in Q_v\), the two values
	\(F_v(t_v\pm z)\) sum to at most \(2g_v\). Convexity, and then
	the marking bounds \(t_q-t_v\in\gamma Q_v\) and \(g_q\le g_v\),
	give
	\begin{equation}\label{eq:anchor-budget}
		\begin{aligned}
			F_v(t_v+\xi z)&\le 2\xi g_v &&(z\in Q_v,\ 0\le\xi\le1),\\
			a_q&=s_q-f_I(t_v)-\pi\cdot(t_q-t_v)
			=g_q+F_v(t_q)\le(1+2\gamma)g_v .
		\end{aligned}
	\end{equation}
	Here \(a_q\) is the normalized allowance of the assigned query.
	For the thin crop let \(u=t_q-t_v\) and let \(x\) be an integral
	witness for \(q\). By \eqref{eq:anchor-budget}, \(x\) has normalized
	cost below \(a_q\), and \(t_v-u/\gamma\) has a fractional witness
	of normalized cost at most \(2g_v\).
	For every vector \(\delta\) with
	\(|\delta_i|\le |x_i-x_i^0|\), the vectors \(x+x^0\pm\delta\)
	lie in \([0,2]^I\). Adding \(\gamma\) times the reflected
	fractional witness and dividing by \(2+\gamma\) produces a pair
	at \(t_v\pm A_I\delta/(2+\gamma)\), whose average normalized
	cost is at most \((1+4\gamma)g_v/(2+\gamma)<g_v\).
	Consequently the whole deviation zonotope lies in
	\((2+\gamma)Q_v\subseteq 3Q_v\).
	
	Choose a maximum-determinant basis in \(Q_v\), with dual
	functionals \(\lambda_1,\ldots,\lambda_r\). Each satisfies
	\(|\lambda_j(u')|\le1\) for \(u'\in Q_v\), and the gauge of
	a column \(a_i\) with respect to \(Q_v\) is at most
	\(\sum_{j=1}^r|\lambda_j(a_i)|\).
	Applying these functionals to the deviation zonotope bounds the
	sum of \(|x_i-x_i^0|\) times the column gauges by \(3r\).
	For a heavy endpoint, \(H_j a_i\notin R_v\).
	Since \(H_j a_i\in H_jP_I\) and \(S_v\subseteq R_v\),
	the \(Q_v\)-gauge of \(a_i\) is greater than \(1/H_j\).
	Thus the witness changes fewer than \(3rH_j\) heavy endpoints.
	Every subset sum of these signed changes lies in
	\(3Q_v\cap3rH_jP_I\subseteq3rS_v\).
	Each changed endpoint column itself lies in \(3S_v\), because
	its singleton deviation also lies in \(P_I\).
	Branch on the heavy fractional labels and leave all light labels
	in the child. The reference on the removed labels is then integral.
	These bounds give the required physical crop and
	allow all relevant endpoint directions to be found by enumerating
	\(3R_v\cap\Lambda_I\). Light directions satisfy
	\(H_ja_i\in R_v\), so the same enumeration finds every light
	type when \(H_j\ge1\).
	
	\subsection{Parameters and buffer schedule}\label{sec:scales}
	
	Fix rational constants \(0<\theta<1\) and \(\beta>0\), and set
	\(\nu=\lceil\wm^\theta\rceil\). The final enlargement of
	\(H_{\min}\) in \cref{thm:ledger} depends on this prescribed
	\(\beta\), before any queries are processed. Fix a
	prospective rank bound \(D=1+\lceil 4/\theta\rceil\) and a depth bound
	\(J=D+\lceil\log_2(2+\wm)\rceil+2\). A rank-zero
	state is terminal. The root has normalized volume at most \(\wm\).
	The volume bounds in \cref{thm:ledger,lem:lift-volume} justify
	\(D\) and \(J\) by induction, so the geometric and structural
	constants can be fixed for ranks through \(D\) before recursion.
	
	Choose the constant in the progression theorem sufficiently small
	for all these ranks as in \cref{lem:count-gap}. Set \(k_0\) to a
	uniform integer bound on the outlier and reserve counts, with
	\(\nu\le k_0=\Ot(\nu)\). Such a bound is available before the first
	query. In fact, the scalar input range in \cref{lem:count-gap} is
	polynomial in \(\wm\), and the two counts are bounded by a constant
	times \(\nu\log^2(\wm+2)\). Take \(\kappa=3k_0\), the same for every
	threshold core in the recursion. For \(\wm\) below the fixed cutoff
	in the rank, progression, and geometric estimates, one may expand
	the bounded-size kernel of \cref{lem:kernel} and solve it by ordinary
	capacity dynamic programming after reading the input.
	
	Choose \(C_*\ge \max_{1\le r\le D}C_r\) to bound the rounding constants, and put
	\[
	\xi_*=2^{-10},\qquad
	\eta=\frac{\xi_*}{10000C_*(J+1)},\qquad
	\gamma=C_*\eta,\qquad b=\eta/4,\qquad
	\tau_{\gamma}=\frac{1+\gamma}{1-\gamma}.
	\]
	Use \(E=(260+6J)k_0\) and \(K_*=E/\gamma\). In the thin test, an anchor
	is deep only if \(C_r K_*P_I\subseteq R_v\), and hence
	\(K_*P_I\subseteq S_v\). The scale decrements are
	\[
	\Delta_\xi=\gamma+(\tau_{\gamma}-1)\xi_*+E/K_*,
	\qquad
	\xi_j=\xi_*/4+(J-j)\Delta_\xi .
	\]
	For \(0\le j\le J\), we have \(\xi_*/4\le\xi_j<\xi_*/2\).
	For \(j<J\) this gives
	\(\xi_j\ge\gamma+\tau_{\gamma}\xi_{j+1}+E/K_*\).
	Choose \(H_{\min}\ge\max(K_*,JE)\) and enlarge it by the slab factor
	in the proof of \cref{thm:ledger}. With
	\(\Delta_H=2E+2\gamma H_{\min}\) and
	\(H_j=H_{\min}+(J-j)\Delta_H\), we have
	\[
	H_j-H_{j+1}\ge\gamma H_j+E,\qquad H_0\le4H_{\min}.
	\]
	Indeed, \(\Delta_H-\gamma H_0-E=(1-2\gamma J)(E+\gamma H_{\min})\ge0\).
	All scales are rational and can be computed at polynomial bit cost.
	The ratio \(H_j/K_*\) is polylogarithmic, \(H_j=\Ot(\nu)\), and
	\(bK_*=\Omega_D(\nu)\).
	
	\subsection{Paying for the windows before filtering}
	\label{subsec:source-halo}
	
	The coefficient box bounds the work of convolution. Source
	construction uses the physical window after cropping, including
	the offsets needed by future LP queries. The following bound
	charges its enlarged volume to lattice points at the parent.
	
	\begin{lemma}[a source halo charged to a parent]
		\label[lemma]{lem:source-halo}
		Let \(\Gamma\subseteq\Lambda\) have rank \(d\ge1\), let \(P\) be the
		symmetric hull of generators of \(\Gamma\), and let
		\(K\subseteq\operatorname{span}_{\R}\Gamma\) be compact. Suppose that
		for some \(t,b_0\in\Lambda\), some centrally symmetric convex \(S\),
		and positive \(C,\alpha\),
		\[
		b_0+K+(6+d/2)P\subseteq t+CS .
		\]
		Then, for \(k=\max(1,\lceil 4C/\alpha\rceil)\),
		\[
		1+\frac{\operatorname{vol}_d(K+6P)}{\det\Gamma}
		\le (1+k^{\operatorname{rank}\Lambda})
		|(t+(\alpha/2)S)\cap\Lambda|.
		\]
	\end{lemma}
	
	\begin{proof}
		An independent set of \(d\) actual generators gives, by rounding
		its real coefficients, a lattice point within \((d/2)P\) of every
		point of the span. Taking a translation-equivariant nearest lattice
		point in the \(P\)-norm gives a measurable fundamental domain of
		\(\Gamma\) inside \((d/2)P\). The translates meeting \(K+6P\) have
		centers in \(K+(6+d/2)P\). They cover \(K+6P\) and each has
		volume \(\det\Gamma\). After adding \(b_0\), their centers belong
		to \((t+CS)\cap\Lambda\). Inequality \eqref{eq:lattice-dilation}
		bounds the number of centers by
		\(k^{\operatorname{rank}\Lambda}|(t+(\alpha/2)S)\cap\Lambda|\).
		The last set contains \(t\), which pays for the additional one.
	\end{proof}
	
	In a thin branch let \(F\) be the heavy fractional labels and \(H\)
	the other heavy labels. For each \(i\in H\) set \(e_i=x_i^0\).
	For each branch \(b\in\{0,1\}^{F}\) use the offsets
	\[
	t_b=\sum_{i\in H}e_i a_i+\sum_{i\in F}b_i a_i,\qquad
	s_b=\sum_{i\in H}e_i c_i+\sum_{i\in F}b_i c_i.
	\]
	Run the transfer on \((t_q-t_b,s_q-s_b)\) with the signed toggles
	on \(H\) and the light pool as child. Store \(b\) with the endpoint
	template. Before building the table, keep at most
	\(\lceil 3rH_j\rceil\) cheapest eligible toggles per signed
	column. The deviation bound limits the total number of heavy
	toggles of any required witness, so this cap preserves a choice
	of no larger cost at the same displacement. The sums on \(H\)
	come from the anchor's witness
	aggregates by subtracting fractional contributions and light endpoint
	sums, including any light tight label at coefficient one. Remote
	heavy endpoints stay in the parent template. Thus the translated
	window is centered at \(t_v-t_b\), and each witness uses the branch
	matching its choices on \(F\).
	
	For a thin source, the physical crop and the bounded displacement
	of assigned centers put the prefilter targets, after one fixed
	translation, in \(t_v+O_r(1)S_v\).
	The full halo lies in an additional constant dilation because
	\(H_jP_{\mathrm{light}}\subseteq R_v\subseteq C_rS_v\).
	For a lower-rank child, first apply its lattice membership test. If
	the candidate set is nonempty, pick any compatible candidate \(t_0\)
	as a base point. Candidate differences and the child halo are in
	\(O_r(1)S_v\cap\operatorname{span}_{\R}\Gamma\). Round this
	symmetric section inside the child span and translate it by \(t_0\).
	This gives an explicit constant-facet window. The translation in
	Lemma~\ref{lem:source-halo} is integral. This step uses lattice
	compatibility and does not require \(t_0\) to survive the LP test.
	There are at most \(2^r\) fractional branches.
	
	For a deep trial or restoration, the physical crop gives displacement
	\(O(E)P_I\), and the child hull lies in \(P_I\). After an integral
	translation, the child targets and their full halo lie in
	\(t_v+\gamma S_v+O(E+H_j)P_I\). Since \(K_*P_I\subseteq S_v\),
	they therefore lie in
	\[
	t_v+\bigl(\gamma+O((E+H_j)/K_*)\bigr)S_v .
	\]
	Round this dilation to a parallelotope and apply
	Lemma~\ref{lem:source-halo}. A restored sublattice uses its own
	covolume in this calculation. If multiple parents reach it, each
	source keeps its window. Taking the convex hull of all such windows
	would introduce uncharged intervening regions.
	
	For a lift, first enumerate the eligible integer heights in the
	uniform strip, then form the slanted prism. Its horizontal projection
	is in the same enlarged parent window as for a deep trial. In the
	sheared coordinates, adding the full child halo increases its height
	by \(O_r(H_j/\nu)\). The fixed navigation halo adds only a
	constant multiple of the child hull. The projected lattice is a
	sublattice of the old lattice and the vertical kernel has a
	positive integer period. In coordinates with the vertical unit
	vector last, the covolume is the projected covolume times that
	period. The normalized prism volume is therefore bounded by the
	normalized volume of the enlarged old window times a polylogarithmic
	factor. Lemma~\ref{lem:source-halo} bounds the first factor by the
	paid mass, up to polylogarithms. The original strip before adding
	the halo has only \(O_r(1)\) integer heights per horizontal
	candidate. Enumeration thus has the same order of cost as for the
	horizontal candidates.
	
	\Cref{lem:lift-volume} uses the same shear to bound the normalized
	column volume of the lifted pool.
	
	A boundary target gives a different kind of source. For its minimal
	face \(F\), an exposing functional fixes every nonparallel type at
	one of its endpoints. Let \((t_{\mathrm{fix}},c_{\mathrm{fix}})\)
	be their target and cost. With the remaining pool \(I_F\) we have
	\[
	f_{I_F}(y)=f_I(y+t_{\mathrm{fix}})-c_{\mathrm{fix}}
	\]
	on the affine face, with value \(+\infty\) at points of that affine
	space outside the face. The embedding sends the restricted child
	hull into the parent hull, so it sends each actual child offset into
	the existing parent halo. Restrict the parent atlas and pull back its prices and separators.
	For an infeasible point in the face affine span, the restricted
	separator is nonzero because it strictly separates the point from
	the nonempty face. Restrict the basic witness to free labels. Cache the free pool, its
	lattice, and the fixed offsets by the canonical minimal face.
	For each translated target \(y=t-t_{\rm fix}\), test
	\(y\in\Lambda_{I_F}\) using the cached basis before retaining it.
	This test is charged to the face query even when it rejects.
	Any integral face witness uses the forced endpoints and a subset
	of the free pool, so the lattice test preserves every necessary
	target. The positive LP gap is unchanged by the fixed cost shift.
	In atlas mode this metadata is available from the line, planar,
	or grid atlas (\cref{lem:line,lem:planar,lem:grid-atlas}). In
	late mode the grouped oracle can report the free types in its
	charged type scan (Lemma~\ref{lem:grouped}). Inherited face queries
	do not create a new full-halo atlas.
	Under the fixed translation, the face buffer lies in the parent
	buffer. Since \(\xi_j<1\), its points lie in the relative
	interior of that minimal face. Grouping all centers for the same
	face makes the sum of face-child masses at most \(M\).
	
	\Cref{alg:process-state} puts these cases together for one stored state.
	The deep branches are justified in
	\cref{lem:trial-restoration,lem:high-trial-center,lem:witness-strip}.
	
	\begin{algorithm}[H]
		\caption{Process a stored state}
		\label{alg:process-state}
		\begin{algorithmic}[1]
			\Require State \((I,j,\ell,\mathcal Q)\) with a merged feasible batch
			\Ensure Terminal witnesses \(\mathcal W\) and retained exits \(\mathcal E\), with source pointers
			\Begin
			\If{\(\rank(\Lambda_I)=0\)}
			\State \Return the empty selection with an attaining source, with \(\mathcal E=\emptyset\)
			\EndIf
			\State \(\mathcal W\gets\emptyset\), \(\mathcal E\gets\emptyset\)
			\State Group boundary queries by their minimal face \(F\subsetneq Z_I\).
			\State Translate each face group and put its lattice-compatible entries in \(\mathcal E\).
			\State Mark the interior anchors \(\mathcal V\) in decreasing order of \(g_q=s_q-f_I(t_q)\).
			\ForAll{\(v\in\mathcal V\)}
			\If{\(C_r K_*P_I\not\subseteq R_v\)}
			\State Enumerate the thin directions and generate each fractional branch into \(\mathcal E\).
			\Else
			\State Build the canonical-core trial pool \(I_\circ\) and filter its source to obtain \(\mathcal Q_\circ\).
			\State \(\mathcal H_v\gets\{q\in\mathcal Q_\circ\mid a'_q\ge g_v/4\}\), with \(a'_q\) the normalized residual allowance.
			\If{\(\mathcal H_v\ne\emptyset\)}
			\State Choose \(q_0\in\mathcal H_v\) and compute its affordable fiber \(\mathcal K\) once.
			\State Apply the point-or-thin procedure to \(\mathcal K\).
			\Statex On a point, round and repair by reserve DP into \(\mathcal W\). On a thin direction, send the clipped lift to \(\mathcal E\).
			\ElsIf{\(\mathcal Q_\circ\ne\emptyset\)}
			\State \(\Gamma\gets\Lambda_{I_\circ}\), \(R\gets I\cap\Gamma\)
			\State Rebuild the table on \(I\setminus R\) and filter the arithmetic source into \(\mathcal E\).
			\EndIf
			\EndIf
			\EndFor
			\State \Return \((\mathcal W,\mathcal E)\)
			\End
		\end{algorithmic}
	\end{algorithm}
	
	For each anchor the marking scan, prefilter candidate scans, lattice
	membership tests, new source volume, and caller-level queries total
	\(\Ot(\mu_v)\), with the bit factor for the rounding calls. The
	coefficient boxes for witnessed tables are also \(\Ot(\mu_v)\)
	in cardinality. Apply \cref{lem:source-box} after rounding an origin-symmetric
	dilate of \(S_v\). For a lift
	the merge on old targets uses the horizontal box and then enumerates
	the bounded strip. The distinction between box entries and
	physical-window entries ensures that an all-rejected batch is still
	charged in full.
	
	For the actual child masses the bound is independent of the
	navigation halo's height. In a thin child, the buffer displacement
	lies in \(H_jP_{\mathrm{light}}\), which is contained in \(R_v\). The integral
	reference therefore sends the whole buffer into
	\(t_v+O_r(1)S_v\), so \eqref{eq:lattice-dilation} bounds its mass by
	\(O_r(b^{-r}\mu_v)\) per fractional branch.
	For a lift, the horizontal projection of a child \(Q\)-body
	belongs to the \(Q\)-body of its attaining parent query: extend
	each opposite-point witness by the same removal assignment.
	Intersecting its projection with the horizontal clipping hull
	puts the projected buffer displacement in \(O_r(1)S_v\).
	The remaining displacement from the removal table is \(O(E)P_I\),
	also in \(O_r(1)S_v\) by deepness. By the uniform-strip condition,
	only \(O_r(1)\) integer heights contribute at each projected
	lattice point of the actual buffer. The same \(O_r(b^{-r}\mu_v)\)
	bound follows. Summing over anchors gives a total thin and face
	mass of at most \(A_{\mathrm{geo}}M\), including the face bound
	\(M\) above, and a total lift mass of at most \(A_{\mathrm{lift}}M\).
	Both factors depend polylogarithmically on \(\wm+2\) through
	\(b\), independently of \(H_j/K_*\).
	
	\subsection{Coalescing an arithmetic phase}
	\label{subsec:phase}
	
	Fix the root pool \(I_0\) of a full-rank arithmetic phase and write
	\(\Gamma_0=\Lambda_{I_0}\). A lattice \(\Gamma\) always represents
	\(I_\Gamma=\{i\in I_0\mid a_i\in\Gamma\}\).
	Let \(e_i(\pi)\) be one if \(\pi\cdot a_i>c_i\) and zero otherwise.
	For \(R\subseteq I\) define endpoint extension at a price \(\pi\) by
	\[
	U_{I,R}(y;\pi)=y+\sum_{i\in I\setminus R} a_i e_i(\pi).
	\]
	Here \(\partial f_R(y)\) is the set of prices supporting the fractional
	value at \(y\). The strict inequality defining \(e_i(\pi)\) fixes
	the same tie convention at every source. For a threshold \(\tau\ge0\) let
	\(L_{I,\pi}(\tau)=\{i\in I\mid |c_i-\pi\cdot a_i|\le\tau\}\).
	A \(\kappa\)-resilient pool has the same lattice after deleting any
	at most \(\kappa\) labels. The canonical core
	\(\mathcal R_\kappa(Y)\) is the union of all resilient subpools of
	\(Y\). This union is resilient, because after any permitted
	deletion the survivors generate the lattice of every constituent
	subpool. In particular, the core is unique and monotone.
	
	\begin{definition}[certified phase]
		\label[definition]{def:certified-phase}
		Each processed lattice \(\Gamma\subseteq\Gamma_0\) has full rank,
		equals \(\Lambda_{I_\Gamma}\), and has index at most
		\(B_{\mathrm{ind}}\) in \(\Gamma_0\). The associated buffer
		satisfies \(D_\Gamma\subseteq\Gamma\cap\operatorname{dom}(f_{I_\Gamma})\).
		For each nonroot \(\Gamma\) and \(y\in D_\Gamma\), one predecessor
		\(\Pi\supsetneq\Gamma\) satisfies both of the following conditions
		for every \(\pi\in\partial f_{I_\Gamma}(y)\):
		\[
		U_{I_\Pi,I_\Gamma}(y;\pi)\in D_\Pi,
		\qquad
		\Gamma=\Lambda_{\mathcal R_\kappa(L_{I_\Pi,\pi}(\tau))}
		\quad\text{for some }\tau\ge0 .
		\]
		The predecessor is independent of the price. The threshold may
		depend on the price.
	\end{definition}
	
	\begin{example}[an endpoint extension depends on the price]\label[example]{ex:price-extension}
		Let \(R\) have two weight-\(2\) labels of costs \(0\) and \(4\).
		At the interior target \(y=2\), the LP value is zero, and the
		supporting prices form \(\partial f_R(2)=[0,2]\). Let \(I\)
		also contain one weight-\(1\) label of cost \(1\). The prices
		\(1/2\) and \(3/2\) give
		\[
		U_{I,R}(2;1/2)=2,\qquad U_{I,R}(2;3/2)=3.
		\]
		A certificate that covers only one of these images cannot be
		used to trace the child at an arbitrary common price.
	\end{example}
	
	\begin{lemma}[phase mass bound]
		\label[lemma]{lem:phase-mass}
		Let \(\mathcal L\) be the set of processed lattices in a certified
		phase and \(M_\Gamma=|D_\Gamma|\). For \(s\ge1\),
		\[
		\sum_{\Gamma\in\mathcal L}M_\Gamma^s
		\le (1+\lfloor\log_2 B_{\mathrm{ind}}\rfloor)M_{\Gamma_0}^s .
		\]
	\end{lemma}
	
	\begin{proof}
		If \(\pi\) supports \(f_R\) at \(y\), any optimum at \(y\) together
		with the excluded endpoints minimizes the reduced cost on \(I\)'s
		box. Hence \(\pi\) supports \(f_I\) at \(U_{I,R}(y;\pi)\).
		Conversely, every price at the extended target supports the value
		at \(y\) by comparing feasible points extended with these same fixed
		endpoints. More explicitly, the value at the image is
		\(f_R(y)+c_X^\top e_X(\pi)\), where \(X=I\setminus R\), and at
		\(y'+A_Xe_X(\pi)\) the value is at most \(f_R(y')+c_X^\top e_X(\pi)\).
		A new supporting price at the image gives an inequality between these
		two translated points using the original fixed endpoints, even if
		its own tie choices differ. Extensions at a fixed price compose by
		partitioning the excluded labels.
		
		For any choice of supporting prices \(\pi_y\), the map
		\(y\mapsto U_{I,R}(y;\pi_y)\) is injective. If two images are equal,
		their difference dotted with \(\pi_{y_1}-\pi_{y_2}\) is zero.
		The contribution from \(y_1-y_2\) is nonnegative by the two
		supporting inequalities. Each excluded label also gives a
		nonnegative contribution, strictly positive if its chosen endpoint
		differs, by the fixed tie rule. Thus the endpoints agree and
		\(y_1=y_2\).
		
		Every point \(y\in D_\Gamma\) admits a supporting price. Choose
		one price \(\pi_y\) and keep it fixed while composing extensions
		along the predecessor chain to a root point \(u\). Strict
		inclusion makes the chain terminate. The excluded pools partition
		\(I_0\setminus I_\Gamma\), so the root image is
		\(U_{I_0,I_\Gamma}(y;\pi_y)\), independently of the chosen chain.
		Choose a common supporting price \(\pi_*\) at \(u\). The reverse
		support implication shows that \(\pi_*\) supports every point on
		every predecessor chain reaching this \(u\). By certification, each
		lattice along any such chain belongs to the root's nested threshold
		family for \(\pi_*\). To verify this, write
		\(C(\tau)=\mathcal R_\kappa(L_{I_0,\pi_*}(\tau))\)
		and \(\Gamma(\tau)=\Lambda_{C(\tau)}\). Restricting to
		\(I_0\cap\Gamma(\sigma)\) leaves \(C(\tau)\) unchanged for
		\(\tau\le\sigma\), because \(C(\tau)\subseteq C(\sigma)\).
		For \(\tau\ge\sigma\), the restricted core contains \(C(\sigma)\)
		and has all its columns in \(\Gamma(\sigma)\). Its lattice is
		therefore \(\Gamma(\min(\tau,\sigma))\) in either case. Induction
		along the chain now gives the claim.
		
		Strict inclusions in this nested family at least double the
		index. Each lattice contributes at most one owner of \(u\) by
		injectivity for the fixed pair \((I_0,I_\Gamma)\). Thus
		\(\sum_{\Gamma\in\mathcal L}M_\Gamma\) is at most
		\((1+\lfloor\log_2 B_{\mathrm{ind}}\rfloor)M_{\Gamma_0}\).
		The injectivity also gives \(M_\Gamma\le M_{\Gamma_0}\), yielding
		the asserted higher moments.
	\end{proof}
	
	\Cref{alg:arithmetic} processes the lattices in increasing order of
	index so that each receives all incoming queries before it is processed.
	
	\begin{algorithm}[H]
		\caption{Coalesce one certified arithmetic phase}
		\label{alg:arithmetic}
		\begin{algorithmic}[1]
			\Require Phase root \((I_0,j_0,\ell,\mathcal Q_0)\) and local certified transfers
			\Ensure Terminal witnesses \(\mathcal W\) and nonarithmetic exits \(\mathcal E\), with source pointers
			\Begin
			\State \(\Gamma_0\gets\Lambda_{I_0}\), \(\mathcal W\gets\emptyset\), \(\mathcal E\gets\emptyset\)
			\State Initialize the queue with \((\Gamma_0,j_0,\mathcal Q_0)\), keyed by index and HNF.
			\While{the queue is nonempty}
			\State Extract the least-index lattice \(\Gamma\) and its accumulated batch \(\mathcal Q_\Gamma\).
			\State Set \(j_\Gamma\) to the maximum incoming depth.
			\State For each target \(t\) in the batch, retain \(q_t\in\operatorname*{arg\,max}_{q\in\mathcal Q_\Gamma,\ t_q=t}(s_q)\) and its source.
			\State Process \((I_0\cap\Gamma,j_\Gamma,\ell,\{q_t\}_{t})\) as in \cref{alg:process-state}.
			\ForAll{retained proper arithmetic child on lattice \(\Gamma'\subsetneq\Gamma\)}
			\State Insert their batches at \(\operatorname{HNF}(\Gamma')\), preserving the source pointers.
			\EndFor
			\State Add terminal witnesses to \(\mathcal W\) and the other retained exits to \(\mathcal E\).
			\EndWhile
			\State \Return \((\mathcal W,\mathcal E)\)
			\End
		\end{algorithmic}
	\end{algorithm}
	
	\begin{lemma}[certifying the merged buffer]\label[lemma]{lem:merged-certificate}
		The buffers produced by \cref{alg:arithmetic} satisfy
		\cref{def:certified-phase}. For each nonroot buffer point, one
		attaining incoming source gives both certificates at every
		supporting price. Source construction and tests remain charged
		where they were incurred.
	\end{lemma}
	\begin{proof}
		Strict index growth puts all arrivals before processing. The
		largest allowance preserves the decision and retains an attaining
		source. For a merged buffer point \(y\), choose a covering query
		and its stored predecessor before considering any price. Both
		buffer scales decrease with depth, so the maximum incoming depth
		keeps \(y\) inside that query's incoming buffer. This argument is
		used in the analysis and requires no enumeration of buffer points.
		
		Let \(Y\) be the saved assignment on the removed pool \(X\) at
		parent depth \(j\). Write \(y=t_q-A_XY+z_b\). Extending the
		fractional solutions by \(Y\) gives
		\(Q_{\mathrm{child}}\subseteq Q_q\), so
		\(z_b\in \tau_{\gamma}\xi_{j+1}Q_v\cap H_{j+1}P_I\).
		At every supporting price \(\pi\) at \(y\), its extension is
		\(U_{I_\Pi,I_\Gamma}(y;\pi)=t_q+z_b+A_X(e_X(\pi)-Y)\).
		By \cref{lem:restored-collar} the last term is in \(EP_I\).
		Thus the displacement of the extension from \(t_v\) lies in
		\[
		(\gamma+\tau_{\gamma}\xi_{j+1}+E/K_*)Q_v
		\cap
		(\gamma H_j+H_{j+1}+E)P_I .
		\]
		By \cref{sec:scales}, this intersection lies in \(\xi_j Q_v\cap H_jP_I\).
		The extension is an integer parent lattice point.
		The same collar lemma identifies the plateau
		lattice at \(\pi\), giving the second certificate for the same
		predecessor. Each phase uses its own root pool and ledger of
		sources, since two phase roots can have different available labels.
	\end{proof}
	
	\subsection{Global moments and total source volume}
	
	The total cost depends on how the normalized column volume changes
	along a recursion path.
	
	For any nonempty current pool,
	\(\Psi(I)\ge1\), since the symmetric hull contains the crosspolytope
	on an independent set of actual columns. It is nonincreasing on
	restrictions, including rank loss. To see the latter, suppose a
	subpool spans dimension \(d<r\). Extend its span by \(r-d\) original
	columns independent modulo the span. The join of the subpool hull
	with the symmetric crosspolytope on these added columns belongs to
	the parent hull. Integrating the scaled sections of this join gives
	\[
	\frac{r!}{2^r}\operatorname{vol}_r(\text{join})
	=
	\frac{d!}{2^d}\operatorname{vol}_d(P_{\mathrm{sub}})
	\,\bigl|\det(\bar a_1,\ldots,\bar a_{r-d})\bigr|,
	\]
	where bars denote orthogonal projections onto the complementary
	space. The subpool lattice covolume is at least that of the parent
	lattice intersected with the subspace. The projected determinant
	is at least the covolume of the projected parent lattice. Their
	product is at least the parent covolume, proving monotonicity.
	In a strict full-rank arithmetic restriction the new covolume at
	least doubles. In a phase rooted
	at \(I_0\), the factor \([\Gamma_0:\Gamma]\) is at most
	\(\Psi(I_0)/\Psi(I_\Gamma)\le\Psi(I_0)\), because the hull only
	shrinks in fixed coordinates. Since \(\Psi(I_0)=O(\wm)\), each
	phase has logarithmic index chain length. The number \(T\) of
	lifts on any path is bounded in terms of \(\theta\). More explicitly, every lift from rank at most
	\(D\) multiplies \(\Psi\) by at most \(C_D/\nu\le\wm^{-\theta/2}\)
	for sufficiently large \(\wm\), by \cref{lem:lift-volume}. All other
	reductions are nonincreasing. A path with more than \(2/\theta\)
	lifts would have \(\Psi<1\). Up to the first such lift the rank is
	at most \(1+\lceil2/\theta\rceil<D\), so the estimate applies
	without presupposing the rank bound. The arithmetic depth is less
	than \(J\) because a proper arithmetic edge at least halves
	\(\Psi\), and taking a maximum incoming depth preserves
	\(2^j\Psi\le\Psi(I_{\rm root})\) by the bound for the incoming
	source of maximum depth.
	
	The number of current types is \(O_r(\Psi)\). Indeed the fundamental
	domain of the lattice used in Lemma~\ref{lem:source-halo} lies in
	\((r/2)P_I\). Its distinct translates centered at the type columns
	are disjoint up to boundaries and lie in \((1+r/2)P_I\). Their
	total volume gives the bound. The same argument shows that a
	nonempty atlas source has \(V(K,I)=\Omega_r(m)\), since \(K+6P_I\)
	contains a translate of \(6P_I\).
	
	\begin{theorem}[global source bounds]
		\label[theorem]{thm:ledger}
		For every fixed rational \(\beta>0\), choose the buffer constants
		as in the proof below. Consider one decision invocation on a
		rank-one kernel with initial source volume charge \(\Ot(\nu\wm)\)
		and normalized volume \(O(\wm)\). Let \(\mathcal S\) denote its
		processed states and \(M_u\) the mass of state \(u\). For \(s\in
		\{1,\rho\}\), where \(1<\rho<2\),
		\[
		\sum_{u\in\mathcal S} M_u^s
		=\Ot\bigl((\nu\wm)^s\wm^\beta\bigr).
		\]
		The sum of construction charges \(V\) of all newly generated
		nonface sources and all caller-level query counts \(Q\), assigned
		to their answering sources, is \(\Ot(\nu\wm^{1+\beta})\) up to the
		polynomial bit factor. This includes prefilter queries, rejected
		batches, and queries inherited through domain faces. After \(\ell\)
		lifts each pool has rank at most \(\ell+1\) and normalized volume
		\(O_D(\wm/\nu^\ell)\), so it has \(O_D(\wm/\nu^\ell)\) types.
		Neither rank loss nor face inheritance resets \(\ell\).
		The root charge also covers an entirely rejected initial batch.
	\end{theorem}
	
	\begin{proof}
		By \cref{lem:merged-certificate}, we may apply the phase mass
		bound to the actual merged buffers. Contract each arithmetic
		phase to a single macro node. Let
		\(L=O(\log(\wm+2))\) be the maximum phase factor of
		Lemma~\ref{lem:phase-mass}. Put
		\(G=L(A_{\mathrm{geo}}+1)\), \(F=L A_{\mathrm{lift}}\), and
		increase them to at least one. For a macro node whose root has
		rank \(r\), accumulated lift count \(\ell\), mass \(M\), and
		normalized volume \(\Psi\), set
		\[
		\Phi_s=\Omega^r Z_*^{T-\ell}M^s\Psi^\beta,
		\qquad
		\Omega=12\lceil G\rceil^2,\qquad
		Z_*=12\Omega\lceil F\rceil^2 .
		\]
		The total mass of same-rank thin exits and rank-losing exits is
		at most \(GM\) per family, by the phase bound with \(s=1\).
		The constants entering \(G\) come from the phase and child-mass
		bounds and are independent of \(H_{\min}\). Let \(C_{\rm slab}\)
		bound the slab constant through rank \(D\). Choose
		\(H_{\min}\ge C_{\rm slab}K_*\Omega^{\lceil1/\beta\rceil}\)
		before processing any queries, in addition to the requirements
		of \cref{sec:scales}. Then every same-rank thin exit has volume
		ratio at most \(C_{\rm slab}K_*/H_j\le\Omega^{-1/\beta}\).
		This adds only a fixed power of \(\log(\wm+2)\) to \(H_j/K_*\).
		A rank-losing exit costs one factor \(\Omega\)
		in the potential and does not increase volume. Each family has
		total potential at most \(\Phi_s/12\), using
		\(\sum_{x\in X} M_x^s\le(\sum_{x\in X} M_x)^s\) for each exit
		family \(X\), and \(s\le2\).
		Lifted exits have total mass at most \(FM\), gain one factor
		\(\Omega\), and lose one factor \(Z_*\). Their normalized volume
		does not increase for sufficiently large \(\wm\), so their
		potential is also at most \(\Phi_s/12\). Smaller \(\wm\) can be
		handled by an exact algorithm with a fixed parameter-dependent
		constant. Thus summing potential over the macro tree costs at
		most a constant times the root potential. All its prefactors
		except \(M^s\Psi^\beta\) are polylogarithmic, because \(r,T\)
		are bounded. Counting internal states adds only the factor \(L\).
		
		For every processed state, the window analysis and anchor packing
		bound the newly generated source volume and the caller-level query
		count by \(\Ot(M)\) with a polynomial bit factor. Candidate scans,
		lattice tests, and every LP test of a compatible candidate are
		charged to the generating anchor, including those of a batch that
		leaves no retained query. Building a source atlas is charged at
		construction. Its internal seed search and neighbor traversal are
		part of that charge. Each later query is assigned to the source
		whose oracle answers it. In particular, an inherited face query
		uses the original source through its stored affine map and adds
		one call to that source's count. It does not build a new atlas.
		This assignment preserves the total number of caller-level calls.
		The root halo has construction charge \(\Ot(\nu\wm)\) even when
		the root test fails. Adding it and summing the state charges
		proves the asserted bound on construction volumes and query counts.
		
		The dimension starts at one and increases only on lifts, so the
		rank after \(\ell\) lifts is at most \(\ell+1\). Each lift reduces
		volume by a factor at least \(\nu/C_D\). All restrictions are
		volume-nonincreasing, giving
		\(\Psi=O_D(\wm/\nu^\ell)\), since the total number of lifts is
		bounded. The type bound \(m=O_D(\Psi)\) gives the stated count.
	\end{proof}
	
	The work outside the LP oracles splits into two terms. The
	witnessed local tables and merges at an anchor cost
	\(\Ot(\mu_v^\rho)\) with the bit factor, and the sum of these
	powers within one state is at most \(M^\rho\).
	For structural work at a deep anchor, the paid body contains
	\((bK_*/2)P_I\). The fundamental-domain covering used above
	gives
	\[
	\mu_v\ge
	\frac{\operatorname{vol}_r((bK_*/2-r/2)P_I)}{\det\Lambda_I}
	=\Omega_r(\nu^r\Psi(I)),
	\]
	where the deep scale constant is chosen large enough.
	By \cref{lem:deep-charge}, the compressed progression
	construction, core computations, and deep scans fit within
	\(\Ot(\nu\mu_v)\) with the bit factor.
	Explicit label extraction, lattice tests, pool edits, and
	source bookkeeping obey the same bound. Thin anchors enumerate
	the active column directions in the local window and look them
	up using the original-weight readout. There are \(\Ot(\mu_v)\)
	such directions and at most \(\Ot(\nu)\) explicit labels per
	direction because \(H_j=\Ot(\nu)\). There are at most \(2^r\)
	fractional-label branches. Thus thin structural work also costs
	\(\Ot(\nu\mu_v)\). In atlas mode, face metadata and free-list
	views are charged to their atlas construction or actual query.
	A late face type scan is part of its grouped LP query charge. These assignments leave
	the structural budget \(\Ot(n+\nu\sum_{u\in\mathcal S}M_u)\) with its bit factor.
	The grouped LP calls at deep anchors remain part of this structural
	budget. The grouped calls in place of an atlas are counted separately
	among the source queries.
	
	\begin{lemma}[geometry costs by source]\label[lemma]{lem:geometry-total}
		Put \(X=\nu\wm\) and fix an integer \(k_{\rm scan}\ge2\).
		Use line and planar atlases for sources born with at most one lift,
		the grid atlas of \cref{lem:grid-atlas} for those born with
		\(2\le\ell<k_{\rm scan}\) lifts, and grouped LP for later sources.
		Write \(u_+=\max(0,u)\). For any fixed \(\eta_{\rm cut}>0\),
		the work for all sources at grid level \(\ell\) is, up to a
		polynomial bit factor, at most
		\[
		\Ot\left(
		X\wm^{\beta+\eta_{\rm cut}}
		\bigl(1+\wm^{(1-\ell\theta)_+(1-1/(\ell+2))}\bigr)
		\right).
		\]
		The line and planar sources cost \(\Ot(X\wm^\beta)\).
		Grouped source queries cost
		\(\Ot((1+\wm/\nu^{k_{\rm scan}})X\wm^\beta)\), including face
		queries handled in grouped mode.
	\end{lemma}
	
	\begin{proof}
		For each source \(s\), let \(I_s\) be its pool, \(V_s\) its
		construction volume, and \(Q_s\) the number of calls it answers.
		By \cref{thm:ledger}, \(\sum_s(V_s+Q_s)=\Ot(X\wm^\beta)\).
		An inherited call counts toward the original source. Forming a
		face view uses the compressed list pointers and an affine map
		charged to the query that creates it. The maps compose along a
		strictly rank-decreasing chain of length at most \(D\).
		\Cref{lem:grid-atlas} gives the factor
		\(1+\Psi(I_s)^{r/(r+1)+\eta_{\rm cut}}\) on \(V_s+Q_s\). Since
		\(r\le\ell+1\), \(\Psi(I_s)=O_D(\wm/\nu^\ell)\), and
		\(\nu\ge\wm^\theta\), this factor is
		\(O_{D,\eta_{\rm cut}}(\wm^{\eta_{\rm cut}}(1+\wm^{(1-\ell\theta)_+(1-1/(\ell+2))}))\).
		This also covers a bounded-volume source. The number of levels
		is fixed.
		
		For the line and planar atlases, construction is nearly linear in
		\(V_s\) and queries have polylogarithmic cost. A grouped source
		spends a polynomial bit factor times \(1+m\) per call. A pool
		served in that mode has at least \(k_{\rm scan}\) accumulated
		lifts. Rank and type restrictions only decrease its type count.
		Thus each such call costs at most a polynomial bit factor times
		\(\Ot(1+\wm/\nu^{k_{\rm scan}})\). Summing over their caller-level
		counts proves the claim. Affordable-image LPs at deep anchors
		are charged separately by \cref{lem:deep-charge}.
	\end{proof}
	
	\section{Local Optimization Oracles}
	\label{sec:geometry}
	
	The source bounds of \cref{thm:ledger} separate geometric volumes
	and actual query counts from the cost of the oracle interface.
	The line and planar interfaces use local subdivisions. The grid
	atlas extends tile navigation to any fixed rank and looks up tiles
	by target location. At the scan threshold, the grouped LP instead
	visits the surviving types on each call.
	All interfaces use a common integer objective.
	
	\subsection{A common objective across pools}
	
	All prices and costs used to compare pools refer to one labeled
	objective. Atlas navigation requires this objective to be generic
	on the stored column systems. Before recursion, take
	\(H=\lceil C_D\wm\rceil\), with \(C_D\) chosen from the normalized
	volume and reboxing bounds of \cref{thm:ledger,lem:rebox} for
	all ranks through \(D\).
	
	\begin{lemma}[uniform integer perturbation]
		\label[lemma]{lem:perturb}
		Let the virtual labels be $j\in [U]$, ordered by nondecreasing integral cost $c_j$ within each original weight class. Fix a rank bound $D\ge1$ and an integer ambient coordinate bound $H\ge 1$. There is a choice $c'_j=A c_j+d_j$ with the following properties.
		\begin{enumerate}
			\item The ordering of distinct integral selection costs is preserved, and $c'_j$ is strictly increasing within each original weight class.
			\item For every subset of the labels, every assignment of integer columns of rank at most $D$ with coordinates in $[-H,H]$, and every price $\pi$, the columns of the labels satisfying $c'_j=\pi\cdot a_j$ are independent.
			\item The additional bit length is $O_D(\log(H+U+2))$. Costs and prefix sums on a constant-cost input run are available from its ordinal endpoints in $\operatorname{poly}_D(\log(H+U+2))$ arithmetic operations.
		\end{enumerate}
	\end{lemma}
	
	\begin{proof}
		Set $L=D!(2H)^D$, $S=(D+1)LU^D$, $R=2S+2$, and
		\[
		d_j=\sum_{h=0}^D j^hR^h,
		\qquad A=1+2\max\bigl(U,(D+1)L\bigr)d_U.
		\]
		The perturbation of any selection is between $0$ and $Ud_U<A/2$. This proves the selection-cost ordering. The formula for $d_j$ also proves the within-class ordering.
		
		For a minimally dependent set of $s\le D+1$ columns, cofactors
		on an independent row set give an integral circuit $z$ with
		$\|z\|_\infty\le L$. A zero column uses the one-term circuit.
		Write
		\[
		T_h=\sum_{j\in\operatorname{supp}(z)}z_jj^h,\qquad
		z\cdot d=\sum_{h=0}^{D}T_h R^h .
		\]
		We have $|T_h|\le S$, and the Vandermonde matrix on the distinct
		label IDs has rank $s$. The highest nonzero $T_h R^h$ dominates
		the sum, so $z\cdot d\ne 0$. Also $|z\cdot d|\le(D+1)Ld_U<A/2$.
		If $z\cdot c=0$ then $z\cdot c'=z\cdot d\ne0$. Otherwise the
		integer $z\cdot c$ gives $|Az\cdot c|\ge A$, again implying
		$z\cdot c'\ne0$. Thus no price makes a dependent set tight.
		The circuit is invariant under invertible coordinate changes.
		
		On each constant-cost run, summing $c'_j$ reduces to sums of the first $D$ powers of the ordinal. Subtract the deleted intervals or labels in the current view using their stored aggregates. The formula gives the claimed bit bound.
	\end{proof}
	
	When $U=\wm^{O(1)}$ and $H=\wm^{O_D(1)}$, the lemma adds
	$O_D(\log(\wm+2))$ bits. The common objective is fixed at
	initialization. In a residual integral cost query, $c(x)<s$ for an
	integer $s$ is equivalent to $c'(x)<As$. All fractional subproblems,
	including gap and price comparisons, use $c'$. Signed coordinates
	at a normalizing price \(\pi\) preserve genericity:
	\(\sigma_i(c'_i-\pi\cdot a_i)=\lambda\cdot(\sigma_i a_i)\)
	is equivalent to \(c'_i=(\pi+\lambda)\cdot a_i\), and sign changes
	preserve independence. The grouped LP oracle also handles ties
	in temporary systems such as the progression image.
	
	\subsection{Exact optimization by scanning the types}
	
	The grouped LP works on an immutable pool version \(I\). For each
	column type \(a\in\mathcal A\), its \(N_a\) available labels are in
	nondecreasing cost order. Write \(a_i\) for the column of label \(i\)
	and \(C_a(k)\) for the sum of the first \(k\) available costs of type \(a\).
	We assume polylogarithmic count, select, and prefix-sum access with
	polynomial overhead in the bit length. In the notation of
	\cref{sec:model}, grouping equal columns gives
	\(Z_I=\sum_{a\in\mathcal A}[0,N_a a]\). The oracle evaluates
	\(f_I\) using the supplied costs, which include the common
	perturbation for stored pools.
	
	\begin{lemma}[grouped exact LP]
		\label[lemma]{lem:grouped}
		In fixed dimension $k$, let $I$ have $m$ types and $U$ available labels. Let $B$ bound the bit length of each column entry, cost, target coordinate and query coefficient. Integer data and rational data obtained using bounded-bit common denominators are allowed. Put $L=B+\log(U+2)$. In $\Ot(m+1)\operatorname{poly}_k(L)$ bit operations one can either return a strict domain separator for $t\notin Z_I$ or return $f_I(t)$, an optimal price, and a basic optimal witness with endpoint prefixes and at most $k$ actual fractional labels. One can also identify the minimal face of $Z_I$ containing $t$ and compute its forced endpoint counts, free-type list and fixed target and cost offsets within the same time bound.
	\end{lemma}
	
	\begin{proof}
		We use exact optimization from strong separation~\cite{GLS88}.
		Each oracle optimizes in at most $k+2$ variables. Each endpoint
		inequality is a sum of at most $U$ bounded-bit contributions
		with common denominators, so its encoding length is
		$\operatorname{poly}_k(L)$. Determinant bounds give a box of radius
		$2^{\operatorname{poly}_k(L)}$ containing an optimum whenever the
		optimum is finite. This gives $\operatorname{poly}_k(L)$ oracle
		calls, also on lower-dimensional faces. A rank scan removes
		redundant equalities and separates targets outside the column span.
		
		The domain test uses the support function
		\[
		h_{Z_I}(v)=\sum_{a\in\mathcal A}N_a\max(0,v\cdot a).
		\]
		A type scan evaluates it and supplies a maximizing endpoint.
		This endpoint gives a separating plane for the hypograph of
		$v\mapsto v\cdot t-h_{Z_I}(v)$. Maximize the function over
		$[-1,1]^k$. A positive optimum yields a strict separator for $t$.
		At a feasible $t$, the dual objective is
		\[
		g_t(\pi)=\pi\cdot t+\sum_{i\in I}\min(0,c_i-\pi\cdot a_i).
		\]
		Maximize $g_t$ over $\R^k$.
		For $k_a=\#\{i\text{ of type }a\mid c_i<\pi\cdot a\}$, the endpoint-prefix expression yields the valid upper plane
		\[
		g_t(\pi')\le \sum_{a\in\mathcal A} C_a(k_a)+\pi'\cdot\left(t-\sum_{a\in\mathcal A} k_aa\right).
		\]
		The bound is tight at $\pi$ and supplies strong separation for the hypograph. Thus each evaluation uses one count and prefix-sum operation per type.
		
		At an optimal price, negative reduced costs force selection and
		positive ones force omission. Let $b$ be the target after the
		selected prefixes are subtracted, and let $N_a^0$ be the number
		of tied labels of type $a$. The remaining constraints are
		\[
		\sum_{a\in\mathcal A}a y_a=b,\qquad 0\le y_a\le N_a^0
		\quad(a\in\mathcal A).
		\]
		When the tied columns are independent, solve for their coefficients
		directly. Otherwise give the types indexed by $j\in[m]$ costs
		$d_j=\sum_{h=0}^{k}j^hR^h$, where $R>2(k+1)C(m+1)^k+1$
		and $C$ bounds the cofactors of a minimal dependence. The circuit
		argument of \cref{lem:perturb} makes the auxiliary tight columns
		independent. A type scan evaluates its grouped dual
		\[
		\lambda\cdot b+\sum_{a\in\mathcal A}N_a^0
		\min(0,d_a-\lambda\cdot a).
		\]
		At its optimal price, fix all nontight types at their prescribed
		endpoints and solve the remaining independent system. Strong
		duality guarantees counts in the prescribed ranges. Realize each
		$y_a$ by a prefix of $\lfloor y_a\rfloor$ tied labels and, if
		needed, the next label at coefficient $y_a-\lfloor y_a\rfloor$.
		The fractional labels have independent columns, giving a basic
		solution at copy level.
		
		To compute the face record, we construct an exposing normal.
		For a feasible target let
		\[
		N_{Z_I}(t)=\{v\in\R^k\mid h_{Z_I}(v)-v\cdot t\le 0\}.
		\]
		Its intersection with $[-1,1]^k$ has the same span as the cone and admits strong separation by an endpoint of $Z_I$. Starting with the zero span, optimize both signs of a basis of its orthogonal complement over this intersection. A nonzero optimum outside the current span extends it. Repeating at most $k$ times gives cone vectors spanning the cone. Their sum belongs to the relative interior of the cone, since every supporting linear functional that is zero on this sum and nonnegative on the cone must vanish on each spanning vector. A relative-interior normal exposes precisely the minimal face containing $t$. A scan at this normal identifies forced labels by the sign of its scalar product with the type column, and the zero-sign types are free. A zero cone means that the whole domain is the minimal face. All witnesses and face records refer to the queried pool version.
	\end{proof}
	
	\begin{lemma}[affordable image optimization]
		\label[lemma]{lem:affordable}
		In the model of \cref{lem:grouped}, let
		$U_Ix=\sum_{i\in I}u_i x_i$ be a fixed-dimensional image
		with $u_i$ constant on each type. Let $a_i$ be a fixed
		linear image of $u_i$, so that $A_Ix=\sum_{i\in I}a_i x_i$
		is the target map. For a target $y$ and budget $b$, put
		\[
		\mathcal K=\{U_Ix\mid A_Ix=y,\ c^\top x\le b,\ 0\le x\le 1\}.
		\]
		In $\Ot(m+1)\operatorname{poly}_k(L)$ bit operations one can detect emptiness, or optimize a given linear functional over $\mathcal K$ and return an optimizing vertex of $\mathcal K$ with a compressed labeled witness and optimal dual multipliers for the original objective. The encoding length of this vertex is bounded by a polynomial in the fixed body data, independently of the objective.
	\end{lemma}
	
	\begin{proof}
		Test emptiness by minimizing $c^\top x$ on the fiber $A_Ix=y$.
		Minimize $\psi(U_Ix)$ on the affordable fiber using the
		fixed-dimensional dual. Write $\beta$ for the target multiplier
		and $\tau\ge0$ for the budget multiplier. The reduced coefficient
		on label $i$ is
		\[
		\psi(u_i)-\beta\cdot a_i+\tau c_i.
		\]
		For $\tau>0$, the negative coefficients form one cost prefix
		per type. For $\tau=0$, they occupy entire types. The separation
		argument of \cref{lem:grouped} applies with one additional dual
		variable.
		
		Impose the first objective's optimal value as a linear equality.
		On the resulting face, successively optimize each coordinate
		of $U_Ix$ and impose its optimal value. All the added equalities
		are functions of the same image, so their rank is bounded by
		the image dimension. The last face is a single image point $z$,
		hence a vertex of $\mathcal K$. Minimize $c^\top x$ subject
		to $U_Ix=z$ and $0\le x\le 1$. The minimum is at most $b$
		because $z\in\mathcal K$, and \cref{lem:grouped} returns its
		witness even when the first dual ties entire types.
		
		A projected vertex of the fixed body is the image of a vertex of its defining box-constrained polytope. Such a vertex has at most $\operatorname{rank}(A_I)+1$ fractional coordinates, by the active-constraint criterion for vertices. The integer endpoint contributions have encoding length polynomial in $L$. Cramer's rule on the remaining system bounds the encoding length of the vertex image by a polynomial in the fixed body data. This permits successive support calls whose objectives are defined using previously returned vertices, as in fixed-dimensional simplex rounding.
	\end{proof}
	
	The affordable-image oracle also applies in temporary reduced-cost coordinates. At an old price $\pi$, split a permanent type into the labels with $c_i<\pi\cdot a_i$ and those with $c_i\ge\pi\cdot a_i$. The selected part is complemented and has reversed columns, while the other part keeps its columns. The normalized costs are $\pi\cdot a_i-c_i$ on the complemented part and $c_i-\pi\cdot a_i$ on the other part. They are sorted in reverse order and forward order, respectively. Each part is thus a prefix or suffix view of the same current pool. Its normalized cost sums are computed from original cost sums and counts. The dual scan visits at most two such parts per permanent type. Complement the selected part back to obtain the original label coefficients.
	
	\subsection{Navigation in a local atlas}\label{subsec:atlas-navigation}
	
	The line and plane admit fast specialized atlases. In higher rank we
	use the same tile navigation with a different index. The atlas is
	built on an immutable pool whose tight columns are independent at
	every price, as in \cref{lem:perturb}.
	
	\begin{lemma}[constructing and navigating tiles]\label[lemma]{lem:atlas-navigation}
		Let a positive-rank generic pool \(I\) have \(m\) nonzero types,
		column lattice \(\Lambda_I\), and symmetric hull \(P_I\). Work in
		its \(r\)-dimensional span. Let \(K\) be a nonempty compact convex
		rational polytope of constant description complexity, and put
		\[
		V=1+\frac{\operatorname{vol}_r(K+6P_I)}{\det\Lambda_I}.
		\]
		In \(\Ot(m)\operatorname{poly}(B)\) time one can find an
		actual-column basis matrix \(B_I\) and a box \(D_I\) in this basis
		with \(P_I\subseteq D_I\subseteq(2^r-1)P_I\), and find a first
		tile meeting \(\operatorname{int}(K+D_I)\) in its interior or
		certify \(K\cap Z_I=\emptyset\). From this first tile a traversal
		finds \(O_r(V)\) tiles covering \(K\cap Z_I\), with \(O_r(V)\)
		neighbor attempts. Each attempt uses \(\operatorname{poly}(B)\)
		tests of current prefix inequalities and changes at most two
		prefixes. The test must return a violated inequality if one exists.
		A containing tile gives the value, a supporting price, and a
		compressed basic optimum by a fixed-dimensional linear solve.
	\end{lemma}
	
	\begin{proof}
		Choose the columns of \(B_I\) successively by maximizing the
		squared norm of the projection onto the orthogonal complement of
		the chosen span. This takes \(r\) scans using rational comparisons.
		For the \(j\)-th orthogonalized column \(q_j\), maximality gives
		\(|q_j\cdot a|\le \|q_j\|^2\) for every actual column \(a\).
		Backward substitution then bounds coordinate \(j\) of \(B_I^{-1}a\)
		by \(1+\sum_{i=j+1}^r 2^{r-i}=2^{r-j}\) in absolute value.
		Use these half-widths for \(D_I\). Each basis column belongs to
		\(P_I\), proving the second inclusion. In particular the computable
		index
		\[
		\widehat\Psi_I=\frac{|\det B_I|}{\det\Lambda_I}
		\]
		satisfies \(1\le\widehat\Psi_I\le\Psi(I)\le C_r\widehat\Psi_I\)
		for a rank-dependent constant \(C_r\). The first volume comparison
		uses the crosspolytope of the chosen columns, and the second uses
		\(D_I\). Compute the index by expressing the columns of \(B_I\)
		in an integer basis of \(\Lambda_I\) and taking the absolute
		determinant.
		
		Put \(\widehat K=K+D_I\). To find a common interior target, maximize
		\(\lambda\ge0\) subject to \(t_0\pm\lambda e_j\in Z_I\cap\widehat K\)
		for \(1\le j\le r\). The window constraints have constant size.
		Strong separation for the domain is supplied by the support-function
		method of \cref{lem:grouped}, using one type scan per support
		evaluation. The variables \((t_0,\lambda)\) have fixed dimension.
		Exact optimization therefore uses \(\operatorname{poly}(B)\) scans.
		If \(K\) contains any feasible point, that point is interior to
		\(\widehat K\), and nearby interior points of the full-dimensional
		domain yield \(\lambda>0\). Thus infeasibility or zero maximum
		certifies an empty intersection with \(K\).
		
		At a common interior target \(t_0\), solve its grouped dual and
		lexicographically optimize over the optimal price face. This
		face is bounded since \(h_{Z_I}(v)>v\cdot t_0\) for \(v\ne0\).
		The returned vertex has \(r\) tight labels. Indeed, with fewer
		tight columns a nonzero direction orthogonal to all of them
		preserves the reduced-cost signs for sufficiently small moves in
		both directions. The dual objective is affine on this segment,
		and optimality makes it constant, contradicting extremality.
		The price vertices have polynomial bit length by the determinant
		bounds for the tight systems, so the lexicographic optimization
		can use an explicit box containing them in its interior. A final
		scan finds the actual
		tight labels and initializes all strict prefix counts \(k_a\).
		
		Let \(z=\sum_{a\in\mathcal A}k_a a\) be the target of the selected
		prefixes. For tight labels \(b_1,\ldots,b_r\), the tile and its
		value function are
		\[
		T=z+\sum_{j=1}^r[0,a_{b_j}],\qquad
		f_I(t)=\sum_{a\in\mathcal A}C_a(k_a)+\pi\cdot(t-z)\quad(t\in T).
		\]
		These tiles are the projections of full-dimensional lower faces
		of the lifted zonotope with generators \((a_i,c_i)\). Lower faces
		lie on the graph of the value function and hence project to a
		face-to-face complex. The price-vertex argument gives coverage of
		the domain interior. The closed tiles also cover its boundary.
		The seed tile contains \(t_0\), so its interior meets
		\(\operatorname{int}\widehat K\) even if \(t_0\) lies on a tile face.
		
		To cross a facet, fix its departing tight label \(b\) at the
		facet endpoint \(\varepsilon\in\{0,1\}\). Increase its type's
		prefix count by one if \(\varepsilon=1\). Keep the other \(r-1\)
		tight labels fixed by tracing a rational price ray \(\pi+\alpha u\)
		orthogonal to their columns, with
		\(\operatorname{sign}(u\cdot a_b)=2\varepsilon-1\). The ray starts
		inside the applicable prefix region
		\[
		c_{a,k_a}\le \pi\cdot a\le c_{a,k_a+1}\quad(a\in\mathcal A),
		\]
		where missing bounds are omitted. The departing inequality becomes
		strict, retained tight constraints remain equalities, and the
		other inequalities start strict. For a constraint \(v\cdot\pi\le b'\)
		with \(v\cdot u>0\), its blocking ratio is
		\((b'-v\cdot\pi)/(v\cdot u)\). All such ratios are positive.
		Reduced numerators and denominators are bounded by \(2^L\) for
		\(L=\operatorname{poly}(B)\). Test the ray at \(2^{L+1}\). If it
		is feasible the ray is unbounded. Otherwise bisect between zero
		and this infeasible point until the interval width is less than
		\(2^{-2L-2}\). The infeasible endpoint is past the first ratio
		and before any distinct later one. A violated row there supplies
		an actual blocking label and its exact ratio.
		
		A finite first exit has exactly one new tight label by genericity
		and the \(r-1\) retained tight labels. The new column is independent
		of the retained ones because its directional derivative is nonzero.
		An upper blocker is the next unselected label. A lower blocker is
		the last selected label, whose count must be decremented when
		storing the new strict prefixes. The resulting tile shares the
		facet with the former tile, since the entering label stays at its
		previous endpoint on that facet. For an unbounded ray, every type
		with positive derivative is fully selected and every type with
		negative derivative is fully omitted. The facet is then on a
		supporting hyperplane of \(Z_I\). This accounts for unsuccessful
		attempts as well as finite exits. At most the departing and
		entering types change counts, and backtracking reverses these changes.
		
		The traversal accepts exactly the tiles whose interiors meet
		\(\operatorname{int}\widehat K\), a test by a constant-size rational
		LP. Deduplicate by the sorted actual tight-label IDs. The accepted
		adjacency graph is connected because a path in the convex common
		interior can avoid the codimension-two skeleton and cross facets
		transversely. Every feasible target in \(K\) belongs to the closure
		of an accepted tile, since it is interior to \(\widehat K\).
		Each accepted tile lies in \(K+D_I+rP_I\) and has volume at least
		\(\det\Lambda_I\). The interiors are disjoint. Dilation about
		any point of \(K\) bounds the containing volume by a constant
		multiple of \(\operatorname{vol}_r(K+6P_I)\), which proves the
		count. Each tile has \(2r\) facets, and all their attempts are
		charged, including rejected, unbounded and previously visited
		neighbors. Keep persistent count roots and update the two prefix
		aggregates at each change. A containing tile recovers the
		coefficients of its tight labels by solving its column system
		for \(t-z\). The stored prefixes give all other coefficients.
	\end{proof}
	
	\begin{algorithm}[H]
		\caption{Enumerating the tiles of one source}
		\label{alg:planar}
		\begin{algorithmic}[1]
			\Require Generic positive-rank pool \(I\), window \(K\), and a prefix-inequality oracle
			\Ensure Tiles covering \(K\cap Z_I\), with their basic witness data
			\Begin
			\State Construct the basis box \(D_I\) and put \(\widehat K\gets K+D_I\).
			\State Find a common interior target of \(Z_I\) and \(\widehat K\).
			\If{there is no common interior target}
			\State \Return an empty tile collection
			\EndIf
			\State Obtain its seed tile \(T_0\) and initialize the prefix state.
			\State Record \(T_0\) with its witness data.
			\ForAll{facets of tiles reached by depth-first search from \(T_0\)}
			\State Fix the departing label at the facet endpoint and trace the price ray.
			\If{there is a finite first blocker}
			\State Obtain the adjacent tile \(T'\) and its strict prefix root.
			\If{\(T'\) is new and its interior meets \(\operatorname{int}\widehat K\)}
			\State Record \(T'\) with its witness data and recurse.
			\EndIf
			\EndIf
			\State Roll back the prefix changes.
			\EndFor
			\State \Return the recorded tiles
			\End
		\end{algorithmic}
	\end{algorithm}
	
	An oracle for the parent domain supplies a compressed view of each
	minimal face, its exposing normal, its fixed target and cost, and the
	lattice of its free types. A pure face restriction inherits the source
	through the affine embedding of \cref{subsec:source-halo}. The basic
	witness is the parent prefix selection restricted to the free types,
	together with its surviving tight labels. To obtain its aggregates,
	subtract the fixed endpoints from the full parent basic selection,
	including tight labels at coefficient one. For prefix aggregates,
	first add the contributions of removed tight labels and then subtract
	the fixed totals. At most \(r\) corrections are needed. Prices and
	strict domain separators restrict to the face span. A subface is
	found by querying the original domain at the embedded target, so its
	view is again one original free-type view. Support directions in
	the face span are extended to the original coordinates. These pure
	inheritance chains have length at most the source rank. The face pool,
	fixed endpoints and canonical face key can be shared by identical
	immutable pools. Each query keeps its own source-window pointer and
	composed affine embedding when batches coalesce. Its cost is charged
	to the source that answers it.
	
	\Needspace{8\baselineskip}
	\subsection{Output-sensitive atlases in ranks one and two}
	
	In rank two, the traversal can use the order of boundary edges
	and the line inequalities in the two-dimensional price space.
	
	\begin{lemma}[planar atlas]\label[lemma]{lem:planar}
		Let a generic pool of rank two and a window \(K\) satisfy the
		conditions of \cref{lem:atlas-navigation}. For its normalized
		window volume \(V\), one can construct an atlas in
		\(\Ot(m+V)\operatorname{poly}(B)\) bit operations and
		\(\Ot(m+V)\) arithmetic values and pointers besides the pool.
		A query on \(K\) costs
		\(O(\log(m+1)+\log(V+1))\operatorname{poly}(B)\) and returns
		a strict domain separator or the exact value, a supporting
		price, a basic witness and a minimal-domain-face record.
	\end{lemma}
	
	\begin{proof}
		Set \(u=(1,1+\max_{a\in\mathcal A}|a_1|)\). No nonzero integral
		column is orthogonal to \(u\), since for \(a_2\ne0\) the second
		term of \(u\cdot a\) has magnitude greater than \(|a_1|\).
		Group the types by primitive integer direction, taking the
		representative \(p\in\mathcal P\) with \(u\cdot p>0\), and write
		\(a=\lambda_a p\). Then
		\[
		z_0=\sum_{\substack{a\in\mathcal A\\\lambda_a<0}}N_aa,\qquad
		L_p=\sum_{\substack{a\in\mathcal A\\a\parallel p}}N_a|\lambda_a|,
		\qquad Z_I=z_0+\sum_{p\in\mathcal P}[0,L_pp].
		\]
		The unique minimizer of \(u\) on \(Z_I\) is \(z_0\).
		Sort the representatives counterclockwise within the open
		half-plane \(u\cdot p>0\) by determinant comparisons. From \(z_0\),
		traverse the edges \(L_pp\) in that order and then their negatives
		in the same order. This is the sequence of endpoint changes as the
		maximizing normal rotates from \(-u\) to \(u\) and back to \(-u\).
		At each step one group flips
		its endpoints. Maintaining persistent counts stores vertex
		witnesses and their endpoint sums in \(\Ot(m)\) time and space,
		as each type flips twice. An edge record stores its free group,
		its lattice by a gcd, its largest free column in that direction,
		and the fixed endpoints. A fan triangulation locates a target's
		minimal face or a violated edge inequality in
		\(O(\log(m+1))\) comparisons. A static convex hull of the signed
		types supports queries on \(P_I\).
		
		Use \cref{lem:atlas-navigation} for the seed and tile traversal.
		To test the prefix inequalities \(v\cdot\pi\le b\) and retrieve
		a violated row, maintain three planar dynamic hulls. Store
		\(v/b\) for \(b>0\) and test whether the maximum scalar product
		with \(\pi\) is at most one. For \(b<0\), store \(v/b\) and test
		whether the minimum is at least one. For \(b=0\), store \(v\)
		and test whether the maximum is at most zero. Keep constraint
		IDs and multiplicities. The dynamic hull operations
		of~\cite{OvL81} take polylogarithmically many exact comparisons
		per change or test. Thus the \(O(V)\) attempts and their
		rollbacks take \(\Ot(V)\operatorname{poly}(B)\) time.
		
		Clip the accepted tiles to \(\widehat K\). Their planar
		subdivision has \(O(V)\) complexity since the tiling is
		face-to-face and the window has constant complexity. Sorting
		and merging the edges, triangulation, and point
		location~\cite{Kirkpatrick83} cost \(\Ot(V)\) additional operations.
		Query the domain structure first for a face record or strict
		separator. For a feasible \(t\), resolve boundary ties by locating
		\(t+\varepsilon(t_0-t)+\varepsilon^2e_1+\varepsilon^3e_2\)
		symbolically, where \(t_0\) is the common interior seed and
		\(e_1,e_2\) are the coordinate unit vectors. For sufficiently
		small positive \(\varepsilon\) this point is in the interior of
		\(Z_I\cap\widehat K\), off the subdivision and triangulation
		edges. Each orientation test reads the first nonzero
		coefficient, so the number and cost of tests are unchanged.
		The returned tile contains \(t\) in its closure. Evaluate
		its value and witness at \(t\) itself as in
		\cref{lem:atlas-navigation}.
	\end{proof}
	
	\Cref{fig:planar} illustrates the facet step in \cref{alg:planar}.
	Crossing the highlighted facet retains one tight label and fixes
	the other. The first blocking label sets the next price.
	
	\PlanarFigure

	The same interface is available in dimension one with interval length in place of area. This case is used for early sources whose type columns span a line.
	
	\begin{lemma}[interval atlas]
		\label[lemma]{lem:line}
		For a generic rank-one pool with nonzero columns in lattice coordinates and a compact interval $K$, put
		\[
		V=1+\frac{\operatorname{length}(K+6P_I)}{\det\Lambda_I}.
		\]
		One can construct an atlas with the outputs and the construction and query bounds of Lemma~\ref{lem:planar} for this value of $V$.
	\end{lemma}
	
	\begin{proof}
		Work in scalar coordinates for the column lattice. The left endpoint
		of the domain and its reference selection are
		\[
		z_{\min}=\sum_{\substack{a\in\mathcal A\\a<0}}N_a a,
		\qquad
		x_i^0=\begin{cases}1,&a_i<0,\\0,&a_i>0.\end{cases}
		\]
		Orient each label away from this reference. Its length is \(|a_i|\)
		and its marginal slope is \(\alpha_i=c_i/a_i\). The slopes are distinct
		by genericity. Within a type they appear in forward cost order for
		\(a>0\), and in reverse cost order for \(a<0\). Increasing target
		weight therefore selects the oriented labels in increasing slope order.
		
		The seed can be located directly from these ordered lists. For a
		rational price \(\lambda\), let
		\[
		n_a(\lambda)=\#\{i\in I\mid a_i=a,\ \alpha_i\le\lambda\},
		\qquad
		F(\lambda)=\sum_{a\in\mathcal A}|a|n_a(\lambda).
		\]
		A threshold count in each type evaluates \(F\) in
		\(\Ot(m)\operatorname{poly}(B)\) time. For negative types this is
		a suffix count in the original cost order. Choose an integer
		\(L=\operatorname{poly}(B)\) such that the reduced numerator and
		positive denominator of every slope have absolute value at most
		\(2^L\). Then all slopes lie strictly between
		\(-2^L-1\) and \(2^L+1\), and distinct slopes differ by at least
		\(2^{-2L}\). These bounds follow from the stored cost and column
		encoding bounds and require no label enumeration.
		
		Set \(\widehat K=K+P_I\). If \(Z_I\cap\widehat K\) has positive
		length, take its midpoint \(t_0\) and put \(\tau=t_0-z_{\min}\).
		The midpoint lies in the interior of the domain, so
		\(0<\tau<\sum_{a\in\mathcal A}N_a|a|\).
		Starting with \(\lambda_-=-2^L-1\) and \(\lambda_+=2^L+1\),
		bisect until the interval is narrower than \(2^{-2L}\). Maintain
		the invariant
		\[
		F(\lambda_-)\le\tau<F(\lambda_+).
		\]
		Bisection takes \(O(L)\) iterations. The invariant and the separation bound imply that
		\((\lambda_-,\lambda_+]\) contains exactly one slope
		\(\alpha_j\). Retrieve its actual label by taking the first slope
		greater than \(\lambda_-\) in each nonexhausted type and choosing
		the least answer. A final threshold scan computes the strict
		reduced-cost prefix counts and their total target
		\[
		k_a=\#\{i\in I\mid a_i=a,\ c_i<\alpha_j a\},
		\qquad z=\sum_{a\in\mathcal A}k_a a.
		\]
		The prefix record and label \(j\) define the seed segment
		\(z+\{a_jx\mid 0\le x\le1\}\). For \(t\) on this segment, the
		coefficient \(x_j\), value \(f_I(t)\), and supporting price \(\pi\) satisfy
		\[
		x_j=\frac{t-z}{a_j},\qquad
		f_I(t)=\sum_{a\in\mathcal A}C_a(k_a)+\alpha_j(t-z),
		\qquad \pi=\alpha_j.
		\]
		This formula also covers negative \(a_j\), whose oriented coefficient
		is \(1-x_j\). When \(t_0\) is a breakpoint, the strict upper
		inequality in the invariant chooses the segment to its right.
		The seed construction costs \(\Ot(m)\operatorname{poly}(B)\) in
		total and never enumerates the virtual copies.
		
		Initialize the strict prefix counts and the previous and next slopes
		of each type by one scan. Explore both directions with priority queues,
		changing only one type at a step. Stop in each direction when a segment
		fails to meet the interior of \(\widehat K\). Every accepted segment
		has length at least \(\det\Lambda_I\), has disjoint interior from the
		other accepted segments, and lies in \(K+2P_I\). Thus there are
		\(O(V)\) such segments and at most two failed attempts. Persistent
		count roots preserve the endpoint witnesses, and sorting the segment
		endpoints supports binary-search queries.
		
		The domain endpoints supply their minimal-face records and strict
		separators. The maximum absolute type column supplies support and
		separation for \(P_I\). If \(Z_I\cap\widehat K\) has zero length,
		then \(K\cap Z_I=\emptyset\), because a point of \(K\) is interior
		to \(\widehat K\). This gives the same construction, query, and
		storage bounds as in \cref{lem:planar}.
	\end{proof}

	\subsection{An atlas indexed by a grid}\label{subsec:grid-atlas}
	
	The source volume \(V\) bounds how many tiles to build.
	For a given target, only tiles near its grid cell are relevant.
	To search these tiles we use blocks of affine sign tests.
	
	\begin{lemma}[blocks for affine sign tests]\label[lemma]{lem:cutting-blocks}
		Fix \(r,p\ge1\) and \(\eta>0\). A block has \(h\) rows with distinct
		identifiers and \(p\) forms per row. In test \(j\), all rows are
		evaluated at a common query point \(x_j\). A test is affine in at
		most \(r\) variables or homogeneous linear in at most \(r+1\).
		For each test, fix a relation on the signs \(\{-1,0,1\}\). Given the
		query points, the block returns disjoint stored row lists whose union
		is the set satisfying all \(p\) tests on the same row. It can return
		one such row or combine their stored aggregates in
		\(\operatorname{polylog}(h+2)\operatorname{poly}(B)\) time. Listing
		the rows takes additional time proportional to the output length.
		For a failure budget \(\zeta\in(0,1)\), the construction time is
		\[
		O_{r,p,\eta}(h^{r+\eta})
		\operatorname{poly}\bigl(B+\log(1/\zeta)\bigr).
		\]
		Construction fails with probability at most \(\zeta\).
		Here \(B\) bounds the rational input and query encoding lengths,
		including \(\log(h+2)\). The aggregates may include row counts,
		endpoint target and cost sums, and HNF bases for the column lattice.
		For these aggregates the row data are integral or use common
		denominators of polynomial bit length, so sums and bases have
		polynomial encoding length as well.
	\end{lemma}
	
	\begin{proof}
		For an affine form on \(x\), homogenize the query to \((x,1)\).
		Divide this vector by the absolute value of a largest-magnitude
		coordinate, breaking ties by index. Fix that coordinate to \(+1\)
		or \(-1\) and use the remaining coordinates in a unit box as the
		chart. The division preserves every sign. We build all the constantly
		many charts for each test. Homogeneous tests can use the same charts,
		with one additional chart for a zero query vector. In that chart all
		first-test signs are zero, so store the continuation on the remaining
		tests if zero satisfies the first relation. Otherwise store the empty
		answer.
		
		We first describe one cutting for a chart on \(N\) rows. Pad to
		dimension \(r\) if needed. Fix a sufficiently large integer \(R\),
		independent of \(h\), and sample \(s_R=O_r(R\log(R+2))\) rows.
		Triangulate the sample arrangement clipped to the unit box by the
		bottom-vertex construction of~\cite[Lemma~3.2]{EHKS20}, including
		the box hyperplanes. It has \(O_r(s_R^r)\) full-dimensional
		simplices. This bound holds for degenerate arrangements. Since the
		sample size is constant, we can enumerate the cells and their
		vertices by exact rational feasibility and fixed-size systems of
		equations, then triangulate each cell using a common lexicographic
		order of the vertices. On the finite vertex set this is the order of
		a generic linear functional, so the bottom-vertex bound applies.
		Duplicate sampled hyperplanes need only be inserted once, and
		constant forms create no hyperplanes.
		
		The sample hits every set of more than \(N/R\) rows crossing the
		interior of an \(r\)-simplex with constant probability. To see this,
		on any list of \(u\) rows a crossing set is determined by the sign
		vectors at the simplex's \(r+1\) vertices. There are
		\(O_r((u+1)^{r(r+1)})\) such traces, by the arrangement bound for
		sign vectors including equality cells. The usual double-sampling
		argument gives the claimed sample size. Indeed, if the first
		sample misses a range of mass greater than \(1/R\), an independent
		sample contains \(\Omega(s_R/R)\) of its rows with probability
		at least \(1/2\). Condition on the combined samples and split them
		randomly into equal halves. The chance that these hits all lie
		in the second half is at most \(2^{-\Omega(s_R/R)}\). A union
		bound over the polynomially many traces makes the first sample
		an appropriate net with probability at least \(1/2\) for a
		sufficiently large constant in \(s_R\).
		The argument uses row multiplicities, so it also covers repeated
		hyperplanes. One can sample by reducing
		\(\lceil\log_2 N\rceil+1\) uniform bits modulo \(N\). This distribution
		gives each row probability at least \(1/(2N)\), which only changes
		the constant in the net size.
		
		No sampled hyperplane crosses the interior of a simplex in its
		arrangement. We check the conflict sizes by scanning the \(N\)
		rows and testing each form at the simplex vertices. A form changes
		sign in the interior precisely when its vertex values include
		both a positive and a negative value. If a conflict list exceeds
		\(N/R\), resample. Each trial takes \(O_{r,R}(N)\operatorname{poly}(B)\)
		time. At every recursive node we use the original forms in the
		same bounded charts. Thus the bit lengths of simplex vertices
		and of exact sign and membership tests are polynomial in \(B\)
		independently of recursion depth.
		
		For a query point on a boundary, choose the first closed simplex
		containing it and the relative interior of its unique minimal face
		containing the point. Preprocess every face of every simplex. On
		the relative interior of a face, the sign is constant unless its
		vertex values have both signs. The all-zero case has sign zero
		throughout the face. A row with both signs at face vertices also
		crosses the containing full-dimensional simplex, so each face
		has at most \(N/R\) conflicting rows. This convention handles
		closed inequalities and equality tests using the actual query
		point.
		
		For each face, recurse on the conflicting rows with the same
		remaining tests. Recurse on the rows whose constant sign satisfies
		the first test with that test removed. Discard the other rows.
		The two retained sets are disjoint. At a query, follow both of
		these branches for the selected face, using the same query
		coordinates and the remaining tests at each child. If no tests
		remain, store the row list and its aggregates. For blocks of
		constant size, a direct scan returns singleton lists. Each
		terminal list stores actual row identifiers. Along a query the
		terminal lists are disjoint, and empty lists can be skipped.
		We combine sums and counts by addition and combine lattice bases
		by taking their HNF. This gives the requested aggregate and an
		actual representative when the answer is nonempty.
		
		There are at most \(b_R=O_{r,p}((R\log(R+2))^r)\) branches of each
		kind in a node, including charts and faces. Ignoring bit factors
		and sampling retries, let \(T_j(N)\) be construction work for \(j\)
		remaining tests. With \(T_0(N)=O(N)\), it satisfies
		\[
		T_j(N)\le b_R T_j(\lfloor N/R\rfloor)
		+b_R T_{j-1}(N)+O_{r,R,p}(N).
		\]
		Choose \(R\) so \(b_R/R^{r+\eta}<1\). Induction on \(j\) and \(N\)
		gives \(T_j(N)=O_{r,p,\eta}(N^{r+\eta})\), and this also bounds the
		number of construction nodes and stored row references. The query
		recurrence is
		\[
		Q_j(N)\le Q_j(\lfloor N/R\rfloor)+Q_{j-1}(N)+O_{r,R,p}(1).
		\]
		Its solution is \(O_{r,R,p}(\log^j(N+2))\), including the number
		of returned list pointers. A scan of the constantly many simplices
		at a node suffices for the exact containment test.
		
		Take \(M=O_{r,p,\eta}(h^{r+\eta})\) as a deterministic bound on the
		number of construction nodes. At each node cap the number of trials
		at \(\lceil\log_2(M/\zeta)\rceil+1\) and report failure if they all
		fail. The sample bound holds conditionally on the rows at that node,
		so a union bound proves the asserted probability and gives the
		stated worst-case construction time.
	\end{proof}

	The same sign-test blocks handle tile containment and the domain
	and navigation operations on groups of types.
	
	\begin{lemma}[grid-indexed local atlas]\label[lemma]{lem:grid-atlas}
		Let \(I,K,V\) satisfy \cref{lem:atlas-navigation}, and fix
		\(\eta>0\). For a positive integer block width \(h\), the
		construction and \(Q\) caller queries together cost
		\[
		\Ot_{r,\eta}\!\left((m+V)h^{r+\eta}
		+(V+Q)(1+\Psi(I)/h)\right)\operatorname{poly}(B).
		\]
		Construction fails with probability at most the requested
		budget \(\zeta\in(0,1)\). Its polynomial bit factor includes
		\(\log(1/\zeta)\). Queries return a strict
		domain separator or the exact value, a supporting price, a basic
		witness and a minimal-domain-face record. The records support
		inherited queries and free-column hull support and separation in
		the same bound. Listing free types incurs their output length.
		With \(h\) the least integer satisfying \(h^{r+1}\ge\widehat\Psi_I\),
		the cost is
		\[
		\Ot_{r,\eta}\!\left((V+Q)
		(1+\Psi(I)^{r/(r+1)+\eta})\right)\operatorname{poly}(B).
		\]
	\end{lemma}
	
	\begin{proof}
		Partition the type records into groups of at most \(h\). A record
		stores its column, multiplicity, full cost sum and immutable
		type pointer. Index \(v\cdot a>0\) and \(v\cdot a=0\) using
		\cref{lem:cutting-blocks}. For the positive types, aggregate the
		full target and cost sums. For the zero types, aggregate a lattice
		basis by HNF and retain the nonempty canonical list pointers.
		The selected endpoint attaining the support value is represented
		by the positive lists. Scanning the groups evaluates \(h_{Z_I}(v)\)
		and supplies this endpoint in
		\(\Ot(1+m/h)\operatorname{poly}(B)\) time. The procedure of
		\cref{lem:grouped} uses these evaluations to separate an exterior
		target or to obtain a normal exposing the minimal face.
		
		For that exposing normal \(v\), the zero-type lists give a
		compressed pool and the positive aggregate gives the fixed
		endpoints. Its face key is the canonical affine space
		\(t+\operatorname{span}\{a\mid v\cdot a=0\}\) together with the pool
		identifier. Indeed \(F=Z_I\cap\operatorname{aff}(F)\). Merge
		free lattice bases in fixed dimension. Each canonical list is
		nonempty and disjoint from the others in the view, so traversing
		the view costs its output length after the charged query. A
		specified type can be tested with its original dictionary entry
		and the exposing normal, without enumerating the view.
		
		For support on the symmetric hull of the free columns, index
		the two tests \(v\cdot a=0\) and \(u\cdot(\pm a)>c\) on the same
		record, handling each sign separately. The first uses at most
		\(r\) parameters. The second is homogeneous in \((u,c)\).
		When this vector is nonzero, divide by the largest absolute
		coordinate and fix that coordinate to \(+1\) or \(-1\). Each
		of these finitely many charts has at most \(r\) free coordinates
		and uses affine tests. The zero parameter vector is answered
		directly. Test emptiness by the free-type count. For a nonempty
		set and \(u\ne0\), the candidate support values have numerators
		and denominators bounded by \(2^{\operatorname{poly}(B)}\).
		Bisect over a rational interval containing them, maintaining
		one threshold with a passing row and another with none. When
		the interval is narrower than the gap between distinct possible
		values, a passing row at the lower threshold attains the
		maximum. Evaluate it exactly on that signed column. For \(u=0\)
		any free column attains zero. Strong separation in the face
		span follows by the same support-to-separation procedure.
		
		Use \cref{lem:atlas-navigation} for the seed and the traversal.
		A navigation record is one actual inequality adjacent to the
		current type prefix, carrying the label of its adjacent copy.
		Each group has at most \(2h\) such records. Index strict violation
		of an inequality and scan the groups for a violated row. After
		changing a prefix, rebuild only its group, including on rollback.
		The \(O_r(V)\) attempts therefore cost
		\(\Ot(V(h^{r+\eta}+1+m/h))\operatorname{poly}(B)\), including
		all failed searches and the persistent witness records.
		
		As the traversal accepts a tile, index it in the grid of
		half-open unit cells in the coordinates \(B_I^{-1}t\). Compute its coordinate bounding
		box from its constant number of vertices. Register it in every
		cell whose index in each coordinate runs from the floor of the
		lower endpoint through the floor of the upper endpoint,
		inclusive. The basis-coordinate bound in
		\cref{lem:atlas-navigation} makes each side length \(O_r(1)\),
		so there are \(O_r(1)\) registrations per tile. Construct only
		the buckets that receive registrations, in a dictionary keyed by
		the integer cell index.
		
		All tiles registered to a cell are contained in a fixed-size
		expansion of that cell in basis coordinates, even if a
		bounding-box registration includes a tile disjoint from the
		cell. In the original coordinates the expanded box has volume
		\(O_r(|\det B_I|)\). The disjoint tile interiors each have volume
		at least \(\det\Lambda_I\), so there are at most
		\(O_r(\widehat\Psi_I)\) registrations in one bucket. Divide each
		bucket's registered tiles into groups of at most \(h\). A row
		is one actual tile. Use its \(2r\) closed facet inequalities as
		the tests of \cref{lem:cutting-blocks}, all on the same query
		point. Total registrations are \(O_r(V)\), so construction of
		these group indexes costs \(\Ot(Vh^{r+\eta})\operatorname{poly}(B)\).
		
		A target query first tests the domain and returns its separator
		if infeasible. Otherwise its index \(\lfloor B_I^{-1}t\rfloor\)
		locates one bucket. Scan that bucket's groups to find an
		actual containing tile, and use its stored witness. The closed
		inequalities and inclusive floor range cover both tile and grid
		boundaries. A feasible target in \(K\) always has a containing
		registered tile by \cref{lem:atlas-navigation}. An empty bucket
		or unsuccessful scan takes at most the same time as a successful
		lookup. Face queries use the inherited embeddings described
		above. They are charged to the answering source with its own
		\(\Psi(I)\), even when the queried child has a smaller rank or
		volume.
		
		Preprocessing the type groups and enumerating the initial seed
		costs \(\Ot(mh^{r+\eta})\operatorname{poly}(B)\). The
		polynomially many exact support evaluations per caller query
		and the one bucket scan cost
		\(\Ot(1+(m+\widehat\Psi_I)/h)\operatorname{poly}(B)\) per
		query. The bounds \(m=O_r(\Psi(I))\) and \(m=O_r(V)\) follow
		from the lattice volume estimates of \cref{thm:ledger}. This
		proves the first cost bound. Give the \(i\)-th index build a
		failure budget \(\zeta/(2i^2)\), where \(\zeta\) is the
		atlas budget, and abort this construction upon failure. The
		truncated builds of \cref{lem:cutting-blocks} have a union bound
		below \(\zeta\), also for the adaptively chosen groups. The
		logarithm of the number of builds is polynomial in \(B\).
		Finally \(h\) is obtained by integer comparison and
		\(h=\Theta_r(\Psi(I)^{1/(r+1)})\). Substitution gives the
		second cost bound.
	\end{proof}
	
	\section{Compressed Transfers and Lattice Cores}
	\label{sec:transfer}
	
	A transfer supplies the child targets and witnesses that the local
	oracles query. Its convolution runs in an integer-coordinate box, and
	the physical crop determines which outputs enter the oracle window.
	Immutable pools let the transfer recover individual labels and let
	the lattice tree locate actual labels without expanding the pool.
	
	\subsection{An integer box with few lattice points}
	\label{sec:coefficient-box}
	
	\begin{lemma}[discrete boxing]
		\label[lemma]{lem:box}
		Let \(\Lambda\subset\R^r\) be a full-rank rational lattice in fixed
		dimension, let \(R=\{x\in\R^r\mid \|Tx\|_\infty\leq 1\}\) for an
		invertible rational map \(T\), and let \(c\geq 1\) be rational. In
		polynomial bit time one can find a basis
		\(\lambda_1,\ldots,\lambda_r\) of \(\Lambda\) and nonnegative integer
		widths \(\Delta_1,\ldots,\Delta_r\) such that
		\[
		\begin{aligned}
			cR&\subseteq \mathcal B
			:=\Bigl\{\sum_{i=1}^r \xi_i\lambda_i
			\mathrel{\Bigm|} \xi\in\R^r,\ |\xi_i|\leq\Delta_i
			\text{ for every }i\Bigr\},\\
			|\mathcal B\cap\Lambda|
			&=\prod_{i=1}^r(2\Delta_i+1)
			\leq C_r c^r |R\cap\Lambda|.
		\end{aligned}
		\]
		The same conclusion holds for a centrally symmetric rational body \(R\)
		with a supplied rational map \(T\) satisfying
		\(B_2\subseteq TR\subseteq C'_rB_2\), where \(B_2\) is the Euclidean
		unit ball. In this case \(C_r\) can also depend on \(C'_r\).
	\end{lemma}
	
	\begin{proof}
		Apply LLL reduction~\cite{LLL82} with Lov\'asz parameter \(3/4\) to
		\(T\Lambda\), clearing denominators and undoing this scalar change
		afterwards. Write \(b_1,\ldots,b_r\) for the reduced basis and \(d_i\)
		for its Gram--Schmidt lengths. Size reduction gives
		\(|\mu_{ij}|\leq 1/2\) for \(j<i\). The Lov\'asz condition gives
		\(d_{i+1}^2\geq d_i^2/2\), so \(d_j\leq 2^{(i-j)/2}d_i\) for
		\(j<i\). Orthogonality gives \(\|b_i\|_2\leq A_r d_i\) for a constant
		\(A_r\).
		
		Let \(b_i^*\) be the coordinate functional of the reduced basis,
		represented as a Euclidean vector. In orthonormal Gram--Schmidt
		coordinates the basis matrix is \(D U\), where
		\(D=\operatorname{diag}(d_1,\ldots,d_r)\) and \(U\) is unit upper
		triangular with all off-diagonal entries bounded by \(1/2\) in absolute
		value. The entries of \(U^{-1}\) are bounded in terms of \(r\) by the
		finite series for \((I+(U-I))^{-1}\). The \(i\)-th row of
		\(U^{-1}D^{-1}\) only uses \(j\geq i\), for which
		\(d_j\geq 2^{-(j-i)/2}d_i\). It follows that
		\(\|b_i^*\|_2\leq A'_r/d_i\).
		
		Set \(\lambda_i=T^{-1}b_i\). For a parallelotope take
		\(\Delta_i=\lceil c\|b_i^*\|_1\rceil\), since the coefficient
		\(\xi_i(x)=b_i^*\cdot Tx\) has support radius \(\|b_i^*\|_1\) on \(R\).
		For the supplied rounding use
		\(\Delta_i=\lceil c C'_r\|b_i^*\|_1\rceil\), increasing \(C'_r\) to a
		fixed integer if necessary. These are rational computations using the
		inverse basis matrix. Thus \(cR\subseteq\mathcal B\), and
		\(2\Delta_i+1\leq C''_r c\max(1,1/d_i)\).
		
		For the reverse count, set
		\(q_i=\lfloor 1/(r\|b_i\|_2)\rfloor\) in the analysis. Every combination
		\(\sum_{i=1}^r z_i b_i\) with \(|z_i|\leq q_i\) lies in \(B_2\subseteq TR\)
		by the triangle inequality. They are distinct points of \(T\Lambda\).
		Since \(2\lfloor u\rfloor+1\geq\max(1,u)\) for \(u\geq 0\), each
		factor \(2q_i+1\) is at least
		\((rA_r)^{-1}\max(1,1/d_i)\). Multiplying proves the required bound.
		The reduced vectors form a basis of the entire lattice, so the product
		for the number of array cells is exact. The rational transformations
		and integer widths have polynomial bit length by the bit complexity of
		LLL and rational matrix inversion.
	\end{proof}
	
	\begin{lemma}[reboxing a column system]\label[lemma]{lem:rebox}
		Let \(m\) distinct integer columns with bit length at most \(L\)
		generate a rank-\(r\) lattice \(\Lambda\), and let \(P\) be
		their symmetric column hull. Put
		\(\Psi=r!\vol_r(P)/(2^r\det\Lambda)\).
		In \(\Ot(m+1)\operatorname{poly}_r(L)\) bit operations one can
		give integral lattice coordinates in which every column
		satisfies \(|a_j|\le q_j\), with integers \(q_j\ge 1\) and
		\[
		\prod_{j=1}^r q_j = O_r(\Psi).
		\]
		The transformation is invertible on the real column span and both directions have polynomial encoding length.
	\end{lemma}
	
	\begin{proof}
		An incremental normal-form scan gives a basis of \(\Lambda\). Work in its coordinates and start with any independent \(r\)-tuple of actual columns. Given the matrix \(A\) of this tuple, maximize the absolute value of each coordinate of \(A^{-1}a\) over the current types by a scan. If some coordinate exceeds \(2\), replace the corresponding column of the tuple by a maximizer. Cramer's rule shows that the determinant grows by a factor greater than \(2\) at each replacement. It starts at a nonzero integer and is bounded by a fixed-dimensional determinant in the input. There are therefore \(\operatorname{poly}_r(L)\) scans. At termination \(P\subseteq R=A[-2,2]^r\subseteq 2rP\).
		Apply \cref{lem:box} to \(R\) and \(\Lambda\). The widths can be
		taken to be at least one because \(P\) contains a spanning set of
		lattice columns. Use the fundamental domain in \((r/2)P\) from
		the proof of \cref{lem:source-halo}. Its translates centered at
		points of \(R\cap\Lambda\) are disjoint up to measure zero, so
		\[
		|R\cap\Lambda|\det\Lambda \le \vol_r\bigl((2r+r/2)P\bigr).
		\]
		The product bound in \cref{lem:box} now yields \(\prod_{j=1}^{r}q_j=O_r(\Psi)\). Its basis identifies \(\Lambda\) with \(\Z^r\) and gives the required invertible map. The arithmetic uses only fixed-dimensional bases, rational matrix inversion, and the bit bounds in \cref{lem:box}.
	\end{proof}
	
	The following estimate relates the array size to the points already paid for by the source.
	
	\begin{lemma}[array size charged to a source]\label[lemma]{lem:source-box}
		Let \(\Lambda\) have rank \(r\), and let \(S\subseteq R\subseteq aS\)
		be origin-symmetric convex bodies in its real span, with \(a\geq1\).
		If \(t\in\Lambda\), \(\alpha>0\), and
		\(\mu=|(t+\alpha S)\cap\Lambda|\), then
		\[
		|R\cap\Lambda|\leq \max(1,\lceil 2a/\alpha\rceil)^r\,\mu .
		\]
		If \(R\) and \(\Lambda\) satisfy the hypotheses of \cref{lem:box}
		and \(a/\alpha\) is at most polylogarithmic, the resulting array for
		\(R\) has \(\Ot(\mu)\) cells. Given the original-weight readout and
		the current type dictionary, one can retrieve the types in the
		array with \(\Ot(\mu)\) dictionary and arithmetic operations.
	\end{lemma}
	
	\begin{proof}
		Set \(k=\max(1,\lceil 2a/\alpha\rceil)\). Since \(t\in\Lambda\),
		translation by \(t\) preserves \(|R\cap\Lambda|\). Apply
		\eqref{eq:lattice-dilation} with \(C=a\) and \(2a/k\le\alpha\).
		Under the array hypotheses, \cref{lem:box} with \(c=1\) gives
		\(\Ot(\mu)\) cells. Enumerate these cells and use the
		original-weight readout to look up their vectors in the current
		type dictionary, without scanning a larger parent dictionary.
	\end{proof}
	
	\subsection{Witnessed tables and the geometric crop}
	\label{sec:table}
	
	A finite removal-table entry stores the cost and a witness for a subset
	of a specified labeled pool. The index is the exact column sum of the
	witness. A table for a removal pool and a prescribed family of required
	assignments has the following two guarantees. Each finite entry is
	attained by an actual subset of that pool. For each required assignment
	at an index retained by the caller, the entry at that index is finite
	and has cost at most the cost of the assignment. A table may improve
	the cost using a subset outside the required family. This one-sided
	guarantee suffices because replacing an assignment with one of the same
	column sum and smaller cost preserves the residual target and increases
	the remaining cost allowance.
	
	Let \(\mathcal D=\prod_{j=1}^r[-\Delta_j,\Delta_j]\) be the
	integer-coordinate box. It has \(M=\prod_{j=1}^r(2\Delta_j+1)\)
	cells.
	For a labeled removal pool \(I\), write \(a_i\) and \(c_i\) for the
	signed column and cost of label \(i\). The family we need to dominate is
	\[
	\mathcal F_{\mathcal D}(I)
	=\Bigl\{Y\subseteq I\mathrel{\Bigm|}
	\sum_{i\in X}a_i\in\mathcal D
	\text{ for every }X\subseteq Y\Bigr\}.
	\]
	Each label has one fixed sign. When a transfer starts from a fixed
	endpoint assignment, toggling a selected label reverses its column
	and negates its cost, while toggling an unselected label keeps both.
	The endpoint assignment itself is carried as a separate affine
	translation of the target and allowance.
	
	\begin{lemma}[witnessed signed-box table]
		\label[lemma]{lem:signed-table}
		Assume full witnessed integer \((\min,+)\)-convolution of length \(N\)
		takes \(\Ot(N^\rho)\) time times its bit-length and error
		amplification factors, where \(\rho>1\).
		In fixed dimension, one can construct a table \(D\) on
		\(\mathcal D\cap\Z^r\) such that every finite entry has an
		actual labeled subset as a witness. With probability at least
		\(1-\eta\), simultaneously for
		every \(z\in\mathcal D\cap\Z^r\),
		\[
		\min_{\substack{Y\subseteq I\\ \sum_{i\in Y}a_i=z}}
		\Bigl(\sum_{i\in Y}c_i\Bigr)
		\;\leq\;D(z)\;\leq\;
		\min_{\substack{Y\in\mathcal F_{\mathcal D}(I)\\ \sum_{i\in Y}a_i=z}}
		\Bigl(\sum_{i\in Y}c_i\Bigr).
		\]
		An empty minimum is \(+\infty\). For explicit labels, the time is
		\(\Ot(|I|+M^\rho)\) times the bit-length and error amplification
		factors. For a compressed pool, suppose its nonzero vectors in
		\(\mathcal D\) can be enumerated in \(\Ot(M)\) time and its \(q\)
		cheapest current labels of a specified nonzero vector retrieved in
		\(\Ot(q+1)\) time. If zero columns are present, also supply their
		optimal total cost and a compressed attaining assignment in
		\(\Ot(M)\) time. Then the time is \(\Ot(M^\rho)\) with the same
		factors, excluding construction of the pool access structures.
		The number of materialized labels, stored entries, and witness
		pointers is \(\Ot(M)\) in the compressed case, with polylogarithmic
		dependence on \(1/\eta\).
	\end{lemma}
	
	\begin{proof}
		Discard columns outside \(\mathcal D\), since a required choice
		cannot use such a column even as a singleton. Fix every negative-cost
		zero label to be selected and every other zero label to be omitted.
		Keep their cost and compressed witness separately. Partition the
		remaining labels into coordinate orthants with a fixed convention
		for zero coordinates. Reflect each group to the nonnegative box
		\(\mathcal D^+=\prod_{j=1}^r[0,\Delta_j]\).
		
		Delete zero-width coordinates within an orthant. Index the remaining
		coordinates by \(1,\ldots,r'\), use radices \(b_j=\Delta_j+1\geq2\),
		and put
		\[
		E(u)=\sum_{j=1}^{r'} u_j\prod_{i=1}^{j-1}b_i,\qquad
		d(u)=\sum_{j=1}^{r'} u_j,\qquad
		T=\prod_{j=1}^{r'} b_j-1\leq M-1.
		\]
		For \(T=0\), only the separately handled zero labels can contribute.
		Otherwise \(E\) bijects the lattice points of the box with
		\([0,T]\cap\Z\), omitting the fixed zero coordinates.
		A scalar subset of weight at most \(T\) uses at most
		\(\lfloor T/E(u)\rfloor\) labels of a nonzero vector \(u\).
		Keep that many of the cheapest eligible labels, or all if fewer
		exist. Substitution within a vector preserves the sum and can
		only decrease cost. Injectivity gives at most
		\(\sum_{w=1}^T\lfloor T/w\rfloor=O(T\log(T+1))\) retained labels.
		
		Write \(J\) for the retained labels of this orthant. Set
		\(C_*=\sum_{i\in J}|c_i|\) and \(K=2C_*+1\).
		Give label \(i\) scalar weight \(E(u_i)\) and cost \(Kd(u_i)+c_i\).
		When a subset has vector sum \(A\) and scalar weight \(E(z)\)
		for \(z\in\mathcal D^+\), carrying its coordinates produces \(z\).
		Each carry from position \(j\) decreases the digit sum by \(b_j-1\).
		There is no overflow, since \(E(z)\leq T\). Thus
		\(d(A)\geq d(z)\), with equality exactly when \(A=z\).
		For instance, in radix two two copies of \((1,0)\) have the
		same encoded weight as \((0,1)\), but a larger digit sum.
		At weight \(E(z)\), minimizing the augmented cost first minimizes
		\(d(A)\), then the unaugmented cost. All integer values in this
		calculation have bit length \(O_r(B+\log(M+1))\), where \(B\)
		bounds the input costs and coordinates.
		
		Compute all exact scalar weights through \(T\) in one call to
		the all-capacities 0--1 reduction of~\cite[Corollary~14]{CMWW19},
		using its convolution circuit with witnessed minimum gates in place
		of maximum gates. The underlying arrays in their Algorithms 2--4
		are indexed by exact weight. The following implementation details
		give the witness and storage bounds. Use layered color
		coding~\cite{Bringmann17} with dyadic layers \([a,2a)\).
		For a layer let \(L=\lfloor T/a\rfloor\), take
		\(\lambda=\lceil C\log((T+2)/\eta)\rceil\) for a sufficiently large
		constant \(C\), and hash labels independently to \(b\) buckets,
		where \(b\) is the next power of two at least \(\max(1,L/\lambda)\).
		Fix an optimal scalar subset for each attainable exact weight.
		The expected number of its labels in any bucket is at most
		\(\lambda\). A Chernoff bound and a union bound over weights,
		layers, and buckets give occupancy at most \(k=16\lambda\)
		for all these subsets with failure probability at most \(\eta/4\).
		In a bucket, use \(2k^2\) colors and repeat independently
		\(O(\log((T+2)/\eta))\) times. The probability that any fixed set
		of at most \(k\) labels is colorful in one repetition is at least
		\(1/2\). A color leaf permits the empty choice and the cheapest
		label at each weight. Combine colors by convolution up to
		\(C_0=\min(T,2ak)\), and take the entrywise minimum over the
		repetitions. The same union bound gives simultaneous coverage
		of all designated bucket subsets with additional failure
		probability at most \(\eta/4\). Merge buckets at height \(h\)
		with cap \(\min(T,2^h C_0)\), then merge layers up to \(T\).
		Positive weights ensure that the designated subsets survive
		their scalar caps.
		
		Each finite gate entry keeps its two child indices and the sum
		of their weights, digit sums, and original costs. Check the
		reported indices and augmented value against those children.
		Each gate joins disjoint label pools, and an entrywise minimum
		keeps only one witness. These checks ensure that every retained
		entry has an actual witness even if a gate fails. The color
		leaves and all bucket levels together store \(\Ot(bC_0)=\Ot(T)\)
		entries per layer. Indeed \(b=O(L/\lambda)\) when \(L>\lambda\),
		and \(b=1\) otherwise. At each bucket-tree level the product of
		the number of nodes and their caps is \(O(bC_0)\). Layer merges
		add only polylogarithmic factors. Thus total convolution input
		length is \(\Ot(T)\), each gate has length \(O(T+1)\), and their
		total time is \(\Ot(T^\rho)\) with the stated bit and error
		factors. At the final entry for weight \(E(z)\), require digit
		sum \(d(z)\) and use the stored original cost and witness.
		This test is performed on aggregate values. Witness labels are
		recovered by following the pointers only when needed.
		
		Reflect the orthant tables back and merge them by witnessed
		convolution. At each signed merge translate input digit \(j\)
		by \(\Delta_j\), use radix \(4\Delta_j+1\), and subtract
		\(2\Delta_j\) from the output digit. This encoding has no carry,
		uses \(O_r(M)\) cells, and allows a crop to \(\mathcal D\) at
		each signed merge. Verify the child indices and values here as
		well, then add the separate zero-column cost and witness.
		For \(Y\in\mathcal F_{\mathcal D}(I)\), every reflected orthant
		sum lies in \(\mathcal D^+\), and every partial union of its
		orthant choices stays in \(\mathcal D\). Replacing an orthant
		choice by a cheaper witness with the same vector sum preserves
		these partial sums. This proves the two inequalities on the
		joint success event. Use a dimension-dependent fraction of
		\(\eta\) for each orthant, and a summable error allocation over
		all witnessed convolution calls. This gives the stated total
		failure probability. With explicit labels the initial grouping
		and selection cost \(\Ot(|I|)\), and the compressed interface
		spends \(\Ot(M)\) by the truncation bound.
	\end{proof}
	
	We next combine a removal table with the allowances of the parent queries.
	
	\begin{lemma}[witnessed allowance merge]\label[lemma]{lem:allowance-merge}
		Given a source allowance table and a witnessed removal table,
		assume all finite allowances and costs have a common positive
		integer denominator. Then the
		maximum residual allowance for each target and its attaining source
		and removal witness can be computed by one witnessed convolution
		in a box containing the reflected supports and their sums.
	\end{lemma}
	
	\begin{proof}
		For a batch of source targets \(t\), let \(S(t)\) be the greatest
		strict cost allowance for \(t\), with \(S(t)=-\infty\) for an
		absent target. A finite removal entry \(D(v)\) gives residual target
		\(u=t-v\) and allowance \(S(t)-D(v)\). Hence
		\[
		\begin{aligned}
			A(u)&=\max_{\substack{v\in\operatorname{dom}(D)\\u+v\in\operatorname{dom}(S)}}\bigl(S(u+v)-D(v)\bigr),\\
			-A&=F\star G,\qquad F(t)=-S(t),\quad G(-v)=D(v).
		\end{aligned}
		\]
		Here \(\operatorname{dom}\) denotes the finite support, and the
		empty maximum is \(-\infty\). In integer coordinates the same carry-free
		construction gives one witnessed convolution in a box containing
		the two supports and their sums. Scale both cost arrays by the
		common denominator and divide finite output values by it.
		Every original allowance is half-integral. In a price
		normalization with price denominator \(d_\pi\), the common
		denominator \(2d_\pi\) suffices and has polynomial bit length.
		For each finite output, check the decoded child indices and
		their finite values, and verify \(t-v=u\) and
		\(S(t)-D(v)=A(u)\). Store their source and removal witness.
		These checks and the convolution scan the output box.
		
	\end{proof}
	
	A fixed supporting price changes the allowances by an affine term.
	To apply \cref{lem:allowance-merge} in this normalization, let \(R\)
	be the pool removed in one transfer and \(e\in\{0,1\}^{R}\)
	its fixed endpoint vector. Set \(b=\sum_{i\in R}e_i a_i\) and
	\(d=\sum_{i\in R}e_i c_i\).
	Toggling label \(i\) contributes signed column
	\(\sigma_i a_i\) and reduced cost
	\(\sigma_i(c_i-\pi\cdot a_i)\), where \(\sigma_i=1-2e_i\) and
	\(\pi\) is a fixed rational price for the batch. If \(e_i\)
	minimizes the reduced cost with a fixed tie rule, these toggle costs
	are nonnegative. Clearing a common denominator in \(\pi\) supplies
	integer input to Lemma~\ref{lem:signed-table}. Undo the scaling
	afterwards. Let \(\widehat D(v)\) denote the reduced table value.
	Each witness at \(v\) has actual cost \(d+\pi\cdot v+\widehat D(v)\)
	on \(R\) and sends \(t\) to \(u=t-b-v\). Thus the best reduced
	allowance for \(u\) is
	\[
	A(u)-\pi\cdot u
	=\max_{\substack{v\in\operatorname{dom}(\widehat D)\\u+b+v\in\operatorname{dom}(S)}}\Bigl(
	S(u+b+v)-\pi\cdot(u+b+v)
	-(d-\pi\cdot b)-\widehat D(v)
	\Bigr).
	\]
	This is the same batch convolution after translating the source
	indices by \(-b\) and adjusting their allowances. The identity
	also shows that costs can be compared and witnessed in either
	normalization. If several source prices are used, each batch fixes
	its own price before making this transformation.

	The boxing lemma controls the number of array entries. The array can
	still extend far beyond the intended geometric window. For example,
	take \(\Lambda=\Z^2\) and
	\(R=[-L,L]\times[-\epsilon,\epsilon]\), with \(L\) a positive integer
	and \(0<\epsilon<1/2\). In the standard basis, upward rounding gives an
	array with three rows of lattice points. Their number is
	\(3|R\cap\Lambda|\), whereas the height ratio is \(1/\epsilon\).
	The array is a computation domain. Its physical image cannot serve as
	a constant dilation of \(R\) for source construction. \Cref{fig:clip} shows the crop for this example.
	
	\ClipFigure
	
	For each transfer the caller supplies the physical removal window and
	the physical candidate window. Run the table circuit in integer
	coordinates, with the crops after the signed orthant merges.
	Scan its final removal table and discard entries outside the removal
	window. Combine the retained entries with the incoming targets in a
	box containing the sum or difference of the two supports, according to
	the transfer sign. If the supports already lie in a common coefficient
	box, doubling its widths suffices. Crop the finite witnessed outputs
	to the physical candidate window, then filter for the child lattice
	and span. Only these geometrically retained entries enter the query
	batch. A constant-facet crop takes a constant number of rational tests
	per entry. A crop defined by a current column hull uses its own charged
	membership oracle.
	
	The transfer proof has to place every required removal sum and
	residual target in the two physical windows. With that property, each
	crop preserves the dominance guarantee, and each finite entry still
	has a witness. The source oracle can be built on the independently
	proved query window and its navigation enlargement. Its construction
	cost cannot be charged using the physical volume of the array.
	
	\subsection{Immutable pools of virtual copies}
	\label{sec:pool}
	
	After the proximity reduction, the residual instance has a bounded
	virtual universe of \(U=\wm^{O(1)}\) labels, including its slack
	labels. Within each original weight class, sort the retained original
	records by cost with a fixed tie rule and assign the labels consecutive
	ordinals. A record is a run of equal original cost. The sorted run
	boundaries and run prefix sums require \(O(n)\) words and
	\(\Ot(n)\) arithmetic operations. The perturbation is a
	fixed-degree polynomial in the ordinal. Thus a perturbed prefix sum
	uses a run prefix sum and sums of powers, and the perturbed costs stay
	sorted within each type.
	
	\begin{lemma}[permanent types under restrictions and lifts]\label[lemma]{lem:stable-type}
		Arithmetic and face restrictions preserve the single stored column
		per permanent type. A lift using the signed scalar encoding and a
		linear progression readout also preserves this property after the
		working orientations are undone.
	\end{lemma}
	
	\begin{proof}
		Store the original weight as the permanent key of a type. Arithmetic
		and face restrictions retain the existing column of a type, in their
		current coordinates. A lift must also store one column per permanent
		type. To verify this for the signed scalar encoding of the GAP
		construction, let \(a\) be the type's column and \(\sigma_j\in\{-1,1\}\)
		the temporary orientation of label \(j\). For a linear scalar encoding
		\(E\), let \(\tau_j\) be the sign of the nonzero value
		\(E(\sigma_j a)\) used to make it positive. Each such label presents
		the same positive value \(|E(a)|\), so the multiset GAP interface gives
		it the same coordinate \(u_a^+\). If the lift appends a linear
		readout \(\psi\) of this coordinate, restoring both signs stores
		\[
		\sigma_j\tau_j\psi(u_a^+)
		= \operatorname{sgn}(E(a))\psi(u_a^+).
		\]
		The score used to choose exceptional labels can depend on cost, while
		the stored column stays constant on a permanent type. Retaining the
		linear readout of the original weight lets the dictionary test a
		candidate vector by checking its weight key and its full column.
		
	\end{proof}
	
	\begin{lemma}[compressed pool trace]
		\label[lemma]{lem:pool}
		Suppose each permanent type is represented by an immutable ordered
		sequence with compressed access to costs and interval sums. A live
		view is an ordinal interval with explicitly recorded deletions. The
		operations below support interval restriction, deletion, prefix
		counts and sums, rank and select, thresholding, and access to the
		complement of a restriction followed by explicit deletions. For a
		trace visiting \(E\) type records, changing \(A\) per-type interval or
		coordinate records, deleting or materializing \(D\) explicit labels,
		and storing \(H\) references, the total time and space are
		\[
		\Ot(n+E+A+D+H)\operatorname{poly}(B+\log(U+2)),
		\]
		where \(B\) bounds the bit length of the stored data. Operations and
		materializations repeated in different branches count with
		repetition.
	\end{lemma}
	
	\begin{proof}
		A type view stores its ordinal bounds and a persistent balanced search
		tree of deleted ordinals, augmented with counts and cost sums.
		Restriction changes the bounds. Deleting an ordinal copies a
		root-to-leaf path of logarithmic length. To count or sum labels on an
		interval, subtract the tree's range aggregate from the base interval
		aggregate. Rank follows from counts, and binary search on rank
		implements select in polylogarithmic time. A cost threshold is a
		binary search in the immutable sequence followed by a query in the
		deletion tree. The first or last \(q\) eligible labels can be retrieved
		with \(q\) select queries. Each saved reference names one version.
		
		If the child retains a cost interval and then deletes an explicit set,
		the removed view in the parent consists of the two inherited interval
		tails and those newly deleted labels that were eligible in the
		retained interval. These three lists are disjoint. A small ordered
		tree stores the explicit list. Merging the ordered iterators returns
		the first \(q\) eligible labels in \(\Ot(q+1)\) time after building the
		tree. Whole-type removal keeps a pointer to the parent type view.
		
		A supporting price changes the cost threshold in a type while leaving
		its live view fixed. Labels on either side of the threshold give the
		ordered candidate lists for adding and removing copies. At each
		explicitly visited type, thresholding and obtaining the endpoint sums
		use the operations above. A data structure that supplies an endpoint
		atlas by reference can avoid such visits when its own construction
		has been paid for. Costs for adding a materialized label or reference
		are polylogarithmic. Summing over the recorded visits and operations
		gives the stated bounds.
	\end{proof}
	
	The structural routines use the pool views in two ways.
	If a structural routine gives counts by type, selecting \(e\) actual
	labels costs \(\Ot(m+e)\) for \(m\) visited types. Select from the
	specified cheap view, so the identities satisfy its current eligibility
	conditions. For the lift, the exceptional-label score on one type is
	\(|\alpha_a+\vartheta c_j|\), where \(c_j\) is the ordered perturbed
	cost. The maximum of this convex function over the remaining ordinal
	interval occurs at one of its eligible endpoints. Maintain two
	endpoint iterators per type in a heap. Each selected endpoint advances
	its iterator and the two iterators stop when they meet. The largest
	\(\nu\) scores can be selected in \(\Ot(m+\nu)\) time. Counts and
	aggregates defined by the sign of an affine reduced cost also reduce
	to a constant number of interval operations per visited type.
	
	In an arithmetic phase rooted at \(I_0\), the pool on
	\(\Gamma\) is \(I_0\cap\Gamma\), keeping the labels whose columns
	belong to \(\Gamma\). For a parent pool \(I=I_0\cap\Pi\) with
	\(\Gamma\subseteq\Pi\), filtering the current types by membership
	in \(\Gamma\) gives \(I\cap\Gamma=I_0\cap\Gamma\).
	Cache the resulting view by phase and lattice, with separate views
	for the temporary cheap pools. The generating parent pays for
	constructing the view even if every candidate is later rejected.
	
	\subsection{Lattice searches on compressed types}
	\label{sec:lattice-core}
	
	For a labeled pool \(Y\), let \(\Lambda_Y\) be the integer span of
	its columns. Fix a full-rank parent \(I\) in dimension \(r\ge1\),
	with lattice \(\Lambda=\Lambda_I\), symmetric column hull \(P=P_I\),
	and normalized volume \(\Psi=\Psi(I)\). Use
	\(J\ge\lceil\log_2\Psi\rceil\).
	Every actual-column basis \(B_0\) satisfies
	\([\Lambda:\Lambda_{B_0}]\le\Psi\), because its symmetric
	crosspolytope lies in \(P\). The choice of \(J\) in
	\cref{sec:scales} uses the upper bound on \(\Psi(I)\) for stored
	pools from \cref{sec:global}.
	
	\begin{lemma}[searching a tree of lattice bases]
		\label[lemma]{lem:certificate}
		A pool \(Y\) on \(m\) types admits a balanced tree with the following
		operations, up to polynomial factors in \(B+\log(U+2)\).
		It can be built in \(\Ot(m)\) time and updated after a type-count
		change in \(\Ot(1)\) time. For a character \(\chi\) of the current
		lattice and an integer \(\kappa\ge0\), let
		\[
		F_\chi=\{i\in Y\mid\chi(a_i)\ne0\}.
		\]
		In \(\Ot(\kappa+1)\) time, the tree returns \(F_\chi\) if
		\(|F_\chi|\le\kappa\) and reports \(|F_\chi|>\kappa\) otherwise.
		Given a rational subspace \(H\), the tree
		finds an actual available label whose column is outside \(H\),
		or certifies that none exists, in \(\Ot(1)\) time.
	\end{lemma}
	
	\begin{proof}
		A leaf stores its type column, current eligible multiplicity, and
		ordinal-access view. A node stores an HNF basis for the lattice
		of all live types below it. Merging two nodes takes an HNF of at
		most \(2r\) basis columns. Polynomial bit algorithms for HNF and
		Smith form are available from~\cite{KB79}. A count change only modifies one
		root-to-leaf path. Path copying retains immutable pool versions,
		with the same logarithmic update bound. All these bases have polynomial
		encoding length.
		
		For a support query, discard a node if \(\chi\) vanishes on its
		basis. Otherwise descend: some actual column in that subtree must
		have nonzero character. At a nonzero leaf, add its entire compressed
		multiplicity to the support count. Stop as soon as this exceeds
		\(\kappa\). At most \(\kappa+1\) such leaves and their search paths
		are visited. If the total does not exceed \(\kappa\), retrieve all
		of their actual eligible ordinals. In particular, a high-multiplicity
		leaf does not trigger an expansion. Character tests are exact
		rational integrality tests.
		
		For the subspace query, discard a node precisely when all its basis
		vectors lie in \(H\). If the root fails this test, descend to a
		failing child until an actual type is reached, then retrieve one
		eligible label. This proves both that a label exists and how it is
		found. Temporary decrements and rollback give the same operations
		while selected labels are unavailable.
	\end{proof}
	
	\begin{example}[a character detects the single odd label]
		Let \(Y\) contain five labeled copies of \(2\) and one of \(3\),
		and let \(\kappa=1\). Then \(\Lambda_Y=\Z\). The character
		\[
		\chi\colon\Z\longrightarrow\Q/\Z,\qquad
		\chi(t)=t/2+\Z
		\]
		vanishes on the weight-\(2\) leaf and is nonzero on the weight-\(3\)
		leaf. The search skips all five even labels together and returns
		the actual odd label. Deleting it leaves the resilient lattice
		\(2\Z\).
	\end{example}
	
	The canonical core of~\cite[Lemmas~8.2--8.3]{CMZOpt26} is the
	largest closed subpool whose lattice survives small labeled
	deletions. The tree supplies the searches needed to compute it.
	
	\begin{lemma}[compressed canonical core]
		\label[lemma]{lem:core}
		Let \(Y_0\subseteq I\) have rank \(r\), and let \(\kappa\ge1\)
		be an integer. Assume deleting any at most \(\kappa(J+r)\) labels
		from \(Y_0\) preserves rank \(r\). One can compute the largest
		subpool satisfying
		\[
		Y_*=Y_0\cap\Lambda_{Y_*},\qquad
		\Lambda_{Y_*\setminus F}=\Lambda_{Y_*}
		\quad(F\subseteq Y_*,\ |F|\le\kappa)
		\]
		in \(\Ot(m+\kappa^2\Psi)\operatorname{poly}(B+\log(U+2))\)
		time, where \(m\) is the number of types in \(Y_0\).
	\end{lemma}
	
	\begin{proof}
		Build the tree of \cref{lem:certificate} and maintain its current
		pool \(Y\), initially \(Y_0\), with root lattice \(\Gamma\).
		To test resilience, extract \(\kappa+1\) disjoint actual-column
		bases \(B_1,\ldots,B_{\kappa+1}\) in a temporary tree version.
		For each basis, start with \(H=\{0\}\), repeatedly find a column
		outside \(H\), decrement its label count, and enlarge \(H\).
		The total number temporarily removed before the last extraction
		is at most \(\kappa r\). After at most \(J\) successful permanent
		restrictions, at most \(\kappa J\) labels will be absent. The
		rank hypothesis therefore leaves full rank before each basis is
		built. Within that basis, the labels already removed have columns
		in \(H\), so they cannot be the only columns outside \(H\). The
		tree search therefore finds all \(r\) columns. Roll back the
		temporary changes after obtaining
		the bases.
		
		For each \(1\le h\le\kappa+1\), compute Smith coordinates for
		\(\Gamma/\Lambda_{B_h}\) and enumerate its characters. There
		are at most \(\Psi\) of them by the actual-basis index bound.
		Use the tree to find the exact support of each pulled-back
		character, stopping its search above \(\kappa\). If
		\(1\le|F_\chi|\le\kappa\), delete these labels and let
		\(\Gamma'=\Lambda_{Y\setminus F_\chi}\), obtained from the
		updated root. This lattice is full-rank and proper, since it lies
		in the kernel of a character nontrivial on \(\Gamma\).
		
		The closure invariant needs no restoration scan. Suppose
		\(Y=Y_0\cap\Gamma\). Every survivor lies in \(\Gamma'\), and
		no deleted column does, since \(\Gamma'\subseteq\ker\chi\).
		Also \(\Gamma'\subseteq\Gamma\). Consequently
		\[
		Y_0\cap\Gamma'=Y\setminus F_\chi.
		\]
		The tree thus already represents the next closed pool. All labels
		of a nonzero type belong to the exact support, so permanent
		deletion simply deactivates the reported leaves. Each successful
		restriction at least doubles the index in \(\Lambda\), and an
		actual basis of the surviving pool keeps this index at most
		\(\Psi\). Hence there are at most \(J\) restrictions, justifying
		the cumulative deletion bound.
		
		Conversely, if deleting at most \(\kappa\) labels \(F\) changes
		the lattice, one extracted basis misses \(F\). The full-rank
		survivor lattice contains that basis lattice. A nontrivial
		character of \(\Gamma/\Lambda_{Y\setminus F}\) is therefore
		among the characters enumerated for this basis. Its nonempty
		support lies in \(F\), and the tree finds that support exactly.
		Thus termination certifies resilience.
		
		Every closed resilient subpool \(Z\subseteq Y_0\) survives each
		restriction. Indeed, if \(Z\subseteq Y\), resilience gives
		\[
		\Lambda_Z=\Lambda_{Z\setminus F_\chi}
		\subseteq\Lambda_{Y\setminus F_\chi}=\Gamma',
		\]
		so \(Z\subseteq Y_0\cap\Gamma'\). Induction proves that the
		terminal pool contains every such \(Z\), and hence is the unique
		largest one.
		
		Building the tree costs \(\Ot(m)\). In one iteration, basis
		extraction, rollback, and the successful update cost \(\Ot(\kappa)\).
		The \((\kappa+1)\Psi\) character searches cost \(\Ot(\kappa^2\Psi)\).
		The at most \(J\) iterations are absorbed by
		the polynomial bit factor. The live tree and its eligible leaf
		views are the output compressed pool, so a final scan is unnecessary.
	\end{proof}
	
	\begin{lemma}[canonical-core charge at a deep anchor]\label[lemma]{lem:core-charge}
		Under the rank guarantee of \cref{lem:core}, a canonical-core test
		at a deep anchor with paid mass \(\mu\geq c_r\nu^r\Psi\) and deletion
		threshold \(\kappa=\Ot(\nu)\) costs
		\(\Ot(\nu\mu)\operatorname{poly}(B+\log(U+2))\).
		The generating anchor pays this cost for every tested threshold.
	\end{lemma}
	
	\begin{proof}
		The lattice point count in \cref{sec:global} gives
		\(m=O_r(\Psi)\) for the parent pool. Substituting this
		and \(\kappa=\Ot(\nu)\) into
		\cref{lem:core} gives the stated cost because \(r\geq1\).
		In the application the rank condition follows from affordable depth
		in \cref{sec:restoration-collar}, and the mass lower bound follows
		from the paid deep window in \cref{thm:ledger}. The charge depends
		only on the generating anchor and applies even when its transfer
		has no feasible target.
		
	\end{proof}
	
	\Cref{alg:canonical-core} summarizes the canonical-core construction.
	
	\begin{algorithm}[H]
		\caption{\textsc{CanonicalCore}}
		\label{alg:canonical-core}
		\begin{algorithmic}[1]
			\Require Cheap pool \(Y_0\), ambient lattice \(\Lambda\), and threshold \(\kappa\), under \cref{lem:core}
			\Ensure The largest closed \(\kappa\)-resilient subpool and its lattice
			\State \Call{CanonicalCore}{Y_0,\Lambda,\kappa}
			\Begin
			\State Build the HNF tree \(\mathcal T\) for \(Y\gets Y_0\).
			\While{true}
			\State \(\Gamma\gets\operatorname{RootLattice}(\mathcal T)\)
			\State Find disjoint bases \(B_1,\ldots,B_{\kappa+1}\) by tree descent with temporary decrements, then roll back.
			\State Enumerate characters \(\chi\) of \(\Gamma/\Lambda_{B_h}\), for \(1\le h\le\kappa+1\).
			\State Seek an exact support \(F_\chi\) with \(1\le|F_\chi|\le\kappa\), using \(\mathcal T\).
			\If{no such support exists}
			\State \Return \((Y,\Gamma)\)
			\EndIf
			\State Deactivate the leaves supporting \(F_\chi\) and set \(Y\gets Y\setminus F_\chi\).
			\EndWhile
			\End
		\end{algorithmic}
	\end{algorithm}
	
	\subsection{Recovering an original count vector}
	\label{sec:count-recovery}
	
	Each transfer stores an endpoint template on the removed pool and a
	witness for the chosen toggles. An endpoint template consists of a
	pool version, the associated endpoint profile, and the removal view.
	An attaining target merge saves one predecessor and one table
	witness. Along the selected chain, the removed pools and the final
	active pool are disjoint. Within a selected table witness, each
	nonempty child is a witness in one of two disjoint label pools. A
	traceback can therefore skip empty witnesses and visit only the
	selected nonempty branches. The number of visited table nodes is at
	most the number of explicit selected toggles times the depth of the
	circuit, plus the roots visited along the selected chain.
	
	A terminal solution from \cref{lem:high-trial-center} stores the
	grouped LP's rounded prefix counts by signed type, the explicit
	reserve subset, and the attaining trial source. Rounding drops at
	most \(k\) fractional ordinals. In a remainder type, add the chosen
	positive-orientation count to the saved endpoint count and subtract
	the chosen negative-orientation count. This is the count form of
	\(x_i=e_i+(1-2e_i)\bar x_i\). Only the reserve subset and the
	fractional ordinals need individual labels, and their extraction
	is covered by \cref{lem:deep-charge}.
	
	To recover a count vector, sum the endpoint counts and signed toggle
	counts by permanent weight type, including the slack type and any
	reserve assignments. For each template this uses
	\(\Ot(\wm)\) count operations, by iterating over the at most \(\wm\)
	permanent types and using the saved views. After adding any fixed
	selection from the proximity reduction, the result is the selected
	count for each original weight. Sweep the sorted original runs and
	take the highest-profit copies of each weight up to its recovered
	count. This preserves weight, respects all multiplicities, and
	cannot decrease the original profit. Include the artificial slack
	record in the sweep if recovering counts for the instance augmented
	by slack, and discard its selected count afterwards. With a
	polylogarithmic-length chain and fixed rank, the sweep and the
	template visits take \(\Ot(n+\wm)\) time plus the time to trace
	the explicit toggles. The output has one count per input record.
	
	\section{Deep-Anchor Reductions}\label{sec:deep}
	
	A deep anchor gives an affordable neighborhood of the anchor
	target. We exploit it first to certify a restored lattice at every
	supporting price, then to construct a progression reserve on compressed
	counts. The reserve repairs rounding on an affordable fiber. When the
	fiber is thin, its width controls the heights tested by the lift.
	
	\subsection{Cheap cores and arithmetic witnesses}
	\label{sec:restoration-collar}
	
	We use the affordable-depth and supporting-price estimates
	of~\cite[Lemmas~7.1, 7.2 and~8.4 and Section~12.2]{CMZOpt26}
	to certify the actual assignment saved by the allowance merge.
	
	Fix a deep anchor \(v=(t_v,s_v)\), an optimal price \(\pi\) for
	\(f_I(t_v)\), and the positive gap \(g=g_v\). Write \(Q=Q_v\),
	\(P=P_I\), and \(K_*=E/\gamma\), so \(K_*P\subseteq Q\).
	Let \(e_i\) be the endpoint \(e_i(\pi)\) from
	\cref{subsec:phase}, and set \(\sigma_i=1-2e_i\). A working
	coefficient \(x_i\) corresponds to the stored coefficient
	\(x_i^{\rm orig}=e_i+\sigma_i x_i\). This gives the working
	columns \(\bar a_i=\sigma_i a_i\) and normalized costs
	\(\bar c_i=|c_i-\pi\cdot a_i|\).
	At \(\bar t=t-\sum_{i\in I}e_i a_i\), the minimum fractional
	cost in these coordinates is \(F_v(t)\) from
	\eqref{eq:anchor-budget}. Thus the anchor has target \(\bar t_v\)
	and allowance \(a_v=g\), and each assigned query has target
	\(\bar t_q\) and allowance \(a_q\le(1+2\gamma)g\).
	At threshold \(\vartheta\), the cheap pool is
	\(L_\pi(\vartheta)=\{i\in I\mid\bar c_i\le\vartheta\}\).
	For \(A\subseteq I\) and \(b\ge0\), define the compact convex set
	\[
	T_A(b)=\left\{\sum_{i\in A}\bar a_i x_i
	\ \middle|\ x\in[0,1]^A,\ \sum_{i\in A}\bar c_i x_i\le b\right\}.
	\]
	
	\begin{lemma}[affordable depth]\label[lemma]{lem:affordable-depth}
		Let \(P^\sharp\) be centrally symmetric with
		\(P\subseteq P^\sharp\subseteq Q/K_*\). For \(0\leq\zeta\leq\xi\leq1\), let \(A\subseteq I\) contain
		all but at most \(e\) labels of \(L_\pi(\vartheta)\), where
		\(\vartheta>0\). If \(d=(\xi-\zeta)K_*-2\xi g/\vartheta-e\)
		is nonnegative, then
		\[
		\bar t_v+\zeta Q+dP^\sharp \subseteq T_A(2\xi g).
		\]
	\end{lemma}
	
	\begin{proof}
		By \eqref{eq:anchor-budget},
		\(\bar t_v+\xi Q\subseteq T_I(2\xi g)\).
		Discarding labels with cost greater than \(\vartheta\) removes
		total fractional mass at most \(2\xi g/\vartheta\). Discarding
		the additional \(e\) cheap labels removes at most \(e\) more.
		Since all normalized columns belong to \(P^\sharp\), this gives
		\[
		\bar t_v+\xi Q
		\subseteq T_A(2\xi g)+(2\xi g/\vartheta+e)P^\sharp.
		\]
		In each direction, compare support functions and use
		\(h_Q\geq K_*h_{P^\sharp}\). The resulting lower bound on the support
		function of \(T_A(2\xi g)\) is that of \(\bar t_v+\zeta Q+dP^\sharp\),
		which proves the inclusion.
	\end{proof}
	
	With \(k_0,\kappa,E,J\) from \cref{sec:scales}, define
	\[
	\vartheta_0=\frac{g}{128k_0},\quad
	\varrho=\frac{g}{32},\quad
	\Delta=\frac{\vartheta_0}{J+1}.
	\]
	The buffer schedule gives
	\(K_*\geq1024 E\) and \(\gamma+\tau_{\gamma}\xi_*\leq1/512\).
	Applying \cref{lem:affordable-depth} with \(\xi=1/128\),
	\(\zeta=0\), \(P^\sharp=P\), and \(e=0\), the zonotope of
	\(L_\pi(\vartheta_0)\) contains a translate of
	\((K_*/128-2k_0)P\). The same is true for any larger threshold.
	A rank drop after deleting \(b\) labels would give a nonzero
	direction orthogonal to the remaining columns. Each deleted
	label contributes at most \(h_P\) in that direction to the
	original zonotope width, whereas the contained body has width
	\(2(K_*/128-2k_0)h_P\). Since
	\(2(K_*/128-2k_0)>\kappa(J+r)\), the rank condition of
	\cref{lem:core} holds at every threshold used here.
	
	Compute two adjacent thresholds \(\vartheta_-,\vartheta_+\) on
	the grid from \(\vartheta_0\) to \(2\vartheta_0\) at spacing
	\(\Delta\) for which the canonical-core lattice agrees.
	There are \(J+1\) adjacent intervals, and there can be at most
	\(J\) strict lattice changes because these full-rank lattices
	are nested and their indices are bounded by the normalized
	volume. Thus such a pair exists. Write \(\Gamma\) for the
	common lattice, \(A_c=\mathcal R_\kappa(L_\pi(\vartheta_-))\)
	for the trial core, and \(R=\{i\in I\mid a_i\in\Gamma\}\) for
	the restored pool. Then \(\Lambda_R=\Gamma\). By the deletion
	bound in the proof of \cref{lem:core}, at most \(\kappa J\)
	labels of \(L_\pi(\vartheta_+)\) lie outside \(R\). Evaluate the
	core at the \(J+2\) grid points and compare the resulting HNF
	bases. The full scan costs \(\Ot(\kappa m+\kappa^2\Psi(I))\)
	times the polynomial bit factor, including construction of each
	cheap view.
	
	\begin{lemma}[every-price certificate for a restored source]
		\label[lemma]{lem:restored-collar}
		Let \(q=(t_q,s_q)\) be assigned to \(v\), and let \(Y\) be the
		attaining assignment on \(X=I\setminus R\) for a retained
		positive-gap query
		\[
		(t',s')=(t_q-A_XY,\ s_q-c_X^\top Y).
		\]
		Set \(\tau=(\vartheta_-+\vartheta_+)/2\). For any point
		\(y=t'+z_b\) in the depth-\((j+1)\) buffer of this query and
		any supporting price \(\pi'\in\partial f_R(y)\), we have
		\[
		\Gamma=\Lambda_{\mathcal R_\kappa(L_{\pi'}(\tau))},
		\qquad
		A_X(Y-e_X(\pi'))\in EP.
		\]
		The cheap pool \(L_{\pi'}(\tau)\) uses the stored costs and
		columns of \(I\) with the price \(\pi'\).
	\end{lemma}
	
	\begin{proof}
		The normalized cost of \(Y\) is smaller than \(a_q\), since
		the residual has positive gap and every normalized fractional
		cost on \(R\) is nonnegative. By \eqref{eq:anchor-budget},
		\(a_q\le(1+2\gamma)g\).
		Hence fewer than \(128(1+2\gamma)k_0\) labels of \(X\) with
		normalized cost greater than \(\vartheta_0\) change from their
		preferred endpoints. Among the remaining labels, at most
		\(\kappa J\) can change by the plateau property. As each change
		contributes an element of \(P\), we obtain
		\[
		A_X(Y-e_X(\pi))\in(E/2)P .
		\]
		The argument applies to the stored attaining assignment itself, using only its cost and the count of cheap excluded labels.
		
		Extending a residual fractional solution by \(Y\) shows that
		its paired-displacement body is contained in \(Q_q\).
		Thus \(z_b\in \tau_{\gamma}\xi_{j+1}Q\). Set
		\(\zeta=\gamma+\tau_{\gamma}\xi_*\leq1/512\). Then
		\[
		y-\sum_{i\in R}e_i a_i
		\in \bar t_v+\zeta Q+(E/2)P .
		\]
		Let \(q_{\rm col}=16(J+1)k_0\). Use
		\cref{lem:affordable-depth} with \(\xi=1/64\),
		\(\vartheta=\vartheta_0\), \(P^\sharp=P\), \(A=R\), and \(e=\kappa J\).
		Its radius is \((1/64-\zeta)K_*-4k_0-\kappa J\),
		at least \(E/2+q_{\rm col}\) by \(K_*\geq1024E\).
		Consequently all points \(y\pm q_{\rm col}a_i\) have
		normalized fractional cost on \(R\) at most \(\varrho\),
		including for columns \(i\) outside \(R\).
		The collar contains a neighborhood of \(y\) because the
		columns of \(I\) span the ambient space.
		
		For a supporting price \(\pi'\) at \(y\), the supporting
		inequalities for the two collar points give
		\[
		\left|(\pi'-\pi)\cdot a_i\right|
		\leq\frac{\varrho}{q_{\rm col}}=\frac{\Delta}{4}
		\qquad (i\in I).
		\]
		It follows that
		\(L_\pi(\vartheta_-)\subseteq L_{\pi'}(\tau)
		\subseteq L_\pi(\vartheta_+)\).
		Monotonicity of the canonical core proves the lattice identity.
		For labels of \(X\) with \(\bar c_i>\vartheta_+\), the sign of
		\(c_i-\pi'\cdot a_i\) is the same as at \(\pi\), so their
		preferred endpoints cannot switch. At most \(\kappa J\)
		excluded labels remain. Thus
		\(A_X(Y-e_X(\pi'))\in(E/2+\kappa J)P\subseteq EP\), as claimed.
	\end{proof}
	
	\subsection{Reserve construction on compressed types}
	\label{sec:compressed-reserve}
	
	The progression algorithm acts on multiplicities. It supplies
	an integer coordinate for each type, a linear readout for the old
	columns, and a small reserve whose witnesses can be materialized.
	
	\begin{lemma}[compressed progression and reserve]\label[lemma]{lem:count-gap}
		Let \(I\) be a current pool of rank \(r\ge 1\), with \(m\) permanent
		types, symmetric hull \(P_I\), and normalized volume \(\Psi(I)\).
		Let \(A_c\subseteq I\) be a full-rank \(\kappa\)-resilient cheap core, with
		\(\kappa=3k_0\). At the anchor price write
		\(\bar a_i=\sigma_i a_i\) for the working columns and
		\(\bar c_i=|c_i-\pi\cdot a_i|\) for the normalized costs. Suppose
		\(\bar c_i\le g_v/(64k_0)\) for \(i\in A_c\), and take \(k_0\) to
		bound the outlier and reserve counts below, with \(k_0\ge\nu\).
		Fix \(\theta>0\) in \(\nu=\lceil\wm^\theta\rceil\) and a rank bound.
		The scalar encoding below has range
		\(L=O_r\bigl(\nu^{r-1}\Psi(I)\bigr)\). The following operations take time
		\[
		\Ot\!\left(m\nu+\nu^2 L\right)\operatorname{poly}(B).
		\]
		One can partition \(A_c\) into \(I_{\mathrm{out}}, I_{\mathrm{res}}\),
		and \(I_{\mathrm{rem}}\), with \(|I_{\mathrm{out}}|,|I_{\mathrm{res}}|
		\le k_0\). For the nonoutliers the algorithm gives a symmetric
		integer box \(\mathsf B\subseteq\R^k\), integral vectors \(u_i\in
		\mathsf B\), and an integral linear map \(\phi\) from the progression
		lattice to the current ambient lattice with \(\phi u_i=\bar a_i\).
		Here \(k\) is bounded by a constant. The algorithm returns the
		actual reserve labels and a center \(z_0\) such that
		\[
		z_0+\bigl(c_{\delta_{\rm gap}}\nu\mathsf B\cap\Z^k\bigr)
		\subseteq\left\{\sum_{i\in S}u_i\ \middle|\ S\subseteq I_{\rm res}\right\}.
		\]
		This filling guarantee certifies attainability. The terminal
		correction is constructed in the old coordinates by
		\cref{lem:reserve-recovery}.
		The reserve's total normalized cost is at most \(g_v/64\).
		The remainder spans \(\R^k\) in progression coordinates even
		after \(\nu\) labeled deletions. The readout satisfies
		\(\phi(\Z^k)=\Lambda_{A_c}\).
	\end{lemma}
	
	\begin{proof}
		Apply \cref{lem:rebox} to obtain integral half-widths \(q_j\ge1\)
		with \(\prod_{j=1}^{r}q_j=O_r(\Psi(I))\). The construction scans only the
		current types and returns the coordinate map.
		
		Choose a sufficiently large constant \(C_{\delta_{\rm gap}}\)
		and use radices \(b_i=2\lceil C_{\delta_{\rm gap}}\nu q_i\rceil+1\).
		The encoding of an old vector is
		\[
		\operatorname{enc}(a)=\sum_{j=1}^{r} a_j\prod_{i=1}^{j-1}b_i.
		\]
		Successive reduction modulo the radices proves injectivity on the
		integer box of half-widths \(\lceil C_{\delta_{\rm gap}}\nu q_i\rceil\), including
		all sums of \(C_{\delta_{\rm gap}}\nu\) signed columns. In contrast, a single
		column has
		\[
		|\operatorname{enc}(a)|\le
		\sum_{j=1}^{r}q_j\prod_{i=1}^{j-1}b_i
		=O_r\!\left(\nu^{r-1}\prod_{j=1}^{r}q_j\right).
		\]
		This is the scalar bound \(L\). Constants here can depend on \(\delta_{\rm gap}\).
		Because \(\Psi(I)\le \Psi(I_{\rm root})=O(\wm)\), we can first fix \(\delta_{\rm gap}>0\)
		small enough that \(\nu\ge L^{\delta_{\rm gap}}\) holds for all sufficiently
		large \(\wm\) and all ranks under consideration. Increasing the
		constant in the radices only changes this finite cutoff.
		
		For each working label, let
		\(\tau_i=\operatorname{sgn}(\operatorname{enc}(\bar a_i))\) and use the positive
		encoded value \(\operatorname{enc}(\tau_i\bar a_i)\). The columns are nonzero and
		injectivity on the signed box gives nonzero encodings. For a
		permanent type \(a\), the value is \(|\operatorname{enc}(a)|\), independent of the
		label. Distinct permanent types have distinct positive values
		since equality would give old columns equal up to sign, and the
		linear readout of original weight distinguishes their positive
		weights. The multiplicity of each value is counted from the
		current cheap view. Apply the count version of the common
		progression theorem~\cite[Theorem~3.1 and Corollary~8.8]{CMZFeas26}
		with parameter \(h=\nu\). It returns outlier counts
		\(O_{\delta_{\rm gap}}(\nu\log^2(\nu L+2))\), reserve counts \(O_{\delta_{\rm gap}}(\nu)\),
		one coordinate \(u_a^+\) per live encoded value, and a reserve
		center with the stated filling guarantee. The theorem's robustness
		condition is an output guarantee on the remainder. No restriction
		on the input multiplicities is needed. A single global choice
		\(k_0=C\nu\log^2(H_{\mathrm{amb}}^*+2)\), rounded up, works for
		\(H_{\mathrm{amb}}^*\ge \nu L\) throughout the recursion and for a
		sufficiently large fixed \(C\). The all-outlier outcome is
		impossible because a full-rank \(\kappa\)-resilient pool has more
		than \(\kappa\) labels. We use the deterministic form of the
		construction.
		
		The filling property proves the existence of an integral
		readout \(\phi\) by~\cite[Lemma~3.4]{CMZFeas26}. To see the
		distinction between this proof and its implementation, choose
		an arbitrary reserve representation for each point of the filled
		box, only for the analysis. Equal scalar sums have equal old-vector
		sums by injectivity. Opposite corners of unit squares and consecutive
		collinear triples show that the old-vector sum is affine, with
		integral linear part. Comparing points differing by \(u_a^+\)
		gives \(\phi u_a^+=\tau_i\bar a_i\). These comparisons involve
		only \(O_{\delta_{\rm gap}}(\nu)\) signed old columns, so the fixed
		radices suffice.
		
		To compute the map, choose \(k\) independent actual type vectors
		from the remainder, which spans \(\R^k\), and solve
		\(\phi u_i=\bar a_i\) on this basis. The existence argument
		identifies the unique solution and proves that it is integral.
		Check its identities on all live types in one scan. No reserve
		representation is queried to obtain \(\phi\).
		
		Retrieve actual disjoint reserve and outlier ordinals from the
		cheap views using the returned counts. These lists are accessible
		by actual label IDs, even though the
		filling property need not supply representations of specified
		progression points. For the working
		orientation use \(u_i=\tau_i u_a^+\). Undoing the positive encoding
		orientation complements the reserve choices on each label with
		\(\tau_i=-1\) and subtracts the sum of their positive progression
		vectors from the reserve center. This preserves the symmetric
		filling inclusion. Every retrieved reserve label is cheap, so
		the total normalized reserve cost is at most
		\(k_0g_v/(64k_0)=g_v/64\).
		Deleting outliers and reserve together removes at most
		\(2k_0\le\kappa\) labels, and hence the remainder's old lattice is
		\(\Lambda_{A_c}\), which is contained in \(\phi(\Z^k)\).
		For the reverse inclusion, use the reserve filling property at
		\(z_0\) and \(z_0+e_j\) for each coordinate unit vector \(e_j\).
		These points are in the filled box for sufficiently large \(\nu\),
		since the remainder spans and each box half-width is a positive
		integer. The difference of their old-coordinate reserve sums is
		\(\phi(e_j)\in\Lambda_{A_c}\). This proves the equality without
		needing to find the two reserve subsets. Label retrieval and the old-vector map cost
		\(\Ot(m+k_0)\operatorname{poly}(B)\), within the stated bound.
	\end{proof}
	
	\begin{lemma}[recovering a reserve subset in the old coordinates]
		\label[lemma]{lem:reserve-recovery}
		Let \(R\) be a reserve of \(s\) actual labels in a rank-\(r\) pool,
		with normalized volume \(\Psi\) and its previously computed
		reboxing coordinates. Given an attainable integer
		old-coordinate target \(d\), one can find a subset of \(R\)
		whose columns sum to \(d\) in
		\[
		\Ot\bigl((s+1)^{r+1}\Psi\bigr)\operatorname{poly}(B)
		\]
		time. At a deep anchor, \(s=\Ot(\nu)\) and
		\(\mu_v=\Omega_r(\nu^r\Psi)\), so this costs
		\(\Ot(\nu\mu_v)\operatorname{poly}(B)\).
	\end{lemma}
	
	\begin{proof}
		The empty reserve only attains zero. Otherwise use the integral
		coordinates of \cref{lem:rebox}, with half-widths \(q_j\ge1\)
		and \(\prod_{j=1}^r q_j=O_r(\Psi)\). For a vector \(a\) in these
		coordinates, put
		\[
		\operatorname{enc}_R(a)=\sum_{j=1}^r a_j
		\prod_{\ell=1}^{j-1}(2sq_\ell+1).
		\]
		Set \(S=\sum_{i\in R}|\operatorname{enc}_R(a_i)|\).
		Every reserve sum lies in the box of half-widths \(sq_j\), and
		successive reduction modulo these radices proves injectivity on
		that box. A single column has encoded magnitude
		\(O_r(s^{r-1}\Psi)\), so \(S=O_r(s^r\Psi)\).
		Run ordinary 0--1 reachability DP over \([-S,S]\), processing
		one actual labeled copy at a time. Each update reads the preceding
		layer. When a sum first becomes reachable, save its predecessor
		sum and the current label. Following these first-discovery pointers
		uses strictly decreasing label positions, so no label is repeated.
		Check that \(d\) lies in the stated coordinate box before lookup.
		The attainable target is found at \(\operatorname{enc}_R(d)\),
		and injectivity converts the scalar witness to the exact old-vector
		sum. The work is \(O(s(S+1))\operatorname{poly}(B)\).
		All reserve costs are nonnegative and their total is already paid,
		so any subset attaining \(d\) obeys the same reserve-cost bound.
	\end{proof}

	A lift appends the type-dependent coordinate of
	\cref{lem:stable-type} after undoing the anchor orientation.
	The dual score for choosing exceptional labels is
	\[
	\omega_i
	=\sigma_i\bigl(\operatorname{sgn}(\operatorname{enc}(a_i))\psi(u_{a_i}^{+})
	-\beta\cdot a_i+\tau(c_i-\pi\cdot a_i)\bigr).
	\]
	Within a type, \(|\omega_i|\) is the absolute value of an affine
	function of the ordered costs. The endpoint iterators of
	\cref{sec:pool} find the \(\nu\) largest absolute scores and the
	endpoint profiles in \(\Ot(m+\nu)\) time.
	
	\begin{lemma}[compressed structural work at a deep anchor]\label[lemma]{lem:deep-charge}
		At a deep anchor of paid mass \(\mu_v=\Omega_r(\nu^r\Psi(I))\),
		the core searches, compressed reserve construction, and the
		associated grouped geometry, exceptional-label extraction, reboxing,
		and a possible terminal reserve recovery cost \(\Ot(\nu\mu_v)\operatorname{poly}(B)\).
		This charge covers the structural work before any trial queries
		are filtered.
	\end{lemma}
	
	\begin{proof}
		Counting a polylogarithmic number of canonical-core queries, the
		threshold searches cost \(\Ot(\nu\mu_v)\operatorname{poly}(B)\)
		by \cref{lem:core-charge}. The reserve construction costs
		\(\Ot(m\nu+\nu^2 L)\operatorname{poly}(B)\), and the grouped
		geometry, exceptional labels, and reboxing are within this
		bound. The scalar range satisfies \(L=O_D(\nu^{r-1}\Psi(I))\)
		by \cref{lem:count-gap}, and \(m=O_r(\Psi(I))\). Thus the
		structural charge is \(\Ot(\nu\mu_v)\operatorname{poly}(B)\).
		A terminal call to \cref{lem:reserve-recovery} also costs
		\(\Ot(\nu\mu_v)\operatorname{poly}(B)\). Each anchor runs
		one trial and at most one restoration or high-center procedure.
	\end{proof}
	
	\subsection{Deep trials and the lifted strip}\label{sec:deep-lift}
	
	Use the anchor normalization and the canonical-core plateau
	\([\vartheta_-,\vartheta_+]\) from \cref{sec:restoration-collar},
	with core \(A_c\) at \(\vartheta_-\) and lattice
	\(\Gamma=\Lambda_{A_c}\). Set \(P_{\rm ref}=S_v/K_*\), so
	\(P\subseteq P_{\rm ref}\subseteq Q/K_*\).
	The construction of \cref{lem:count-gap} partitions \(A_c\)
	into outliers \(I_{\rm out}\), a reserve \(I_{\rm res}\),
	and a remainder \(I_{\rm rem}\). The first two have
	at most \(k_0\) labels each. Write \(I_\circ=I_{\rm res}\cup I_{\rm rem}\),
	\(X_\circ=I\setminus I_\circ\), and \(R=I\cap\Gamma\).
	The remainder spans \(\R^k\) in the progression coordinates
	\(u_i\) even after \(\nu\) deletions, and its columns satisfy
	\(\phi u_i=\bar a_i\). The reserve subset sums contain
	\(z_0+(k\mathsf B\cap\Z^k)\). A retained trial entry is \emph{high}
	if its normalized allowance is at least \(g/4\).
	
	We use \cref{lem:affordable-depth} with \(P^\sharp=P_{\rm ref}\).
	Prescribing an endpoint of a deleted label consumes one additional
	unit of radius and translates by that endpoint. This follows from
	the same support-function argument, since \([0,\bar a_i]\) lies in
	either endpoint plus \(P_{\rm ref}\). The prescribed endpoint costs
	are accounted for separately.
	
	\begin{lemma}[trial rejection and proper restoration]\label[lemma]{lem:trial-restoration}
		An empty trial batch certifies that the anchor's assigned queries
		have no integral witness. If the trial has a survivor but no high
		entry, the trial lattice \(\Gamma\) is a proper sublattice.
		Restoring \(R=I\cap\Gamma\) and recomputing the transfer on
		\(I\setminus R\) preserves the existence of an integral witness
		and produces no high restored entry.
	\end{lemma}
	
	\begin{proof}
		The core construction omits at most \(\kappa J\) labels of cost at
		most \(\vartheta_-\). In particular \(X_\circ\) has at most
		\(\kappa J+k_0\) labels of cost at most \(\vartheta_0\).
		An assignment \(Y\) on \(X_\circ\) of normalized cost \(d(Y)<a_q\)
		uses fewer than \(128(1+2\gamma)k_0\) labels costing more than
		\(\vartheta_0\). Every subset of its selected labels therefore
		has weight in \((E/2)P\), since
		\[
		128(1+2\gamma)k_0+\kappa J+k_0\le (130+3J)k_0=E/2 .
		\]
		The same bound applies to \(I\setminus R\), with \(\kappa J\) cheap
		excluded labels. This estimate concerns any actual assignment of
		cost below the allowance, including the assignment attained by
		a cheaper table entry. It does not rely on how the entry was found.
		For an explicit physical crop, use \(ER_P\), where
		\(P\subseteq R_P\subseteq 2rP\) is the parallelotope constructed
		in \cref{lem:rebox}. It contains all required displacements,
		including the final lift removals below. The signed table uses
		a separate integer box covering this crop. The crop precedes
		source construction.
		
		The lattice of the trial pool is \(\Gamma\). In fact the core is
		\(\kappa\)-resilient and the outliers and reserve together contain at
		most \(2k_0\le\kappa\) labels, so \(\Lambda_{I_{\rm rem}}=\Gamma\)
		and \(\phi(\Z^k)=\Gamma\). Every integral completion yields a
		trial residual at the horizontal weight of a required assignment.
		Replacement by an attaining assignment of no larger cost preserves
		the residual weight and its completion in the trial pool. The trial
		residual passes the lattice, feasibility, and positive-gap filters.
		Thus an empty trial correctly rejects this anchor.
		
		Suppose the trial has survivors and none is high. If
		\(\Gamma=\Lambda_I\), the anchor itself with zero
		working deviations on \(X_\circ\) yields a compatible trial residual
		at \(\bar t_v\) of allowance \(g\). Apply
		\cref{lem:affordable-depth} at \(\xi=1/128\), \(\zeta=0\),
		\(\vartheta=\vartheta_-\), eroding by \(\kappa J+k_0\) labels, to obtain
		a fractional trial witness of cost at most \(\varrho/2\).
		The dominating entry survives and is high, a contradiction.
		In the proper restoration branch the table is recomputed on
		\(I\setminus R\). An integral solution splits into an affordable
		assignment on this removal pool and a completion on \(R\), and the
		same bound on partial sums and the dominance guarantee prove
		completeness. Conversely a restored solution extends through its
		attaining assignment on the disjoint removal pool. Only its source
		lattice is carried from the trial. In fact a high
		restored residual would also produce a high trial entry. Extend its
		assignment by zero working choices on \(R\setminus I_\circ\).
		Compatibility and allowance are preserved. The working target
		lies in \(\bar t_v+\gamma Q+(E/2)P\). Apply
		\cref{lem:affordable-depth} with \(A=I_\circ\), \(P^\sharp=P\),
		\(\xi=1/128\), \(\zeta=\gamma\), and \(\vartheta=\vartheta_-\).
		At most \(\kappa J+k_0\) cheap labels are missing from \(I_\circ\).
		Since \(2\xi g/\vartheta_-\le2k_0\), the radius \(d\) in
		\cref{lem:affordable-depth} satisfies
		\[
		d\ge (1/128-\gamma)K_*-2k_0-(\kappa J+k_0)
		\ \ge\ 6E-(3J+3)k_0\ >\ E/2 .
		\]
		The working target thus has a trial fractional witness of cost
		at most \(2\xi g=\varrho/2\). The trial table dominates the
		extended assignment at that target, so the entry survives with
		allowance at least \(g/4\), a contradiction.
		
	\end{proof}
	
	We turn to the remaining outcome, where the trial has a high
	entry. The reserve first gives a fractional completion with room
	for prescribed endpoints. We can then use the image fiber either
	to round an integral point or to obtain a direction for the lift.
	Let \(y_0\) be the working residual target of the chosen high
	entry, with allowance at least \(g/4\). Its attaining assignment
	\(Y_0\) satisfies \(d(Y_0)<a_{q_0}\), because the entry passed the
	positive-gap test. Hence
	\(y_0\in\bar t_v+\gamma Q+(E/2)P\).
	Use \cref{lem:affordable-depth} for the cheap set at \(\vartheta_-\)
	and erase the at most \(\kappa J+k_0\) cheap labels missing from
	\(I_\circ\). Set \(\mu=(1/128-\gamma)K_*-2k_0-2E\).
	With \(\xi=1/128\), \(\zeta=\gamma\), and the center offset just
	established, the affordable-depth bound gives
	\[
	y_0+\mu P_{\rm ref}\subseteq T_{I_\circ}(\varrho/2).
	\]
	Using the reserve center \(z_0\), define the budgeted image
	\(\mathcal J_b\) and the affordable fiber \(\mathcal K\) by
	\[
	\mathcal J_b=z_0+\Bigl\{\sum_{i\in I_{\rm rem}}u_i x_i
	\mathrel{\Bigm|}x\in[0,1]^{I_{\rm rem}},\
	\sum_{i\in I_{\rm rem}}\bar c_i x_i\le b\Bigr\},
	\qquad \mathcal K=\mathcal J_{\varrho}\cap\phi^{-1}(y_0).
	\]
	\begin{lemma}[rounding at a high trial center]\label[lemma]{lem:high-trial-center}
		At a high trial center \(y_0\), fixing the reserve to a witness
		for \(z_0\) and prescribing endpoints on any at most \(\nu\)
		remainder labels admits a fractional completion with remainder cost
		at most \(\varrho\). With no prescribed remainder endpoints,
		\(y_0+(\mu-k_0)P_{\rm ref}\subseteq\phi(\mathcal J_{\varrho/2})\).
		An integral point in \(\mathcal K\) yields an integral witness for
		the assigned parent query.
	\end{lemma}
	
	\begin{proof}
		Fixing the reserve to a witness for \(z_0\) consumes at most \(k_0\)
		units of radius, and prescribing at most \(\nu\) further endpoints
		consumes at most \(\nu\) units. The chosen \(\mu\) is at least
		\(k_0+\nu\) by the scale schedule. Every core label has cost at
		most \(2\vartheta_0\le\varrho/(2\nu)\), and the entire reserve
		costs at most \(g/64\). Starting from the affordable body at budget
		\(\varrho/2\) proves both fractional claims, because the forced
		remainder endpoints cost at most \(\varrho/2\).
		
		Compatibility supplies the ambient affine lattice
		\(\phi^{-1}(y_0)\cap\Z^k\). If \(k=r\), the nonempty fiber is
		the unique affine point, which is integral. For \(k>r\), compute
		Smith coordinates of the affine lattice and use the
		optimization-and-membership form of point-or-thin
		from~\cite[Lemma~10.1]{CMZOptRev26}.
		For a membership query at \(z\), minimize remainder cost at
		\(U_{I_{\rm rem}}x=z-z_0\) and compare with the budget.
		For optimization, \cref{lem:affordable} returns a vertex with
		encoding length \(\operatorname{poly}(B)\) independent of the
		objective. The fixed Smith conversion has polynomial bit length
		and preserves this bound. Thus the polynomially many oracle
		calls take \(\Ot(m+1)\operatorname{poly}(B)\) time on the
		compressed remainder, including the coordinate conversions.
		If the procedure returns a point \(z\in\mathcal K\cap\Z^k\),
		minimize remainder cost at image \(z-z_0\) and round down the at most
		\(k\) fractional labels of a basic optimum. The lost image is
		integral and in \(k\mathsf B\). The filling inclusion shows
		that a reserve subset attains the old-coordinate deficit
		\[
		d=y_0-\sum_{i\in I_{\rm rem}}\bar a_i\lfloor x_i\rfloor.
		\]
		Use \cref{lem:reserve-recovery} to find such a subset of the
		actual reserve. This branch terminates in the old coordinates,
		so it need not recover a prescribed progression representation.
		The old target is exactly \(y_0\), and the total cost is at most
		\(\varrho+g/64<g/4\). The remainder,
		reserve, and removed trial labels are disjoint, so this returns
		an actual solution to the chosen assigned query.
	\end{proof}
	
	If the point-or-thin procedure returns a thin direction,
	we use its optimality multipliers to bound the lifted columns
	after removing a small exceptional set.
	
	\begin{lemma}[lifted columns after exceptional deletions]\label[lemma]{lem:thin-direction}
		Suppose \(\psi\) is integral, nonzero on \(\ker\phi\), and has
		width at most \(F\) on \(\mathcal K\). There are optimal dual
		multipliers \(\beta,\tau\) for minimizing \(\psi\) on \(\mathcal K\).
		Delete the \(\nu\) labels of \(I_{\rm rem}\) with largest scores
		\(\left|\psi(u_i)-\beta\cdot\bar a_i+\tau\bar c_i\right|\).
		Every remaining label \(i\in I'\) satisfies
		\[
		|\psi(u_i)-\beta\cdot\bar a_i|\le 2F/\nu .
		\]
		The lifted columns on \(I'\) span rank \(r+1\).
	\end{lemma}
	
	\begin{proof}
		Here \(F\) depends only on the fixed dimension. The affine-lattice
		functional returned by the procedure extends to an integral
		functional on \(\Z^k\) because \(\ker\phi\cap\Z^k\) is saturated.
		Take the optimal equality and budget multipliers \(\beta\) and
		\(\tau\ge0\) in
		the convention
		\[
		\omega_i=\psi(u_i)-\beta\cdot\bar a_i+\tau\bar c_i,\qquad
		\epsilon_i=[\omega_i<0],\qquad
		\alpha_0=\psi(z_0)-\beta\cdot\phi(z_0)+
		\sum_{i\in I_{\rm rem}}\min(0,\omega_i).
		\]
		The dual optimum is \(m_\psi=\beta\cdot y_0-\tau\varrho+\alpha_0\).
		For any representation of \(z\in\mathcal K\) with remainder vector \(x\),
		\[
		\psi(z)-m_\psi=
		\sum_{i\in I_{\rm rem}}|\omega_i||x_i-\epsilon_i|
		+\tau\Bigl(\varrho-\sum_{i\in I_{\rm rem}}\bar c_i x_i\Bigr)
		\ \le F .
		\]
		Both terms are nonnegative. At a half-budget point this proves
		\(\tau\le2F/\varrho\). Force opposite endpoints on the \(\nu\)
		labels of largest \(|\omega_i|\) and use the affordable completion
		proved above. The sum of their scores is at most \(F\).
		Let \(I_{\rm exc}\) consist of those labels and
		\(I'=I_{\rm rem}\setminus I_{\rm exc}\). Its columns satisfy
		\[
		|\psi(u_i)-\beta\cdot\bar a_i|
		\le|\omega_i|+\tau\bar c_i\le 2F/\nu\qquad (i\in I').
		\]
		They generate a rank-\((r+1)\) lifted lattice because the remaining
		\(u_i\) span \(\R^k\) and \(\psi\) is independent of the row space
		of \(\phi\). The multipliers and aggregate endpoint scores are
		computable by the same fixed-dimensional dual as in
		Lemma~\ref{lem:affordable}. The exceptional scores require only the
		compressed endpoint iterators of Section~\ref{sec:pool}.
	\end{proof}
	
	The bound on individual columns controls the normalized volume.
	To limit the number of candidates, we also need a common strip
	for the heights of all witnesses the lift must preserve.
	
	\begin{lemma}[common strip for a witnessed lift]\label[lemma]{lem:witness-strip}
		Use a fresh removal table on \(I\setminus I'\) with the reserve
		choices free. There is a computable intercept \(b_*\) such that
		every integral completion of an assigned query still satisfies the
		following bound after its removal assignment is replaced by a
		dominating table witness. The bound also holds for every point of
		the corresponding child buffer at scales \(0\le\xi\le\xi_*\).
		Writing \((h,z)\) for the lifted coordinates, it is
		\[
		b_*-256F\le z-\beta\cdot h\le b_*+255F .
		\]
		It suffices to enumerate at most \(511F+1\) integer heights
		per old residual target before LP filtering.
	\end{lemma}
	
	\begin{proof}
		Form one fresh old-coordinate table on
		\(I\setminus I'=X_\circ\cup I_{\rm res}\cup I_{\rm exc}\), a
		disjoint union, and then lift its residuals. The reserve is free
		in this table. Attainability of the known center \(z_0\) supplies
		the comparison assignment in the affordable fiber. The lift
		table computes its own reserve choices with the other removals.
		Since the last two removal pools have
		at most \(2k_0\) labels, every partial sum of an affordable final
		assignment belongs to \((E/2+2k_0)P\). An integral completion
		of the parent query on \(I'\) is unchanged if a cheaper table
		witness of the same old weight replaces its removal assignment.
		The replacement cannot change the appended coordinate of the
		completion, which depends only on the labels in \(I'\).
		
		For each residual query kept after filtering, its actual attaining
		assignment \(Y\) has cost \(d(Y)<a_q\).
		Let \(x\) represent any point in this residual's buffer at scale
		\(0\le\xi\le\xi_*\). A paired witness has cost at most twice the
		residual allowance \(a_q-d(Y)\). Mixing it with a center witness
		in proportion \(\xi\) gives a fractional completion satisfying
		\[
		d(Y)+\bar c_{I'}^\top x
		\le d(Y)+(1+\xi)(a_q-d(Y))
		\le (1+\xi_*)(1+2\gamma)g<2g .
		\]
		For a true integral completion, the same inequality holds at
		\(\xi=0\). In either case combine \(x\) with the exceptional
		labels of the actual removal assignment and use \(z_0\) as the
		reserve translation in \(\mathcal J_b\). Set
		\(R_{\rm p}=(1+2\gamma)K_*+E+2k_0\). The resulting point has
		old-coordinate target \(y\in y_0+R_{\rm p}P_{\rm ref}\).
		Indeed the assigned query targets differ by a vector in
		\(2\gamma S_v\), and the projected child-buffer displacement
		belongs to \(S_v\). The latter follows by extending its two
		defining LP witnesses with the same removal assignment,
		using \(Q_q\subseteq \tau_{\gamma}Q_v\) and \(\xi_*\tau_{\gamma}<1\), and projecting
		the child hull into \(P\).
		The two trial-complement assignments each have drift in
		\((E/2)P\) by the count bound, including the restriction of
		the actual attaining \(Y\) to \(X_\circ\). Replacing the actual
		reserve part by the fixed reserve witness moves the weight by
		at most \(2k_0P\). This proves the claimed radius without
		requiring the actual reserve to equal the provisional one.
		The cost on \(I_{\rm rem}\) is at most \(2g\).
		
		For a feasible pair \((y,b)\), define
		\[
		G(y,b)=\max_{z\in\mathcal J_b,\;\phi z=y}\bigl(\psi(z)\bigr)-\beta\cdot y+\tau b-\alpha_0.
		\]
		The function is concave by mixing representations and is
		nonnegative by the dual decomposition above. Also
		\(G(y_0,\varrho)\le F\). Set \(U_g=4g\). The schedule
		\(K_*=E/\gamma\), \(\gamma\le 2^{-10}/10000\), ensures
		\(\mu-k_0\ge\max(\nu,R_{\rm p}/254)\) by substitution.
		Thus \(y'=y_0-(y-y_0)/254\) is feasible at budget \(\varrho/2\).
		Since \((y_0,\varrho)=(y,U_g)/255+254(y',\varrho/2)/255\),
		concavity gives \(G(y,U_g)\le255F\).
		The dual decomposition therefore bounds the deviation score of
		every such representation, and of its restriction to \(I'\),
		by \(255F\). These are the support-function and reflection
		arguments of~\cite[Sections~9.4--9.5 and~12.2]{CMZOpt26}.
		
		Write \(h=\sum_{i\in I'}\bar a_i x_i\) and
		\(z=\sum_{i\in I'}\psi(u_i)x_i\). Define the intercept
		\(b_*=\sum_{i\in I'}\min(0,\omega_i)\). Then
		\[
		z-\beta\cdot h=b_*+
		\sum_{i\in I'}|\omega_i||x_i-\epsilon_i|
		-\tau\sum_{i\in I'}\bar c_i x_i .
		\]
		Since \(\tau U_g\le256F\), the common strip is
		\[
		b_*-256F\le z-\beta\cdot h\le b_*+255F.
		\]
		For each old residual \(h\), enumerate its at most \(511F+1\)
		integer heights in this interval before LP filtering. The
		proof applies to every true completion and every point of the
		actual buffer, even though the table may choose a new cheaper
		attaining assignment. It imposes no feasibility requirement
		on candidates that lie inside the enumerated strip.
	\end{proof}
	
	\begin{lemma}[normalized volume of the lifted source]\label[lemma]{lem:lift-volume}
		After lifting to \(I'\), the old-coordinate readout and the
		stored costs of the retained labels remain available. The normalized
		volume satisfies
		\(\Psi(I')\le (2(r+1)F/\nu)\Psi(I)\).
		The horizontal folding bound also pays for the full query and
		navigation halos before filtering.
	\end{lemma}
	
	\begin{proof}
		Store column \((a_i,\sigma_i\psi(u_i))\) in the original
		orientation. The stored target is the working target plus
		\(\sum_{i\in I'}e_i(a_i,\sigma_i\psi(u_i))\), a fixed integral
		translation. If its old part is \(h_{\rm orig}\) and the
		working allowance is \(s'\), its allowance in the stored orientation is
		\(s'+\pi\cdot h_{\rm orig}+\sum_{i\in I'}e_i(c_i-\pi\cdot a_i)\).
		Use \cref{lem:rebox} to give the lifted lattice integral
		coordinates. Store its basis matrix to recover the old
		coordinates and the appended height, and compose with the
		old-coordinate readout to recover the input weight. All
		coordinate maps and translations have polynomial bit length.
		
		The shear \((h,z)\mapsto(h,z-\beta\cdot h)\) has determinant one.
		The sheared child hull lies in \(P\times[-2F/\nu,2F/\nu]\).
		The lifted lattice projects into the parent lattice and its
		vertical kernel is generated by an integral vector of length
		at least one. Its covolume is the product of the projected
		covolume and this kernel length, hence at least the parent
		covolume. Comparing the normalized volumes in ranks \(r\)
		and \(r+1\) gives
		\[
		\Psi(I')\le \frac{2(r+1)F}{\nu}\Psi(I).
		\]
		The horizontal folding argument in \cref{subsec:source-halo}
		pays for the query and navigation halos using the same shear
		and covolume product.
	\end{proof}
	
	\section{Correctness and Running Time}\label{sec:runtime}
	
	The deep reductions complete the branch analysis of
	\cref{alg:process-state}. A trial tests a temporary pool
	\(I_\circ\). A proper restoration uses the current parent \(I\)
	to build a fresh table on \(I\setminus R\), where
	\(R=I\cap\Gamma\) is the pool entered on the restored lattice.
	The proof below combines these witness-preserving transitions
	with the global source bounds.

	\begin{theorem}[quantitative transfer]\label[theorem]{thm:transfer}
		Fix a witnessed integer \(\MinConv\) exponent \(1<\rho<2\), a
		rational \(0<\theta<1\), an integer \(k_{\rm scan}\ge2\), and
		constants \(\lambda,a>0\). Define
		\[
		\begin{aligned}
			\Gamma_q(\theta)&=\frac{1+2\theta}{q}+(q-4)\theta,\\
			E_\rho(\theta,k)&=\max\bigl(
			\{\rho(1+\theta),1+2\theta,2-(k-1)\theta\}\\[-2pt]
			&\hspace{6.1em}{}\cup\{2-\Gamma_q(\theta)\mid q=4,\ldots,k+1\}\bigr).
		\end{aligned}
		\]
		With \(\nu=\lceil\wm^\theta\rceil\) and scan threshold
		\(k_{\rm scan}\), \cref{alg:bounded} returns an optimal count
		vector with probability at least \(1-(n+\wm+2)^{-a}\) within
		\(\Ot(n+\wm^{E_\rho(\theta,k_{\rm scan})+\lambda})
		\operatorname{poly}(B)\) bit operations.
	\end{theorem}
	
	\begin{proof}
		The proximity kernel and slack preserve the optimum up to the fixed
		profit offset. Perturbation preserves the ordering of distinct
		integral costs. The root lists give
		\(D_0=\sum_{i\in I_0}|c_i|\), an \(O(B)\)-bit bound on the
		magnitude of any perturbed cost. Selecting only slack makes the
		equality instance feasible. A decision call with strict allowance
		\(z+1/2\) tests whether the minimum integer cost is at most \(z\).
		Binary search on \([-D_0,D_0]\) uses \(O(B)\) decision calls,
		including recovery at the least accepted threshold.
		
		On the joint success event, the branch constructions in
		\cref{sec:global,sec:deep} preserve a feasible residual or return
		a terminal witness whenever the batch contains a yes query.
		Lattice membership and a fractional optimum strictly below the
		allowance are necessary for each integral witness. At a repeated
		target, the largest allowance covers the solutions of all smaller
		ones.
		
		Each retained entry stores its attaining source and assignment
		on the removed labels. A child witness extends to a witness for
		that source by adding this disjoint assignment. Induction on the
		finite recursion gives correctness at the root. The traceback
		and sweep of \cref{sec:count-recovery} recover original counts
		from the optimum in perturbed costs.
		
		Take fixed positive rational \(\beta,\eta_{\rm cut}\) with
		\(\beta+\eta_{\rm cut}\le\lambda\) and choose the buffer constants for this
		\(\beta\). For one decision call, put \(X=\nu\wm\). By
		\cref{thm:ledger}, witnessed convolutions cost
		\(\Ot(X^\rho\wm^\beta)\), and the structural work costs
		\(\Ot(n+\nu X\wm^\beta)\), with polynomial bit factors.
		\Cref{lem:geometry-total} accounts for all atlas work and grouped
		source queries. The line and planar costs are within the structural
		bound. The grouped source-query bound contributes exponent
		\(2-(k_{\rm scan}-1)\theta\) and a term of exponent \(1+\theta\),
		before the loss \(\beta\). At a grid level \(\ell\ge2\) with
		\(\ell\theta\le1\), the exponent before \(\beta+\eta_{\rm cut}\) is
		\[
		1+\theta+\left(1-\frac1{\ell+2}\right)(1-\ell\theta)
		=2-\Gamma_{\ell+2}(\theta).
		\]
		At \(\ell\theta>1\) the bound contributes only the base exponent
		\(1+\theta\). All such base terms are within the structural exponent
		\(1+2\theta\). This yields the stated maximum. Rank-zero tests
		are part of the structural work. \Cref{tab:branch-charges} records
		the charge at each exit. Recovery traverses one attaining chain and
		its selected table branches. The labels it reads were paid for at
		materialization, and the final sweep costs \(\Ot(n+\wm)\).
		
		Each allowance is the initial threshold minus the costs of removed
		labels and uses \(O(B)\) bits. Clear each local price denominator
		for its invocation. Oracle and vertex encodings have polynomial
		bit length by \cref{lem:grouped,lem:affordable,lem:grid-atlas}.
		There are constantly many lifts, and \cref{lem:rebox} bounds
		reboxed column coordinates by \(O_D(\wm)\). The recovery maps
		and the progression and lattice data of \cref{sec:deep} also
		have polynomial bit length. The factor \(\operatorname{poly}(B)\)
		covers this arithmetic and binary search.
		
		Use fresh randomness for transfer tables, convolutions, and atlas
		constructions. Split call \(j\)'s failure budget
		\((n+\wm+2)^{-a}/(4(j+1)^2)\) among its randomized operations.
		Cutting trials check their conflict lists and use a logarithmic
		retry cap. The preceding history fixes the next call's input.
		Summing the budgets for the first failure gives the success bound.
		Cap total work at the bound for a successful trace, including
		bit and retry factors. Call indices before the cap have polynomial
		encoding length. On joint success the cap is never reached.
		Reject invalid witnesses and arithmetic edges. The time bound
		holds on every execution.
	\end{proof}
	
	\begin{table}[H]
		\centering\small
		\setlength{\tabcolsep}{6pt}
		\begin{tabularx}{\linewidth}{@{}l >{\raggedright\arraybackslash}X >{\raggedright\arraybackslash}X@{}}
			\toprule
			Branch & Work charged before or at dispatch & Witness or atlas record passed on \\
			\midrule
			Thin & Direction lookup, endpoint branches, transfer tables, physical crop and filtering & Restricted pool and one attaining choice per target \\
			Deep trial & Canonical core, reserve, trial table and all tests & A rejection, a high center, or a request for restoration \\
			Restoration & A new table on the actual removed pool and a new source window & The full restored pool, the attaining removal witness and its source \\
			Lift & Affordable-fiber calls, horizontal table and strip enumeration & A lifted pool with a type-constant extra coordinate \\
			Face & Queries to the answering source, or a charged late type scan & The inherited atlas and fixed endpoint offsets \\
			\bottomrule
		\end{tabularx}
		\caption{Work charged and records retained by each branch of \cref{alg:process-state}.}
		\label{tab:branch-charges}
	\end{table}
	
	\Needspace{10\baselineskip}
	\subsection{The transfer exponent}\label{subsec:transfer-exponent}
	
	Let \(\delta=2-\rho\) and \(\sigma=\delta/(1+3\delta/2)\).
	To keep the convolution and structural exponents at most
	\(2-\sigma\), require \(\theta\le\theta_c\) and
	\(\theta\le\theta_b\), respectively, where
	\[
	\theta_c=\frac{\delta-\sigma}{2-\delta}
	=\frac{3\sigma^2}{4(1-2\sigma)},\qquad
	\theta_b=\frac{1-\sigma}{2}.
	\]
	Set \(\theta_* =\min(\theta_c,\theta_b)\). The two bounds are
	within \(2-\sigma\) at \(\theta_*\). For geometry, \(\Gamma_q\) is increasing
	in \(\theta\) for each \(q\ge4\). Since \(0<\sigma<2/5\), we have
	\[
	\begin{aligned}
		4q(1-2\sigma)\bigl(\Gamma_q(\theta_c)-\sigma\bigr)
		&= [2-(q+2)\sigma]^2+2[(q-2)^2-3]\sigma^2\ge0,\\
		2q\bigl(\Gamma_q(\theta_b)-\sigma\bigr)
		&=(q-2)^2-(q^2-2q+2)\sigma\\
		&\ge (q-4)(3q-4)/5\ge0.
	\end{aligned}
	\]
	Thus \(\Gamma_q(\theta_*)\ge\sigma\) at every grid level.
	
	Choose a positive rational \(\theta\in[(1-\tau)\theta_*,\theta_*]\),
	where \(\tau=\min(1/2,\epsilon/(2\sigma))\). The convolution and
	structural exponents only decrease. Since \(\Gamma_q\) is affine with
	positive intercept, \(\Gamma_q(\theta)\ge(1-\tau)\Gamma_q(\theta_*)\ge
	\sigma-\epsilon/2\) simultaneously for all \(q\ge4\). Fix
	\(k_{\rm scan}=1+\lceil\sigma/\theta\rceil\), so the scan exponent
	is at most \(2-\sigma\). Use \(0<\lambda\le\epsilon/2\) in
	\cref{thm:transfer}. The resulting exponent is
	\(f(\rho)+\epsilon\), where
	\[
	f(\rho)=2-\sigma=\frac{4(3-\rho)}{8-3\rho}.
	\]
	\Cref{lem:minconv-witness} supplies the witnessed interface from a
	value-only convolution algorithm.
	
	For \(\rho_0=784055881637/400000000000\) from \cref{lem:minconv},
	direct substitution gives \(f(\rho_0)=1.9623885116738\ldots\).
	A separate choice of the internal parameters improves the numerical
	bound. Take \(\theta=1/2400\) and \(k_{\rm scan}=100\). For every
	\(q\ge4\), the arithmetic-geometric mean inequality gives
	\(\Gamma_q(\theta)\ge2\sqrt{\theta(1+2\theta)}-4\theta\). The four
	exponents before losses satisfy
	\[
	\begin{aligned}
		\rho_0(1+\theta)&=1.9609564289692\ldots,\\
		2+4\theta-2\sqrt{\theta(1+2\theta)}&=1.9608248308171\ldots,\\
		1+2\theta&=1.0008333333333\ldots,\\
		2-99\theta&=1.95875.
	\end{aligned}
	\]
	Taking the total loss \(\lambda=1/50000\) leaves the exponent below
	\(1.961\). The maximum-profit reduction below carries the same bound
	to \(\pmx\), completing \cref{thm:main,cor:concrete}.
	
	\subsection{Parameterizing by maximum profit}\label{sec:profit}
	
	Complementing the chosen counts turns a profit target into a capacity. This is the transformation of~\cite[Section~4]{PRW21}, also used in~\cite[Theorem~2]{Bringmann24}.
	
	\begin{lemma}[maximum-profit transfer]\label[lemma]{lem:profit-transfer}
		If Bounded Knapsack on logarithmic-bit integer inputs admits an algorithm that returns an optimal count vector in time \(\Ot(n+\wm^\gamma)\) for a fixed \(\gamma>0\), then it admits one with running time \(\Ot(n+\pmx^\gamma)\) and the same output. The reduction preserves binary multiplicities. If the weight algorithm succeeds with constant probability greater than one half, the profit algorithm can achieve any fixed inverse-polynomial failure probability.
	\end{lemma}
	
	\begin{proof}
		Discard records with nonpositive profit or zero multiplicity and set their output counts to zero. Let \(I\subseteq[n]\) index the retained records. If \(I\) is empty, the zero vector is optimal. Otherwise define
		\[
		S_w=\sum_{i\in I}w_i u_i,
		\qquad
		S_p=\sum_{i\in I}p_i u_i,
		\qquad
		\mathcal X=\prod_{i\in I}\bigl([0,u_i]\cap\Z\bigr).
		\]
		For a target profit \(q\in[0,S_p]\cap\Z\), substitute \(y_i=u_i-x_i\) in the two feasibility inequalities. This gives
		\[
		\begin{aligned}
			&\exists x\in\mathcal X:
			\sum_{i\in I} w_i x_i\le C,
			\quad \sum_{i\in I}p_i x_i\ge q
			\\
			&\hspace{1.5em}\Longleftrightarrow\quad
			\max\left(\sum_{i\in I}w_i y_i
			\ \middle|\ y\in\mathcal X,\
			\sum_{i\in I}p_i y_i\le S_p-q\right)\ge S_w-C.
		\end{aligned}
		\]
		The maximum on the right is Bounded Knapsack on records
		\((p_i,w_i,u_i)_{i\in I}\) with capacity \(S_p-q\).
		Replace each omitted record by \((\pmx,1,0)\).
		The presented instance then has \(n\) records and maximum weight
		\(\pmx\), even if a discarded record supplied the maximum profit.
		Thus the logarithmic-bit promise holds with respect to the
		parameter on the presented input. An attaining vector \(y\) that
		passes the threshold gives the feasible vector \(u-y\) of profit
		at least \(q\) on the retained records.
		
		The predicate is monotone in \(q\). Sorting by profit-to-weight ratio gives an optimal fractional prefix with at most one fractional copy, found by scanning the compressed records. Rounding that copy down gives a feasible profit \(L\) less than one maximum item profit below the fractional optimum. Binary search over the integers from \(L\) through \(\min(S_p,L+\pmx)\) therefore finds the largest feasible \(q\) with a logarithmic number of calls. The witness at this target gives an optimal count vector. The transformed capacities and sums have logarithmic bit length under the input convention, and all auxiliary work takes \(\Ot(n)\) time. For a randomized solver, repeat each call independently and keep the best feasible returned count vector. Amplification and a union bound over the binary search preserve the claimed time and success probability.
	\end{proof}
	
	\begin{example}[a profit threshold as a capacity]
		Take records \((7,2,3)\) and \((4,1,2)\) and capacity \(12\). Their full weight and profit are \(29\) and \(8\). Testing target profit \(3\) asks whether a set of discarded copies can have original weight at least \(17\) while spending at most \(8-3=5\) units of profit. Discarding two copies of the first record and one of the second spends \(5\) and removes weight \(18\), so keeping one of each attains the target. For target \(4\), the discard budget falls to \(4\), for which the maximum removed weight is \(15\). Thus the optimum is \(3\).
	\end{example}
	
	\section{Conclusion}\label{sec:conclusion}
	
	Greedy proximity bounds the residual instance in terms of the
	maximum item weight. Within this kernel, witnessed convolution
	generates exact-weight queries and local linear programs filter
	them by cost. Resilient lattice cores identify the restrictions
	that can be coalesced. Progression reserves give the attainability
	needed for rounding, and a dynamic program on actual reserve labels
	finds the required subset.
	
	The source bounds charge table construction and fractional tests
	to the anchors that generate them, including tests of rejected
	batches. Each lift reduces normalized column volume, reducing
	grid lookup costs until scans of the surviving types are affordable.
	Attaining assignments saved during the recursion recover an
	optimal count vector. Complementing the counts transfers the
	same exponent to the maximum-profit parameter.
	
	\section*{AI Disclosure}
	
	The work began when the author noticed that integer
	\((\min,+)\)-convolution had received a truly subquadratic algorithm,
	while the known bounds for 0--1 and Bounded Knapsack in terms of
	\(n\) and \(\wm\) still had a quadratic weight term. About a year
	earlier, the author had read the near-quadratic algorithms of
	\cite{Jin24,Bringmann24}. Their weight dependence was nearly optimal
	under the \((\min,+)\)-convolution hypothesis. The breakthroughs for
	3SUM and APSP, and their consequences for integer convolution,
	removed the premise of that conditional lower bound~\cite{AV26,CIMY26}.
	This reopened the question of a fixed power saving for small items.
	
	The author drew on the connections among Subset Sum, Knapsack,
	\((\min,+)\)-convolution, and integer programming when directing
	GPT-6 Pro to combine the earlier knapsack algorithms~\cite{Jin24,Bringmann24},
	the new 3SUM techniques~\cite{AV26,CIMY26}, and the geometric and
	additive methods for Subset Sum and integer programming
	in~\cite{CMZFeas26,CMZOpt26}. The author provided guidance,
	encouragement, and emotional support through several rounds of
	intensive work. GPT-6 Pro developed an initial, technically
	involved algorithm for 0--1 Knapsack. Further simplification
	and generalization produced the present bounded algorithm.
	
	The congruence and progression methods of~\cite{Jin24} and the
	explicit lattice techniques of~\cite{CMZFeas26,CMZOpt26} and this
	paper also recall the author's earlier work~\cite{Zho26}.
	The author wrote that manuscript himself. Before submitting it to
	ITCS 2027, he used AI to organize the literature, check the proofs,
	and polish the prose. That paper studies
	convolution through tropical polynomials and proves modular convexity
	theorems with a natural interpolation description. The author views a
	higher-dimensional lattice-convexity theorem as a natural extension
	and sees a connection with the lattice structures used here.
	
	The submission received two scores of 1 (Reject) and one score of 3
	(Weak Accept). Based on the contents of the reports, the author
	suspects that the two Reject reports and their scores came from human
	reviewers with expertise in fine-grained complexity. The report
	assigning a score of 3, which the author suspects was AI-generated,
	spoke highly of the earlier paper's central idea and offered
	constructive suggestions for revision.
	One of the two reviews that scored the submission 1 called for
	an entirely algorithmic presentation, treated the tropical-algebraic
	language as repackaging, and questioned the author's command of
	that language. The author was disappointed to see the language
	that had motivated the work treated as a liability, especially at
	a conference that values conceptual innovation.
	
	The author accepts the criticism in that report and has decided against resubmitting
	the paper, and plans to turn from tropical algebra and lattice convexity
	to the design of faster algorithms. The author hopes this shift will
	make a greater contribution to theoretical computer science. Powerful
	AI systems may soon be able to find faster and more elaborate algorithms
	by combining many techniques, without building the conceptual
	frameworks that could guide their discovery. Readers may spend more
	time understanding those algorithms as their exponents continue to fall.
	
	GPT-6.1 Sol Ultra assisted in preparing this manuscript. The author
	guided its organization, language, and technical presentation.
	The author takes responsibility for the correctness, completeness,
	and integrity of the paper.

	\appendix
	\Needspace{10\baselineskip}
	\section{Proof of the Compressed Proximity Reduction}\label[appendix]{app:kernel}
	
	The input reduction in \cref{sec:model} uses proximity to a greedy
	prefix in the virtual copy order. The proof computes all intervals
	by their endpoints so the reduction can work directly on runs.
	
	\begin{proof}[Proof of \cref{lem:kernel}]
		Sort the records by nonincreasing \(p_i/w_i\), using exact cross
		multiplication and breaking ties by the input order. Within each
		record use its natural copy order. Let \(G\) be the longest
		feasible prefix of this virtual sequence. A scan of the runs
		computes its count vector \(g\), stopping at the first copy that
		would exceed \(C\). Thus \(0\le C-w^\top g\le\wm-1\).
		
		Choose an optimum \(X\) minimizing \(|X\mathbin{\triangle}G|\)
		and write \(A=X\setminus G\), \(D=G\setminus X\). For a set
		of copies \(Y\), let \(w(Y)\) be its total weight.
		Nonempty \(A'\subseteq A\) and \(D'\subseteq D\) of equal
		weight would give an exchange. Replacing \(A'\) by \(D'\)
		preserves feasibility and cannot decrease profit because
		the ratios \(p_i/w_i\) of the copies in \(D'\) are at least
		those of the copies in \(A'\). The exchange would strictly
		decrease the distance to \(G\), a contradiction. At least
		one positive-profit copy is omitted, since the all-copies case
		was handled before the reduction. If \(w(X)\le C-\wm\), an
		omitted copy could be added, contradicting optimality. Thus
		\(w(X)\ge C-\wm+1\).
		
		The initial signed discrepancy \(w(A)-w(D)=w(X)-w(G)\)
		lies in \([-\wm+1,\wm-1]\). In the discrepancy computation,
		remove one remaining copy of \(A\) when the discrepancy is
		positive and one of \(D\) when it is negative. The required
		sign exists, and every update stays in the same interval
		because each weight is at most \(\wm\). A zero discrepancy
		with copies remaining would give an equal-weight exchange
		from the remaining copies. If a discrepancy repeated, the
		copies removed between the two occurrences would give such
		an exchange. Hence the sequence processes all of \(A\cup D\)
		in at most \(2\wm-1\) steps~\cite{PRW21,Bringmann24}.
		
		Within a weight class \(a\), replacing any \(k\) chosen copies
		by the first \(k\) in the inherited order preserves the weight
		and cannot decrease profit. It also maximizes overlap with
		\(G\), which is a prefix within that class. We can therefore
		take \(X\) to be a prefix in each class. If \(N_a\) is the
		number of copies of weight \(a\) and \(g_a\) the number in
		\(G\), the optimum prefix length is within \(2\wm-1\) of
		\(g_a\). Set
		\[
		L_a=\max(0,g_a-2\wm),\qquad R_a=\min(N_a,g_a+2\wm).
		\]
		Fix the first \(L_a\) copies, retain positions \(L_a+1\)
		through \(R_a\), and discard the rest. This preserves the
		optimum just described. A record \(i\) occupying positions
		\(s_i+1,\ldots,s_i+u_i\) has fixed and retained counts given
		by the numbers of integers in the intersections of
		\((s_i,s_i+u_i]\) with \((0,L_a]\) and \((L_a,R_a]\),
		respectively. Compute each count from the interval endpoints.
		
		The retained interval has length at most \(4\wm\). The fixed selection is a subset of \(G\), so
		\[
		C'=C-w^\top f
		=(C-w^\top g)+\sum_{a=1}^{\wm}a(g_a-L_a)
		\le \wm-1+2\wm\sum_{a=1}^{\wm}a.
		\]
		The slack record fills the unused part of \(C'\) with
		weight-one copies at zero profit. Every feasible residual
		selection extends to the equality instance, and deleting
		the slack recovers the residual selection. The scans and
		sorts touch only nonempty runs and weight classes.
	\end{proof}
	
	\Needspace{10\baselineskip}
	\section{Integer Convolution from 3SUM}\label[appendix]{app:minconv}
	
	The numerical bound uses \texttt{Race.bound} at the commit specified
	in~\cite{Wang26}. For each fixed integer \(\kappa\), this theorem gives
	one finite deterministic program for deciding whether three distinct
	positions of an \(N\)-entry list sum to zero, for all
	\(|x_i|\le N^\kappa\). The time exponent is
	\[
	\alpha=\frac{384055881637}{200000000000}=1.920279408185.
	\]
	At any word width of at least \(b_\kappa(\lfloor\log_2 N\rfloor+1)\),
	the program uses \(O(N^\alpha)\) word-RAM steps. Choose this
	logarithmic width. The instruction set uses the constant
	one, modular addition, subtraction and multiplication, indirect
	memory access, a sign branch, and accept and reject instructions.
	A word operation costs \(\operatorname{polylog}(N+2)\) in our
	random-access bit-cost model. For example, a binary trie on word
	addresses implements the initially zero memory by storing only the
	cells accessed, with logarithmically many steps per access.
	The construction below uses this decision interface to report every
	attained index with exponent \(1+\alpha/2\).
	
	\begin{remark}[the distance-product route]
		The APSP and 3SUM breakthroughs already imply a fixed power saving for integer
		\(\MinConv\) via earlier reductions~\cite{BCD14,BIS17,CMWW19,AV26}.
		For example, put \(s=\lceil\sqrt m\rceil\), extend two length-\(m\)
		arrays by \(+\infty\), and form
		\[
		L_{q,r}=A[qs+r],\qquad R_{r,j}=B[j-r]
		\quad(0\le q,r<s,\ -s^2\le j<2m).
		\]
		Writing \(\otimes\) for the min-plus matrix product, we have
		\[
		D[k]=\min_{0\le q<s}(L\otimes R)_{q,k-qs}
		\quad(0\le k<2m-1).
		\]
		The rectangular product splits into \(O(s)\) square products, with
		\(O(m^{3/2})\) additional work. Thus the deterministic distance-product
		exponent \(2.9995\) of~\cite[Theorem~2]{AV26} gives convolution exponent
		\(1.99975\) on polynomially bounded entries. The smaller exponent below
		uses 3SUM decision and includes the dependence on entry bit length.
	\end{remark}
	
	\begin{lemma}[indexed equality from 3SUM decision]
		\label[lemma]{lem:indexed-equality}
		Fix \(1<\alpha<2\). Suppose that, for every fixed magnitude exponent,
		distinct-position 3SUM on \(N\) polynomially bounded integers takes \(O(N^\alpha)\)
		deterministic word-RAM steps on \(O(\log(N+2))\)-bit words.
		The word instructions are simulable in
		\(\operatorname{polylog}(N+2)\) bit cost.
		Given length-\(m\) integer arrays \(a,b\) and targets
		\((v_k)_{k=0}^{2m-2}\), let \(b_0\) bound the input bit length.
		One can report all indices \(k\) satisfying
		\[
		\exists i,j\in\{0,\ldots,m-1\}:\quad i+j=k,\qquad a_i+b_j=v_k.
		\]
		The algorithm uses
		\(\Ot(m^{1+\alpha/2})\operatorname{poly}(b_0+\log(1/\eta))\)
		bit operations. It reports an incorrect set of indices with probability
		at most \(\eta\).
	\end{lemma}
	
	\begin{proof}
		Take \(H=2m+1\) and encode
		\[
		x_i=Ha_i+i,\qquad y_j=Hb_j+j,\qquad z_k=Hv_k+k.
		\]
		Then \(x_i+y_j=z_k\) holds exactly when both required equalities hold.
		First consider inputs of polynomial magnitude in \(m\). We use the
		block-pair reporting schedule of~\cite[Theorem~5]{CMWW19} with equality
		tests. Partition each
		input index set into consecutive blocks of size at most
		\(s=\lceil\sqrt m\rceil\). A block pair \((I,J)\) can contribute only
		indices in \(I+J\), an interval of at most \(2s-1\) indices. Maintain a
		global set of reported indices and let \(T\) be the unreported indices
		in \(I+J\). Test whether any triple \((i,j,k)\in I\times J\times T\)
		satisfies \(x_i+y_j=z_k\). A negative answer finishes the pair. After a positive
		answer, bisect \(T\) with the same decision routine to find one actual
		index \(k\). Report it globally, remove it from \(T\), and test this
		pair again.
		Every pair has at most one final negative test, and there are at most
		\(2m-1\) successful reports over all pairs. Hence the total cost is
		\[
		\Ot\!\left(\left(\left\lceil\frac ms\right\rceil^2+m\right)s^\alpha\right)
		=\Ot(m^{1+\alpha/2}).
		\]
		Constructing the explicit lists costs \(O(s)\) per test and is included.
		This count also covers an input with no successful report.
		
		Each test is reduced to distinct-position 3SUM by the three lists
		\[
		\{16x_i+1\}_{i\in I},\qquad
		\{16y_j+3\}_{j\in J},\qquad
		\{-16z_k-4\}_{k\in T}.
		\]
		The residues modulo \(16\) force one position from each list, and the
		remaining equality is \(x_i+y_j=z_k\). Their total length is \(O(s)\). Pad to \(\Theta(s)\) positions
		with harmless positive entries as below. Since \(m=O(s^2)\), the
		magnitudes are polynomial in the padded length.
		
		For general bit length, reduce \(x_i,y_j,z_k\) modulo a random prime
		\(p\in[L,2L]\), with \(L=\lceil Cm^4(b_0+1)\rceil\). Among the \(O(m^3)\)
		possible differences \(x_i+y_j-z_k\), a nonzero difference has
		\(O((b_0+\log m+1)/\log L)\) prime divisors in this interval.
		After increasing \(C\), a uniform prime preserves all equalities
		with probability at least \(19/20\). For canonical residues,
		include both \(z_k\bmod p\) and \((z_k\bmod p)+p\) in the target list,
		retaining their common original index. The list for one block still has
		\(O(s)\) entries. Padding each decision instance to
		\(N=\Theta(s(b_0+\log m+1))\) positions makes its magnitude polynomial
		in its size. Indeed \(m=O(N^2)\) and \(b_0+1=O(N)\), so
		\(p=O(N^9)\). Choose the padding integers to equal one plus four
		times the largest genuine magnitude. This preserves the decision
		answer and keeps every tagged and padding integer in \(O(N^9)\).
		One fixed \(\kappa\) therefore covers all padded calls.
		The padding changes only the polynomial bit factor. The prime number
		theorem gives density \(\Omega(1/\log L)\) in \([L,2L]\).
		Generate uniform candidates with rejection and use the deterministic
		primality test of~\cite{AKS04}. Cap the combined sampling and testing at
		\(O(\log L)\) trials with a sufficiently large constant, so the chance
		of exhausting the cap is at most \(1/20\). A capped repetition reports
		no hits. Conditional on finding a prime, it is uniform. Each
		repetition uses polynomial bit work per
		sampling step and is correct at all indices with probability at least
		\(9/10\). Take the majority answer at every index over
		\(O(\log(1/\eta)+1)\) independent repetitions. The event that a majority
		of repetitions are correct simultaneously gives the required error
		bound, including for adaptive tests.
	\end{proof}
	
	\begin{proof}[Proof of \cref{lem:minconv}]
		Pad the shorter array to \(m=\max(\ell_A,\ell_B)\) with \(+\infty\).
		Only indices \(0\le k<\ell_A+\ell_B-1\) belong to the original output.
		For nonnegative finite arrays of equal length, expose their binary digits
		from most to least significant. If \(\varepsilon_i,\delta_j\in\{0,1\}\)
		are the next bits, the prefix arrays \(a,b\) become
		\[
		a_i'=2a_i+\varepsilon_i,\qquad b_j'=2b_j+\delta_j.
		\]
		Write \(D\) for the convolution of \(a,b\). The new convolution value
		at \(k\) belongs to
		\(\{2D[k],2D[k]+1,2D[k]+2\}\). Every pair gives the lower bound,
		and a previous minimizing pair gives the upper bound. Apply
		\cref{lem:indexed-equality} to targets \(v_k=2D[k]+t\), for
		\(t=0,1\). The smallest successful value is the new minimum, with
		\(2D[k]+2\) when both tests fail. Starting with zero prefixes,
		one round per input bit recovers every value. Fresh randomness and a union bound over
		the first incorrect round give the requested joint success probability.
		For signed inputs, shift each finite entry by \(U\), where \(U\) bounds
		its absolute value. Replace infinite entries by a sentinel exceeding
		\(4U\). A finite pair has shifted sum at most \(4U\), so finite and
		infinite outputs are separated exactly. Undo the shift after convolution.
		Applying \cref{lem:minconv-witness} gives witnessed convolution with
		exponent \(1+\alpha/2=\rho_0\). The 3SUM decision program is
		deterministic, and the random prime reductions supply the error
		probability.
	\end{proof}

\end{document}